\documentclass[aps,prd,a4paper,twocolumn,amsmath,showpacs,superscriptaddress,nofootinbib,preprintnumbers]{revtex4-1}

\usepackage{verbatim}
\usepackage[T1]{fontenc}
\usepackage[utf8]{inputenc}
\usepackage[american]{babel}
\usepackage{epsfig}
\usepackage{graphicx}
\usepackage{booktabs}
\usepackage{multirow}
\usepackage{dcolumn}
\usepackage{amsmath}
\usepackage{mathtools}
\usepackage{amsfonts}
\usepackage{amssymb}
\usepackage{epstopdf}
\usepackage{bm}
\usepackage{siunitx}
\usepackage{braket}
\usepackage{enumitem}
\usepackage{soul}
\usepackage[table]{xcolor}
\usepackage{color}
\usepackage{transparent}
\usepackage{comment}
\usepackage{pifont}
\usepackage{soul}

\definecolor{navyblue}{rgb}{0.0, 0.0, 0.5}
\definecolor{royalblue}{rgb}{0.25, 0.41, 0.88}
\definecolor{cadmiumgreen}{rgb}{0.0, 0.42, 0.24}
\definecolor{blue-violet}{rgb}{0.54, 0.17, 0.89}
\definecolor{darkviolet}{rgb}{0.58, 0.0, 0.83}
\definecolor{orange(colorwheel)}{rgb}{1.0, 0.5, 0.0}
\definecolor{burgundy}{rgb}{0.5, 0.0, 0.13}

\usepackage{hyperref}
\hypersetup{
    colorlinks=true, 
    linkcolor=royalblue, 
    citecolor=magenta}

\newcommand\eq[1]{Eq.~\eqref{eq:#1}}
\newcommand\sect[1]{Sec.~\ref{sec:#1}}

\usepackage{booktabs}
\usepackage{multirow}
\usepackage{dcolumn}
\usepackage{colortbl}

\begin{document}

\title{Illustrating the consequences of a misuse of $\sigma_8$ in cosmology}

\author{Matteo Forconi}
\email{matteo.forconi@fe.infn.it}
\affiliation{Physics Department and INFN sezione di Ferrara,  Università degli Studi di Ferrara, via Saragat 1, I-44122 Ferrara, Italy}
\author{Arianna Favale}
\email{afavale@roma2.infn.it}
\affiliation{Dipartimento di Fisica and INFN Sezione di Roma 2, Università di Roma Tor Vergata, via della Ricerca Scientifica 1, I-00133 Rome, Italy} 
\affiliation{Departament de Física Quàntica i Astrofísica (FQA), Universitat de Barcelona (UB), c. Martí i Franqués, 1, 08028 Barcelona, Catalonia, Spain} 
\affiliation{Institut de Ciències del Cosmos (ICCUB), Universitat de Barcelona (UB), c. Martí i Franqués, 1, 08028 Barcelona, Catalonia, Spain} 

\author{Adrià Gómez-Valent}
\email{agomezvalent@icc.ub.edu}
\affiliation{Departament de Física Quàntica i Astrofísica (FQA), Universitat de Barcelona (UB), c. Martí i Franqués, 1, 08028 Barcelona, Catalonia, Spain} 
\affiliation{Institut de Ciències del Cosmos (ICCUB), Universitat de Barcelona (UB), c. Martí i Franqués, 1, 08028 Barcelona, Catalonia, Spain} 


\preprint{}
\begin{abstract}
The parameter $\sigma_8$, which represents the root-mean-square (rms) mass fluctuations on a scale of $R_8=8h^{-1}$ Mpc (where $h$ is the reduced Hubble parameter), is commonly used to quantify the amplitude of matter fluctuations at linear cosmological scales. However, the dependence of $R_8$ on $h$ complicates direct comparisons of $\sigma_8$ values obtained under different assumptions about $H_0$, since $\sigma_8$ in such cases characterizes the amount of structure at different physical scales. This issue arises both when comparing $\sigma_8$ values from fitting analyses of cosmological models with differing $H_0$ posteriors, and when contrasting constraints from galaxy clustering experiments that employ different priors on the Hubble parameter. As first noted by Ariel G. Sánchez in Phys. Rev. D 102, 123511 (2020), quantifying the growth tension using $\sigma_8$ can introduce substantial biases and couple the growth and Hubble tensions in an intricate and uncontrolled way. To address these challenges, Sánchez proposed an alternative parameter, $\sigma_{12}$, defined as the rms mass fluctuations at a scale of $12$ Mpc, which is independent of $h$. Although Sánchez's work was published five years ago and other authors have since highlighted the limitations of $\sigma_8$, much of the cosmological community -- including large collaborations -- continues to rely on this parameter rather than adopting $\sigma_{12}$, seemingly due only to historical considerations. In this work, we illustrate the biases introduced by the use of $\sigma_8$ through some clear examples, aiming to motivate the community to transition from $\sigma_8$ to $\sigma_{12}$.  We show that the bias found in models with large values of $H_0$ is more prominent. This artificially complicates the search for a model that can efficiently resolve the Hubble tension without exacerbating the growth tension inferred from galaxy clustering measurements. We argue that the worsening of the growth tension in these models is much less pronounced than previously thought or may even be nonexistent.   
\end{abstract}
\maketitle

\section{Introduction}\label{sec:Intro}

One of the main goals of cosmology is to understand the physical processes driving the formation and evolution of pressure and energy fluctuations of the various cosmic species, from their origin in the inflationary epoch to the present day. These fluctuations are imprinted on the cosmic microwave background (CMB) maps \cite{WMAP:2012fli,Planck:2018vyg,ACT:2020gnv,SPTpol:2025kpo}, offering a snapshot of the existing inhomogeneities at the last scattering surface, and they also shape the large-scale structure (LSS) of the universe, which is measured by galaxy and weak lensing (WL) surveys, see, e.g., \cite{BOSS:2016wmc,Heymans:2020gsg,DES:2021wwk,DESI:2024hhd}, and also through the impact of weak lensing and the integrated Sachs-Wolfe effect on the CMB. The study of these data is a precious source of information and allows us to put constraints on cosmological models, usually breaking degeneracies in their parameter spaces when combined with those obtained with standard rulers, candles and clocks \cite{Huterer:2022dds}.

The amplitude of density perturbations on a scale $R$ is typically quantified using the root-mean-square (rms) of mass fluctuations, 

\begin{equation}\label{eq:def_sig}
\sigma_R^2(z)\equiv <[\delta_R(\vec{r},z)]^2>\,,
\end{equation}
where the average is performed over the volume of the entire universe at a given $z$ and

\begin{equation}
\delta_R(\vec{r},z)=\int d^3r^\prime\,\delta(\vec{r}+\vec{r}^\prime,z)W_R(r^\prime)\,
\end{equation}
is the matter density contrast $\delta=\delta\rho/\bar{\rho}$ smoothed with a normalized spherical top-hat window function of radius $R$, $W_R(r)$. Eq. \eqref{eq:def_sig} can be also written as follows \cite{Kolb:1990vq,Liddle:2000cg},

\begin{equation}\label{eq:sigmaR}
\sigma_R^2 (z)=\frac{1}{2\pi^2}\int dk\, k^2 P(k,z)W^2(kR)\,,
\end{equation}
with $P(k,z)$ the matter power spectrum and 

\begin{equation}\label{eq:WF}
W(kR) =\frac{3}{k^2R^2}\left(\frac{\sin(kR)}{kR}-\cos(kR)\right)
\end{equation}
the Fourier transform of $W_R(r)$. Eq. \eqref{eq:sigmaR} is usually employed to characterize the LSS fluctuations at linear cosmological scales. For historical reasons, the scale $R$ is usually taken to be $R_8=8h^{-1}$ Mpc, where $h=H_0/(100\zeta)$ is the reduced (dimensionless) Hubble parameter, with $\zeta=1\,{\rm km/s/Mpc}$. In this case, the function \eqref{eq:sigmaR} is simply known as $\sigma_8(z)$. However, serious concerns about the use of this quantity have been raised by several authors in the last five years, starting with the work by A.G. Sánchez \cite{Sanchez:2020vvb}. He argued that, due to the dependence of the scale $R_8$ on $h$, the use of $\sigma_8$ can introduce important biases in our interpretation of LSS measurements and the results for $\sigma_8$ obtained from fitting analyses of cosmological models. The processing of galaxy clustering data within the framework of a specific model, while considering different priors on $H_0$, essentially yields values of $\sigma_8$ that describe the amplitude of fluctuations at varying scales. This complicates the direct comparison of the various measurements. On the other hand, fitting analyses of two models with dissimilar posterior distributions of $H_0$ produce derived values of $\sigma_8$ that cannot be directly compared due to exactly the same reason. Therefore, interpreting the values of $\sigma_8$ in these cases as corresponding to the same physical scale is completely inappropriate and can lead to significant errors in our conclusions. In \cite{Sanchez:2020vvb}, Sánchez suggested to solve this issue by replacing $\sigma_8$ with the new parameter $\sigma_{12}$, defined as the rms of mass fluctuations at the linear scale of $R_{12}=12$ Mpc, which does not depend on $h$ and coincides with $R_8$ only if $h=h_*$, with $h_*=8/12\simeq 0.667$. The comparison of $\sigma_{12}$ values obtained from different galaxy clustering experiments or from fitting analyses of models with different posteriors of $h$ can be carried out without problems, since in this case there is no mixing of scales and, therefore, the differences in $\sigma_{12}$ can only arise from changes in the experimental results or the models. Although the debate in the last few years had predominantly clustered around using $\sigma_8$ and derived quantities as $S_8=\sigma_8(\Omega_m/0.3)^{0.5}$ (see e.g.~\cite{Gomez-Valent:2018nib,SolaPeracaula:2020vpg,Heymans:2020gsg,DiValentino:2020vvd,Marra:2021fvf,Hall:2021qjk,Abdalla:2022yfr,Pedreira:2023qqt,Vagnozzi:2023nrq,Sakr:2023bms,Secco:2022kqg,DESI:2024hhd,Simon:2024jmu,RoyChoudhury:2024wri,Liu:2024vlt,Toda:2024uff}), some authors have already made use of the parameter $\sigma_{12}$ in several observational and theoretical works~\cite{Gomez-Valent:2021cbe,eBOSS:2021poy,Sanchez:2021plj,Gomez-Valent:2022hkb,Gomez-Valent:2022bku,Secco:2022kqg,Semenaite:2022unt,Garcia-Garcia:2024gzy,Gomez-Valent:2023hov,Gomez-Valent:2024tdb,Esposito:2024qlo,Gomez-Valent:2024ejh}. Note that the bias under discussion does not affect weak lensing measurements, as the constraints on $S_8$ derived from them are nearly insensitive to the prior on $h$ \cite{Hall:2021qjk}, and it has been already demonstrated that $S_8$ is an optimal parameter for this cosmological probe \cite{Secco:2022kqg,Garcia-Garcia:2024gzy}. However, the WL constraints on $S_8$ should not be interpreted as applying at a fixed physical scale, but rather at a scale of $8/h$ Mpc, which inherently depends on the prior assumed for $h$. Therefore, the physical implications of WL constraints on the amplitude of matter fluctuations can only be meaningfully extracted once the prior on $h$ is specified, as this determines the scale at which the constraints apply. 

While the current standard model of cosmology proficiently explains a wide range of observations at both the background and linear perturbations level, it has faced limitations in addressing certain anomalies and tensions over the past decade \cite{Perivolaropoulos:2021jda,Aluri:2022hzs}, further motivating the exploration of physics beyond the standard paradigm \cite{Abdalla:2022yfr}. The tension between the value of the Hubble parameter measured by SH0ES \cite{Riess:2021jrx} and the CMB value obtained under the assumption of $\Lambda$CDM \cite{Planck:2018vyg} currently stands as the most statistically significant one, already reaching the $\sim 5\sigma$ c.l.\footnote{See, however, \cite{Freedman:2021ahq,Perivolaropoulos:2022khd,Freedman:2023jcz,Perivolaropoulos:2023iqj,Freedman:2024eph,Wojtak:2024mgg,Gall:2024oyl,Sharma:2025ucg} for other distance ladder measurements of $H_0$ in lesser or no tension with $\Lambda$CDM and some critical discussions about the SH0ES measurements \cite{Efstathiou:2020wxn,Perivolaropoulos:2024yxv}.}, but there are other long-standing issues afflicting the $\Lambda$CDM model that also deserve our attention. Some examples include: the CMB anomalies \cite{WMAP:2012fli,Schwarz:2015cma,Planck:2018vyg,Handley:2019tkm,DiValentino:2019qzk,Minami:2020odp,Aluri:2022hzs,Komatsu:2022nvu,Galloni:2022rgg,Diego-Palazuelos:2022dsq,Yeung:2022smn,Giare:2023xoc,Galloni:2023pie,Jung:2024slj,COMPACT:2024cud}; cosmic dipoles in quasars and radio sources \cite{Secrest:2020has,Aluri:2022hzs,Domenech:2022mvt,Mittal:2023xub,Wagenveld:2023kvi,Oayda:2024hnu}; the unexpected large population of extremely massive galaxies at large redshifts $z\gtrsim 5-10$ measured by the James Webb Space Telescope (JWST) \cite{Labbe:2022ahb,Menci:2020ybl,Menci:2022wia,Forconi:2023izg,Forconi:2023hsj,Menci:2024rbq,Menci:2024hop}; and the growth tension \cite{Macaulay:2013swa,Joudaki:2017zdt,Gomez-Valent:2017idt,Nesseris:2017vor,Gomez-Valent:2018nib,Benisty:2020kdt,Wright:2020ppw,Nunes:2021ipq,Nguyen:2023fip,Adil:2023jtu,Sailer:2024coh,Toda:2024fgv,Akarsu:2024hsu,Artis:2024zag,ACT:2024okh,ACT:2024nrz}. Interestingly, most of them impact the description of LSS in one way or another. The growth tension arises from a $2-3\sigma$ mismatch between the Planck best-fit values of $\sigma_8$ and $S_8$ obtained under the assumption of the $\Lambda$CDM model and some LSS measurements extracted from the analysis of redshift space-distortions (RSD) and weak gravitational lensing \cite{DiValentino:2020vvd}\footnote{For some recent results finding no hints of growth tension, see \cite{DESI:2024hhd,Wright:2025xka}.}. Clearly, the discussion on the validity of the use of $\sigma_8$ can have a substantial impact on these matters, also because it might intertwine the Hubble and growth tensions in a non-trivial way, at least when the latter is derived from galaxy clustering measurements. It is therefore of utmost importance to remove this source of potential bias. The solution is actually very simple and consists of replacing $\sigma_8$ with another quantity that can be used to describe the amplitude of matter fluctuations at linear scales, but is independent of $h$. $\sigma_{12}$ is a very good candidate \cite{Sanchez:2020vvb}.

Nevertheless, we believe that, despite the efforts of some authors, the importance of addressing this bias still remains underappreciated within the cosmological community. Relying on the use of $\sigma_8$ solely for historical reasons is no longer justifiable. We devote this paper to improve the situation by means of a series of examples that show the impact of using $\sigma_{8}$ in the light of current data and in the context of several cosmological models, some of them favoring the range of $H_0$ values preferred by SH0ES and others more in accordance with the small estimates obtained in $\Lambda$CDM. 

We structure this work as follows. In Sec. \ref{sec:sec2} we first illustrate the bias introduced by $\sigma_8$ with a very simple but powerful example based on $\Lambda$CDM. Then, we show how it affects the constraints extracted from current data on peculiar velocity and redshift-space distortions. This section will help the reader understand the extent to which $\sigma_8$ can obscure the physical interpretation of results in cosmological studies. We also provide an approximate formula for the bias as a function of the Hubble parameter, which we derive semi-analytically. This section sets the stage for the more involved analysis of Sec. \ref{sec:MR}, in which we study the impact of the bias in the light of current data on CMB, baryon acoustic oscillations (BAO) and supernovae of Type Ia (SNIa), and on different models, including $\Lambda$CDM, late-time dynamical dark energy, some extensions of the standard model with massive neutrinos and/or additional ultra-relativistic species, and early dark energy. All these models are interesting because they can leave a non-negligible imprint on the LSS and some of them can even help to bring the Hubble parameter closer to the SH0ES value. This is why we deem it important to analyze and quantify the bias introduced by $\sigma_8$ in these scenarios.  After outlining the data and methodology used to constrain the various cosmological models, we discuss our results. Finally, in Sec. \ref{sec:conclusions}, we summarize our conclusions. 

\begin{figure*}[htp]
	\centering
	\includegraphics[width=\textwidth]{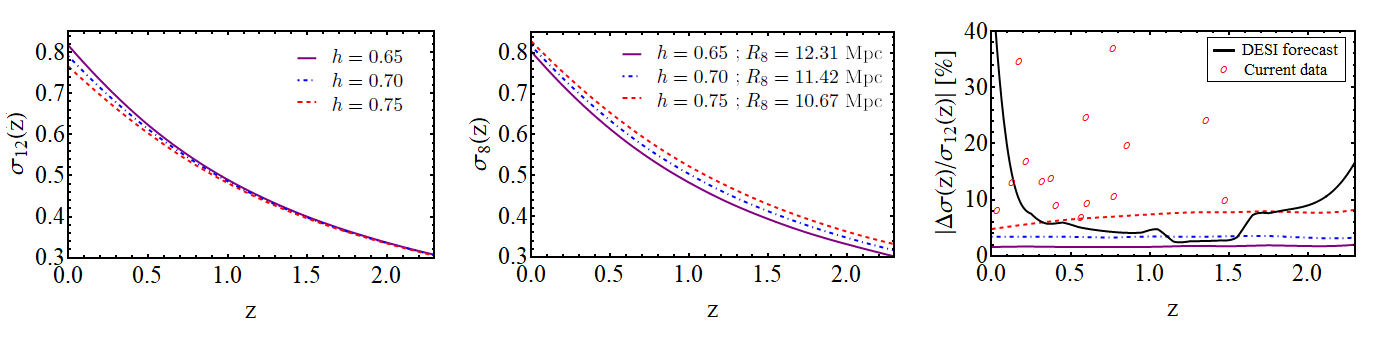}	\caption{{\it Left plot:} Curves of $\sigma_{12}(z)$ obtained in $\Lambda$CDM using the best-fit values of the parameters $\omega_b$, $\omega_{cdm}$, $n_s$ and $A_s$ from the Planck 2018 TT,TE,EE+lowE+lensing analysis \cite{Planck:2018vyg} and considering three different values of $H_0$. Varying the Hubble parameter is in this case equivalent to modifying the value of the cosmological constant. The parameters controlling the primordial power spectrum (and, hence, the initial conditions for the perturbations) as well as the matter and radiation energy densities are the same in the three scenarios. This is why at high redshifts $\sigma_{12}(z)$ tend to the same function, regardless of the value of $H_0$; {\it Middle plot:} The same, but for $\sigma_8(z)$. The separation between the curves at high redshifts is not due to a physical difference in the amplitude of the matter power spectrum, which is the same in all cases, but to the change in the value of the scale $R_8$ that is induced by the shift in $H_0$. As it is shown in the left plot, at small redshifts we expect to have a larger suppression of structure in the universe for larger values of $H_0$ (or, equivalently, for larger values of $\Lambda$). However, $\sigma_8(z)$ exhibits the opposite behavior. This demonstrates the difficulty of interpreting the results given in terms of $\sigma_8$, as they mix real physical effects with spurious effects caused by changes in the scale $R_8$; {\it Right plot:} Relative difference $\Delta\sigma(z)/\sigma_{12}\equiv \sigma_8(z)/\sigma_{12}(z)-1$ for the three scenarios under consideration, which coincides with the relative difference between $f(z)\sigma_8(z)$ and $f(z)\sigma_{12}(z)$. We compare them with the relative uncertainties of current RSD data (red circles) \cite{Avila:2021dqv,Said:2020epb,Simpson:2015yfa,Blake:2011rj,Blake:2013nif,Gil-Marin:2016wya,Mohammad:2018mdy,Song:2008qt,Guzzo:2008ac,Okumura:2015lvp,eBOSS:2020gbb} and mock RSD data from the Dark Energy Spectroscopic Instrument (DESI) \cite{DESI:2023dwi}, in black. Some current data points already exhibit a sensitivity comparable to the differences between $\sigma_{12}$ and $\sigma_8$ and this issue will clearly become even more critical with the advent of new data.}
	\label{fig:plotIntro}
\end{figure*}


\section{The bias of $\sigma_8$}\label{sec:sec2}

{\it An example to understand the impact of the bias.} The problem described in the Introduction can be understood very easily with a trivial example, by comparing the shapes of $\sigma_{12}(z)$ and $\sigma_8$(z) obtained by using different values of the Hubble parameter (or, equivalently, different values of the cosmological constant) in $\Lambda$CDM, while fixing the primordial power spectrum through the parameters $A_s$ and $n_s$, as well as the current baryon and cold dark matter (CDM) energy densities through the reduced density parameters $\omega_{b}$ and $\omega_{cdm}$, which in turn fix their entire cosmic evolution. This is done in the two leftmost plots of Fig. \ref{fig:plotIntro}. Regardless of the value of $H_0$ that we consider in this exercise, the cosmic history in the radiation- and matter-dominated universe is exactly the same in all cases, since the cosmological constant, which is the only differing piece among the various models, only plays a role at $z\lesssim \mathcal{O}(1)$. The energy densities of the other species evolve exactly in the same way, and the initial conditions for the perturbations are also identical. Hence, the amplitude of fluctuations in these epochs of the cosmic expansion are indistinguishable. However, this is not what $\sigma_{8}(z)$ tells us, see the central plot of Fig. \ref{fig:plotIntro}. There is a non-null vertical shift between the various curves even at large redshifts, which is fully counterintuitive and, in fact, unphysical. It is caused by the difference in the scale $R_8$ induced by the change of $h$. In addition, the plot of $\sigma_8(z)$ shows us that the larger is $h$ (and, hence, the larger is $\Lambda$) the larger is the amplitude of matter fluctuations. This result once again challenges our most basic intuition. It is not a manifestation of a real enhancement of the matter perturbations for increasing values of the cosmological constant, which of course does not occur, but a direct consequence of the shift of $R_8$. In that plot we are just comparing the amplitude of fluctuations at different scales, see the values of $R_8$ displayed in the legend. Consequently, accurately interpreting the results through this plot is challenging, if not impossible. It can lead to a deep misunderstanding of the true physical meaning of the results. The left plot of Fig. \ref{fig:plotIntro}, in contrast, demonstrates that $\sigma_{12}(z)$ is actually behaving exactly as we would expect, i.e., the curves start to depart from each other at $z\lesssim  1$, when $\Lambda$ starts to be sizable compared to non-relativistic matter, and larger values of $h$ suppress $\sigma_{12}$ in the universe's accelerated phase. The interpretation of the results displayed for $\sigma_{12}(z)$ in that plot is crystal-clear because now we are comparing the amplitude of fluctuations at the same physical scale.

It is also very illustrative to compute the relative difference between the curves of $\sigma_8(z)$ and $\sigma_{12}(z)$ obtained for a given value of $H_0$, using the setup explained above. This is equivalent to computing the relative difference between the RSD observables $f(z)\sigma_8(z)$ and $f(z)\sigma_{12}(z)$, since the value of the growth rate $f(z)=d\ln\delta/d\ln a$ cancels in the ratio. We show the results in the rightmost plot of Fig. \ref{fig:plotIntro}. Obviously, such a difference grows in absolute value with $|R_8-R_{12}|$. The interesting point is that the relative difference $\Delta\sigma(z)/\sigma_{12}(z)$ obtained for $h=0.75$, which is a value pretty close to the SH0ES measurement, is already comparable to the sensitivity of some current RSD data at $1\sigma$ c.l., and the situation will become even more pressing in the future, with the advent of data from Euclid \cite{EUCLID:2011zbd,EuclidTheoryWorkingGroup:2012gxx} and the next data releases from DESI. In fact, using the forecasted final-year DESI data \cite{DESI:2023dwi}, we find that they will be very sensitive to the differences between $\sigma_8$ and $\sigma_{12}$ even for much smaller values of $h$, lying below the SH0ES measurement, $h\lesssim 0.7$, and in a very broad redshift range which might be as high as $z\sim 2$. Current CMB data, on the other hand, already constrain $\sigma_8$ at the $\lesssim 1-2\%$ level \cite{Planck:2018vyg,ACT:2020gnv}. Thus, the comparison of $\sigma_8$ values from models fitted to CMB data and with substantial differences in the posteriors of $H_0$ can already lead to statistically significant biases, as we will  explicitly see in Sec. \ref{sec:MR}.  

The confusing behavior of $\sigma_8(z)$ can be derived and understood semi-analytically, Taylor-expanding Eq. \eqref{eq:WF} around $R_{12}$ and substituting the result into Eq. \eqref{eq:sigmaR}, thereby establishing a simple relation between $\sigma_8(z)$ and $\sigma_{12}(z)$. The Taylor expansion of Eq. \eqref{eq:WF} at leading order in the difference $\delta h = h-h_*$ reads, 

\begin{align}
W^2(&kR_8)=  W^2(kR_{12})\\& +2W(kR_{12})\left[\frac{\partial W(k R_8)}{\partial R_8}\frac{\partial R_8}{\partial h}\right]_{h=h_*}\delta h+\mathcal{O}(\delta h^2)\,.\nonumber
\end{align}
Plugging this expression into Eq. \eqref{eq:sigmaR} we find, after some algebra,

\begin{equation}\label{eq:eq6}
\sigma_8^2(z)=\sigma_{12}^2(z)[1+9\delta h]-9\delta hI(z)+\mathcal{O}(\delta h ^2)\,,
\end{equation}
with

\begin{equation}
I(z)= \frac{1}{2\pi^2}\int dk\, k^2\,P(k,z)\,W(kR_{12})\left[\frac{\sin(kR_{12})}{kR_{12}}\right]\,.
\end{equation}
We would also like to express the integral $I(z)$ in terms of $\sigma_{12}^2(z)$ to simplify Eq. \eqref{eq:eq6}. Unfortunately, this cannot be done analytically. Nevertheless, for models like $\Lambda$CDM, where the time dependence of the density contrast at the scales of interest during the matter-dominated epoch and beyond can be expressed as

\begin{equation}\label{eq:D}
\delta(k,z) = D_+(z)\delta(k,0)\,,
\end{equation}
one finds numerically that \footnote{Here, $D_+(z)$ represents the growing mode of the solution to the density contrast equation, with $D_+(0)=1$ \cite{Liddle:2000cg}.} 

\begin{equation}
\frac{I(z)}{\sigma_{12}^2(z)}= A\,,
\end{equation}
with $A\simeq 0.77(2)$\footnote{Strictly speaking, $A$ is a function of the parameters of the model and the model itself, but we have checked that it is actually only mildly sensitive to the details of the model, as long as it has a reasonable shape of $P(k,z)$. The values of $A$ are found to be in all cases in the approximate range $A\in (0.75-0.79)$.}. The time-dependence carried by $D_+(z)$ cancels in the ratio. This result allows us to write the relation \eqref{eq:eq6} in the following very compact form,

\begin{equation}\label{eq:bias}
\sigma_8(z)\approx\sigma_{12}(z)[1+1.035\,\delta h]+\mathcal{O}(\delta h^2)\,,
\end{equation}
which is key to understand the results displayed in Fig. \ref{fig:plotIntro}. In the matter-dominated era $\sigma_{12}(z)$ is the same for all the $\Lambda$CDM models under study, so the bias can be estimated using the formula $(\sigma_8(z)-\sigma_{12}(z))/\sigma_{12}(z)=1.035\delta h$, which is independent of the redshift. We see, for instance, that if $h=0.75$, the bias is of order $\sim 8-9\%$. This number is pretty close to the one shown in the rightmost plot of Fig. \ref{fig:plotIntro}. We also see that there is an unphysical shift which increases for larger values of $|\delta h|$, even if the initial conditions and all the relevant energy densities are the same at large redshifts. Moreover, formula \eqref{eq:bias} tells us that the shift in $\sigma_8(z)$ induced by $\delta h\ne 0$ is present at all redshifts. Even if $\sigma_{12}$ decreases for larger values of $h$ at late times, this decrease is much less pronounced than the positive shift caused by the change in the scale $R_8$. This gives rise to the awkward phenomenon observed in the central plot and already discussed above: for larger values of $h$ (or, equivalently, of $\Lambda$, with the other parameters fixed), $\sigma_8$ points to a larger amount of structure. This result, of course, makes no sense and highlights the challenges of extracting meaningful interpretations from the analysis of $\sigma_8(z)$.
\newline
\newline
{\it Bias in the constraints extracted from current data on peculiar velocity and redshift-space distortions.} Peculiar velocities introduce spurious anisotropies in tracer maps giving rise to the fingers of God and Kaiser effects \cite{Jackson:1972,Kaiser:1987qv}. The analysis of these phenomena allows galaxy surveys to measure the quantity $f(z)\sigma_R(z)$, with $f(z)=d\ln\delta/d\ln a$ the growth rate, see, e.g., \cite{Gil-Marin:2016wya}. This measurement is performed under the assumption of a $\Lambda$CDM fiducial model with fixed parameters, typically using a value of $H_0$ close to the Planck/$\Lambda$CDM best-fit value \cite{Planck:2018vyg}. Galaxy surveys report constraints on the quantity $f(z)\sigma_8(z)$. However, for the fiducial value of $h$ employed in these analyses, we have $R_8\approx R_{12}$, so one could treat these data as if they were data on the quantity $f(z)\sigma_{12}(z)$ in first approximation. Indeed, doing this might be much more appropriate, since the RSD measurements encapsulate the clustering information associated with some particular scale close to $R_{12}$ and, therefore, one cannot treat them as if they characterized the amplitude of the power spectrum at different scales dependent on $h$. This is what is actually done in practice when the RSD and peculiar velocity data are interpreted as data on $f(z)\sigma_8(z)$ and employed to constrain non-standard cosmological models with values of $h$ departing significantly from the Planck/$\Lambda$CDM best-fit value\footnote{Note that the bias discussed here is, in principle, unrelated to the model dependence of the RSD measurements, which arises from the choice of a fiducial cosmology. The bias introduced by using $\sigma_8$ can be eliminated - or at least mitigated - by replacing it with $\sigma_{12}$, although the issue of model dependence may persist. See Refs. \cite{Amendola:2022vte,Schirra:2024rjq} for a possible way of solving the model-dependence problem in galaxy clustering studies.}. 

We illustrate now the impact of this bias using real data from several galaxy surveys \cite{Guzzo:2008ac,Song:2008qt,Blake:2011rj,Blake:2013nif,Simpson:2015yfa,Gil-Marin:2016wya,eBOSS:2020gbb,Said:2020epb,Avila:2021dqv,Mohammad:2018mdy,Okumura:2015lvp}, collected, e.g., in Table I of \cite{Toda:2024fgv}. We use the parameterization of the growth rate introduced in \cite{Linder:2005in},

\begin{equation}\label{eq:f}
f(a)=\Omega_m^\gamma(a)\,,
\end{equation}
where $\gamma$ is the so-called growth index, and is a constant. For $\Lambda$CDM, it is well-known that $\gamma_\Lambda\simeq 0.55$ \cite{Linder:2005in}, but it can differ from this value in alternative cosmologies, see, e.g., \cite{Basilakos:2015vra}. This parameterization of the growth rate works very well for a large variety of dark energy (DE) models \cite{Linder:2005in,Cortes:2024yon}. Using Eq. \eqref{eq:f}, we find that the late-time evolution of the matter density contrast takes the form, 

\begin{equation}
D_+(a)=\exp{\left[\int_{1}^a\frac{da^\prime}{a^\prime}\Omega_m^\gamma(a^\prime)\right]}\,,
\end{equation}
with $a_0=1$ the scale factor today. Hence, the RSD observable can be expressed as follows,

\begin{equation}
f(z)\sigma_{R}(z)=\Omega_m^\gamma(z)\sigma_R^0D_+(z)\,.
\end{equation}
where $\sigma_R^0\equiv \sigma_R(z=0)$ and $z=a^{-1}-1$. Under the assumption of a background cosmology, RSD measurements provide constraints on 
$\gamma$ and $\sigma_R^0$. The former governs the evolution of perturbations with cosmic expansion, while the latter determines the present-day amplitude of the power spectrum at the scale $R$. However, how should we interpret RSD measurements from galaxy surveys? Do they truly constrain $\sigma_{8}$, or do they instead characterize the power spectrum amplitude at a specific scale, close to 12 Mpc? Although $\sigma_8\simeq\sigma_{12}$ in the fiducial models commonly used to process survey data, this correspondence does not hold in general. In fact, a measurement of $\sigma_8$ is not a measurement of the amplitude of fluctuations at a concrete scale, but a continuous collection of constraints at different scales $R_8(h)$, one for each possible value of $h$, which is completely unrealistic. In other words, different values of $h$ lead to the same constraint on the amplitude of fluctuations, but linked to different scales. Using RSD data obtained under a fiducial model with $h \simeq h_*=8/12$ to constrain models with significantly different values of $h$ can lead to biased constraints on the amplitude of fluctuations if we mistakenly assume that the RSD measurements constrain $\sigma_8^0$ rather than $\sigma_R^0$ at $R\simeq R_{12}$. 

Let us illustrate this in detail with another example, in which we extract constraints on $\gamma$ and $\sigma_{12}^0$ from the RSD and peculiar velocity data \cite{Guzzo:2008ac,Song:2008qt,Blake:2011rj,Blake:2013nif,Simpson:2015yfa,Gil-Marin:2016wya,eBOSS:2020gbb,Said:2020epb,Avila:2021dqv,Mohammad:2018mdy,Okumura:2015lvp} assuming that the correct background cosmology is $\Lambda$CDM and, hence, that 

\begin{equation}
\Omega_m(z)= \frac{\Omega_m^0(1+z)^3}{1+\Omega_m^0\left[(1+z)^3-1\right]}\,.
\end{equation}
We do so under two different assumptions: 

\begin{figure}[t!]
	\centering
	\includegraphics[width=8.cm]{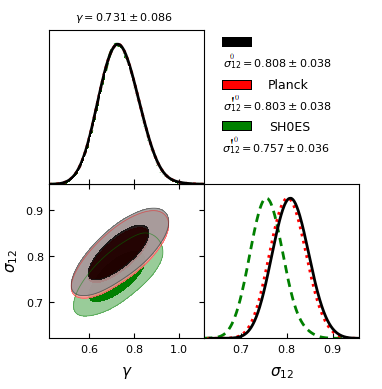}	\caption{Triangle plot with the one- and two-dimensional posterior distributions of the growth index $\gamma$ and the amplitudes $\sigma_{12}^0$ (in black) and $\sigma_{12}^{\prime 0}$ (in red, green)  obtained making use of RSD and peculiar velocity data \cite{Guzzo:2008ac,Song:2008qt,Blake:2011rj,Blake:2013nif,Simpson:2015yfa,Gil-Marin:2016wya,eBOSS:2020gbb,Said:2020epb,Avila:2021dqv,Mohammad:2018mdy,Okumura:2015lvp}, the prior $\Omega_m^0=0.30\pm 0.01$, and two different priors on $H_0$. The constraint on $\sigma_{12}^{\prime 0}$ derived with the prior $H_0=67.36\pm 0.54$ km/s/Mpc from Planck \cite{Planck:2018vyg} is shown in red. No significant bias is found for $\sigma_{12}^0$ in this case, since $h\approx h_*$. The constraint on $\sigma_{12}^{\prime 0}$ obtained with the SH0ES prior $H_0=73.04\pm 1.04$ km/s/Mpc \cite{Riess:2021jrx} is shown in green and exhibits a much larger bias, as expected. See the main text of Sec. \ref{sec:sec2} for details.}
	\label{fig:RSD}
\end{figure}

\begin{enumerate}
    \item Assuming that the data is on the quantity $f(z)\sigma_{12}(z)$.
    \item Assuming, instead, that the data compilation is on $f(z)\sigma_8(z)$. In this case, we need to translate our measurement of $\sigma_8^0$ into a measurement of $\sigma_{12}^0$ making use of Eq. \eqref{eq:bias}. We refer to the resulting value as $\sigma_{12}^{\prime0}$ to distinguish it from the one measured in point 1, which we call simply $\sigma_{12}^0$. Obviously, this translation will depend on $h$, so we need to use a prior on this parameter.
\end{enumerate}
The ratio $\Delta_\sigma\equiv (\sigma_{12}^{\prime0}-\sigma_{12}^0)/\delta \sigma_{12}^0$, with $\delta\sigma_{12}^0$ the uncertainty of $\sigma_{12}^0$, is a measure of the bias caused by interpreting the RSD measurements as measurements of $f(z)\sigma_8(z)$. We employ the prior $\Omega_m^0=0.30\pm 0.01$ and two different priors on $H_0$ to study how the bias depends on the value of the Hubble parameter: $H_0=67.36\pm 0.54$ km/s/Mpc, from the TT,TE,EE+lowE+lensing Planck/$\Lambda$CDM fitting analysis \cite{Planck:2018vyg}; and $H_0=73.04\pm 1.04$ km/s/Mpc, from SH0ES \cite{Riess:2021jrx}. The $H_0$ prior only affects the constraint on $\sigma_{12}^{\prime 0}$, i.e., the constraints on $\gamma$ and $\sigma_{12}^0$ are completely insensitive to it. The results obtained with the Planck and SH0ES priors are displayed in Fig. \ref{fig:RSD}. When the Planck prior is used the bias is quite small ($\Delta_\sigma = -0.13$), as expected, given that $h \approx h_*$. In contrast, when we force $H_0$ to be close to the SH0ES value, $\Delta_\sigma = -1.3$, indicating a substantial increase of the absolute value of the bias. As noted in the example discussed at the beginning of this section, the bias becomes more significant in models with larger values of $H_0$. If interpreted as measurements of $f(z)\sigma_8(z)$, the RSD and peculiar velocity measurements typically lead to smaller estimates of the rms mass fluctuations at 12 Mpc, i.e., $\sigma_{12}^{\prime 0}<\sigma_{12}^0$, and this can incorrectly penalize models that favor larger values of $H_0$, making them to enter (artificially) in conflict with the LSS measurements.  

Under the assumption of a standard cosmological background evolution and the prior of $\Omega_m^0$  employed in this exercise, we constrain the growth index to be approximately $2.1\sigma$ above the $\Lambda$CDM value, $\gamma_\Lambda$, which is consistent with previously published results (see, e.g., \cite{Nguyen:2023fip}). This is a reflection of the growth tension, of course.

With these very simple discussions, we can already understand that it is imperative to correct the bias introduced by the use of $\sigma_8$ if we want to safely analyze the performance of different models, compare the experimental results obtained by different galaxy surveys, and extract meaningful conclusions. This is especially important in the era of the cosmological tensions and, in particular, of the Hubble tension. Since the bias is larger for larger values of $|\delta h|$, we find that it is greater for models that alleviate or solve the $H_0$ tension, since they lead to higher values of $H_0$, close to $73$ km/s/Mpc. This is extremely worrisome because by studying $\sigma_8$ one typically concludes that models that make a good job in relieving the Hubble tension tend to worsen the growth tension, complicating the search for an efficient theoretical framework able to accommodate the data and mitigate the problems afflicting the standard model. This is what happens, for instance, with early dark energy \cite{Gomez-Valent:2022hkb}. In the next section, we quantify the impact of all these effects on a heterogeneous selection of models, performing fitting analyses with current data and comparing the posterior constraints on $\sigma_{8}(z)$ and $\sigma_{12}(z)$. 
\newline
\newline
\noindent {\it A brief note on the absence of $\sigma_8$-bias in weak lensing studies.} As already noted in the Introduction, weak lensing alone is not capable of constraining the Hubble parameter \cite{Hall:2021qjk}, essentially because weak lensing measurements are fundamentally angular. In view of the relevant role WL measurements play in the discussion of the growth tension, it is of utmost importance to remark that the $\sigma_8$-bias discussed above does not have any significant bearing on the constraints derived from weak lensing observations. Indeed, several works have shown that the weak lensing constraints on $S_8$ are largely insensitive to (and hence are stable under changes in) the prior on $h$. However, the physical scale associated to this constraint varies with $h$ and, therefore, as expected, the constraints on the parameter $S_{12}=\sigma_{12}(\omega_m/0.14)^{0.4}$ \cite{Sanchez:2020vvb}, which is linked to the fixed scale of $12$ Mpc, changes with the prior on $h$, see \cite{Secco:2022kqg,Garcia-Garcia:2024gzy}. In weak lensing studies, in fact, the quantification of the tensions can be
safely performed using either $S_8$ or $S_{12}$ as long as they are carried out considering a sufficiently large parameter space, which involves, at least, the parameters $(\Omega_m, S_8)$ or $(\Omega_m,S_{12},h)$ \cite{Secco:2022kqg}. The triad $(\Omega_m,S_{12},h)$, though, might ease the physical interpretation of the WL measurements \cite{Sanchez:2020vvb,Secco:2022kqg,Garcia-Garcia:2024gzy}.


\section{The problem in the light of current data}\label{sec:MR}

In the previous section, we have studied the bias introduced by $\sigma_8(z)$ in a simple and very controlled setting. This approach enabled us to quickly identify some unphysical and counterintuitive results associated with this function that might otherwise have been overlooked. We have seen that the issues caused by the use of $\sigma_8(z)$ disappear when we replace it with $\sigma_{12}(z)$. In this section, we study the problem by performing a full Monte Carlo analysis with state-of-the-art data from CMB, BAO and SNIa and five different cosmological models in order to quantify the impact of the bias induced by $\sigma_8(z)$ in current analyses, which of course requires the estimation of the central values and uncertainties of $\sigma_{12}(z)$ and $\sigma_{8}(z)$.

\subsection{Methodology}\label{sec:method}

We perform Monte Carlo Markov chain (MCMC) analyses using the publicly available package \texttt{Cobaya}~\cite{Torrado:2020dgo}. In particular, we explore the parameter space of the various models 
using the MCMC sampler developed for \texttt{CosmoMC}~\cite{Lewis:2002ah,Lewis:2013hha}, which implements the ``fast dragging” procedure~\cite{neal2005taking} and is tailored for parameter spaces with a speed hierarchy. The convergence of the chains obtained with this procedure is tested using the Gelman-Rubin criterion~\cite{Gelman:1992zz}, and we choose as a threshold for chain convergence $R-1\lesssim 0.02$. The corresponding posterior distributions are analyzed using the \texttt{GetDist} package~\cite{Lewis:2019xzd}. In order to compute the theoretical values of the various cosmological observables we solve the Einstein-Boltzmann set of coupled differential equations making use of modified versions of the code \texttt{CLASS}~\cite{Lesgourgues:2011re,Blas:2011rf} that allow us to obtain as derived parameters the values of $\sigma_{8}(z)$ and $\sigma_{12}(z)$ at the redshifts of interest.

\subsubsection{Datasets}\label{sec:data}

In this work, we employ two different cosmological datasets made of data from CMB, SNIa and BAO to constrain the models described in Sec. \ref{sec:models}. Our data strategy goes as follows. 

First and foremost, we make use of the strong constraining power of CMB observations. As explained in the Introduction, the growth tension arises as a mismatch between the late-time measurements of the matter clustering and the Planck data. Additionally, the discussion surrounding $\sigma_8$, as previously mentioned, is inevitably intertwined with the ongoing issue of the Hubble tension, in which the Planck analysis plays a pivotal role. As a consequence, it is natural to use the CMB temperature and polarization power spectra \textit{plik}TTTEEE+lowl+lowE (hereafter Planck) from the legacy Planck release as reference for CMB observations \cite{Planck:2018vyg,Planck:2019nip}. 

However, to have a better understanding of the impact of the bias introduced by $\sigma_8$ and how it is inextricably connected with the Hubble tension, Planck alone is not sufficient. We also need to take into account the SH0ES measurements.  
 We employ the Pantheon+ compilation, which consists of 1701 light curves of 1550 spectroscopically confirmed SNIa over the redshift range $0.001 < z < 2.26$~\cite{Brout:2022vxf,Scolnic:2021amr}, and use the SH0ES~\cite{Riess:2021jrx} distance moduli of the host galaxies to calibrate the absolute magnitude of SNIa in the second rung of the distance ladder. We refer to this dataset as PPS, for short. By combining it with Planck, we obtain larger values of $H_0$, allowing us to study the impact of the bias in frameworks designed to mitigate the Hubble tension. As per \eq{bias} and as previously explained, these scenarios result in larger values of the bias, making them an interesting case study.

Lastly, we also incorporate BAO data into our analysis in order to add constraining power. More concretely, we rely on DESI BAO measurements~\cite{DESI:2024mwx} from galaxy samples spanning a broad range of redshifts. In particular, we include data from the Bright Galaxy Sample (BGS) in $0.1< z< 0.4$, the Luminous Red Galaxy Sample (LRG) in $0.4< z< 0.6$ and $0.6< z< 0.8$, the combination of LRG and Emission Line Galaxy Sample (ELG) in $0.8< z< 1.1$, ELG data in $1.1< z< 1.6$, a quasar sample in $0.8< z< 2.1$~\cite{DESI:2024uvr} and the Lyman-$\alpha$ forest sample in $1.77< z< 4.16$~\cite{DESI:2024lzq}. For completeness, we also validate our results using BAO measurements from BOSS DR12~\cite{Alam:2016hwk} and eBOSS DR16~\cite{eBOSS:2020yzd}, collectively referred to as the BOSS dataset. We obtain our main results using the two data combinations Planck+BOSS+PPS and Planck+DESI+PPS, and show that both, the fitting results and the conclusions regarding the bias introduced by $\sigma_8$, are completely stable under the choice of BAO data employed in the analysis.

Following the discussion in \sect{sec2}, in our main analyses we prefer to avoid the use of RSD and peculiar velocity data. Nevertheless, we perform a little test using growth data \cite{Guzzo:2008ac,Song:2008qt,Blake:2011rj,Blake:2013nif,Simpson:2015yfa,Gil-Marin:2016wya,eBOSS:2020gbb,Said:2020epb,Avila:2021dqv,Mohammad:2018mdy,Okumura:2015lvp} to reinforce the results presented in the previous section.

\subsubsection{Models}\label{sec:models}

We aim to explore the implications of selecting a different radius for the sphere over which the rms mass fluctuations are computed, particularly in the context of $\Lambda$CDM and some of its extensions. They have been chosen based on their influence on the large-scale structure of the universe and their ability to increase $H_0$, which is of course relevant for the discussion surrounding the Hubble tension and has also an impact on the bias described by Eq. \eqref{eq:bias}. Specifically, we have decided to investigate the following models:

\begin{itemize}

\item {\it $\Lambda$CDM}. Dark energy~\cite{Weinberg:2013agg,Mortonson:2013zfa,Frieman:2008sn,Copeland:2006wr} constitutes approximately $68\%$ of the total energy density in the current universe and is the primary driver of its accelerated expansion~\cite{SupernovaSearchTeam:1998fmf,SupernovaCosmologyProject:1998vns}. For any component to induce cosmic acceleration, its dynamics must exhibit negative pressure,  characterized by an equation of state parameter  $w\equiv p_{\rm DE}/\rho_{\rm DE} <1/3$. The simplest model describing dark energy is $\Lambda$CDM, the standard model of cosmology, which incorporates the so-called cosmological constant, $\Lambda$ \cite{Sahni:1999gb,Carroll:2000fy,Peebles:2002gy,Padmanabhan:2002ji}. The latter gives rise to a constant energy density permeating space homogeneously, with $w_\Lambda=-1$. 

We treat $\Lambda$CDM as a benchmark model, so we also obtain fitting results for it. We consider the six $\Lambda$CDM parameters, i.e., the baryon $\omega_b$ and cold dark matter $\omega_{cdm}$ reduced density parameters\footnote{where $\omega_i\equiv \Omega_i h^2$.}, the angular size of the horizon at the last scattering surface $\theta_*$, the optical depth $\tau$, the amplitude of primordial scalar perturbations $\log(10^{10}A_s)$ and the scalar
spectral index $n_s$. For these parameters we choose flat-prior distributions, varying them uniformly in the ranges listed in Table \ref{tab:priors}. 

\item \textit{$w_0w_a$CDM parameterization.}  Despite the numerous phenomenological successes of the standard model, a rigid cosmological constant might not be the most plausible possibility from a theoretical perspective (see, e.g., \cite{Weinberg:1988cp,Martin:2012bt,Sola:2013gha,SolaPeracaula:2022hpd}) and is afflicted by several substantial tensions with observations \cite{Perivolaropoulos:2021jda}. This has motivated the study of a plethora of cosmological models and parameterizations of dynamical dark energy in the late universe, see \cite{Amendola:2015ksp} and references therein. In this work, we opt to adopt the widely-used $w_0w_a$CDM (or CPL) parameterization~\cite{Chevallier:2000qy,Linder:2002et}, which introduces a linear dependence on the scale factor:
    \begin{equation}
    w(a)=w_0+w_a(1-a)\,.
    \end{equation}
    This parameterization represents a first-order Taylor expansion of $w(a)$ around the present epoch up to first order\footnote{Higher order expansion and different parameterization can be also employed, of course, at the expense of a higher degree of complexity}, see e.g.~\cite{Giare:2024gpk,Najafi:2024qzm} and references therein. and could serve as a proxy of different theories of DE with a well-posed action \cite{Scherrer:2015tra}. The evolution of the dark energy density $\rho_{\rm DE}$ is governed by
    \begin{equation}
        \rho_{\rm DE}(a)=\rho_{\rm DE}^0\text{exp}\left(3\int_a^1\frac{1+w(a^\prime)}{a^\prime}da^\prime\right)\,,
    \end{equation}
    with $\rho_{\rm DE}^0\equiv \rho_{\rm DE}(a=1)$. This expression is obtained by integrating the continuity equation of DE, considering that DE is covariantly self-conserved. Given a fixed amount of the various cosmic species at present, the growth of large-scale structures is more suppressed in the past if dark energy increases with redshift, exhibiting quintessence behavior ($w > -1$), compared to a scenario where DE is phantom ($w < -1$). The $w_0w_a$CDM parameterization allows for both behaviors as well as crossing the phantom divide, depending on the values of $w_0$ and $w_a$. This versatility in influencing structure formation makes the model particularly compelling to study. Moreover, recent observations have shown a preference for dynamical dark energy in the context of the $w_0w_a$CDM parameterization at more than $2\sigma$ \cite{DESI:2024mwx,DESI:2024lzq,DESI:2024uvr}, which does not seem to rely exclusively on the DESI data \cite{Chan-GyungPark:2024mlx,Gomez-Valent:2024ejh}. See also \cite{Liu:2024gfy,Cortes:2024lgw,Pourojaghi:2024tmw,Giare:2024gpk,Shlivko:2024llw,Chan-GyungPark:2024mlx,Wang:2024dka,Gialamas:2024lyw,Notari:2024rti,Chan-GyungPark:2024brx,Wolf:2024eph,Wolf:2024stt} for related discussions and \cite{Sahni:2014ooa,Sola:2015wwa,Sola:2016jky,SolaPeracaula:2016qlq,SolaPeracaula:2017esw,Zhao:2017cud,SolaPeracaula:2018wwm} for earlier hints of dynamical dark energy. In the Monte Carlo analyses of the $w_0w_a$CDM parameterization, we allow the parameters $w_0$ and $w_a$ to vary using the flat priors of Table \ref{tab:priors}.
    
    \item \textit{Neutrino Sector.} By means of $N_{\rm eff}$ and $\sum{m_\nu}$ we parameterize the neutrino sector. $N_{\rm eff}$ is the effective number of neutrino species in the universe~\cite{Mangano:2001iu,Bennett:2019ewm}. In the standard model we have $N_{\rm eff}\simeq 3.043$~\cite{Mangano:2005cc,deSalas:2016ztq,Akita:2020szl,Froustey:2020mcq,Bennett:2020zkv,Cielo:2023bqp} and deviations from this value could signal new physics (e.g. see~\cite{Maggiore:1999vm,deSalas:2015glj,Giare:2022wxq,Barenboim:2024wek}). Notably, as the contribution of neutrinos to the total radiation energy density $\rho_r$ is proportional to the effective number of neutrinos, variation of $N_{\rm eff}$ modifies the sound horizon at recombination. In particular, larger $N_{\rm eff}$ values decrease the horizon and, consequently, require higher values of $H_0$. As anticipated, this dynamics has the counterpart of enhancing the LSS tension when analyzed in terms of $\sigma_8$. We will show, though, that in terms of the more reliable parameter $\sigma_{12}$, the growth of LSS remains close to the one predicted in $\Lambda$CDM.

    \begin{table}
	\begin{center}
		\renewcommand{\arraystretch}{1.5}
		\begin{tabular}{c@{\hspace{0. cm}}@{\hspace{1.5 cm}} c}
			\hline
			\textbf{Parameter}    & \textbf{Prior} \\
			\hline\hline
			$\omega_{\rm b}$         & $[0.005\,,\,0.1]$ \\
			$\omega_{\rm c} h^2$     	 & $[0.001\,,\,0.99]$\\
			$100\,\theta_{*}$     & $[0.5\,,\,10]$ \\
			$\tau$                       & $[0.01\,,\,0.8]$\\
			$\log(10^{10}A_{s})$     & $[1.61\,,\,3.91]$ \\
			$n_{\rm s}$                  & $[0.8\,,\, 1.2]$ \\
            \hline
            $N_{\rm eff}$                   &$[0\,,\,5]$\\ 
            $\sum{m_\nu}$ [eV]                 &$[0.06\,,\,0.6]$\\
            $w_0$                           & $[-2\,,\, 2]$\\
            $w_a$                           & $[-4\,,\, 2]$\\
            $f_{\rm EDE}$                       &$[0.001, 0.5]$\\
            $\log_{10}{z_c}$                       &$[3.0, 4.3]$\\
            $\theta_i$                       &$[0.1, 3.1]$\\
            \hline\hline
		\end{tabular}
		\caption{List of priors for the baseline parameters, as well as for the additional parameters in the alternative scenarios. For further details, refer to Sec. \ref{sec:models}. }
		\label{tab:priors}
	\end{center}
\end{table}

    \begin{figure*}
    \centering
    \includegraphics[width=15cm]{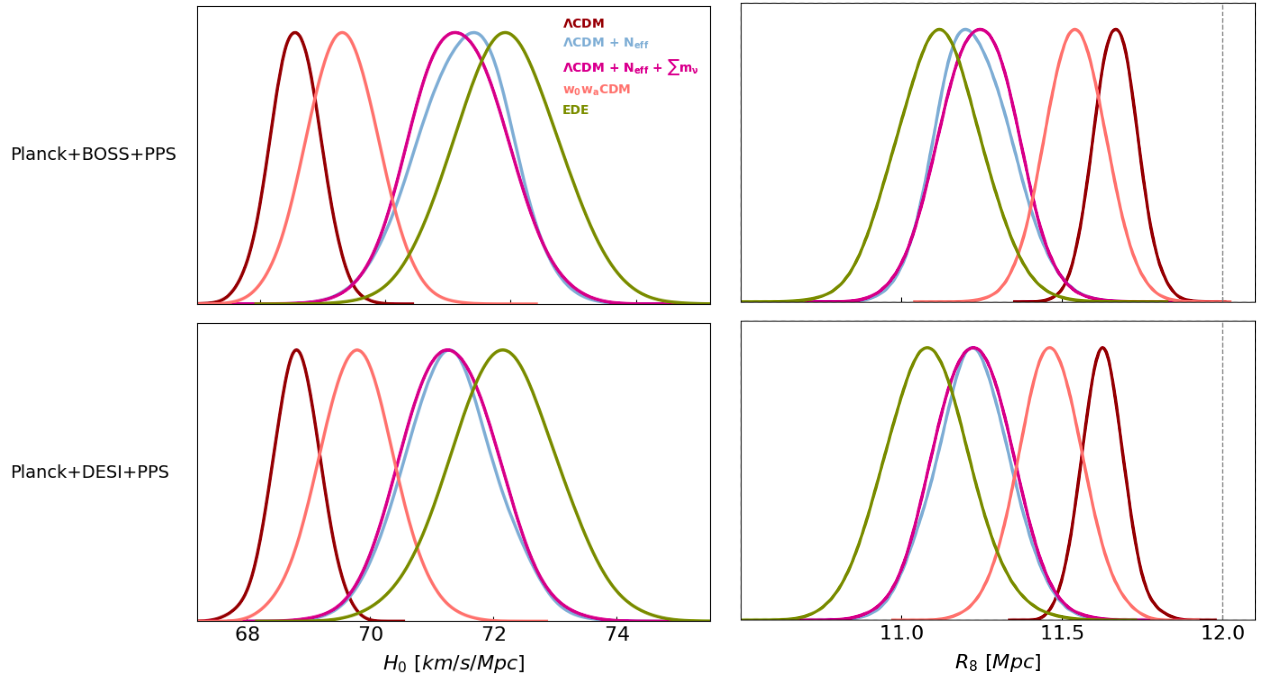}
    \caption{Posterior distributions of $H_0$ (left panels) and $R_8=8/h$ (right panels) obtained from Planck+BOSS+PPS (first row) and Planck+DESI+PPS (second row) for the five cosmological models listed in the legend. The vertical dashed line in the right-hand side plots indicates the reference scale of $R_{12} =12$ Mpc.}
    \label{fig:H0R8}
\end{figure*}
    
    The parameter $\sum{m_\nu}$ refers to the sum of neutrino masses~\cite{Lesgourgues:2006nd,Lattanzi:2017ubx,Loureiro:2018pdz,RoyChoudhury:2018gay,DeSalas:2018rby,RoyChoudhury:2019hls,DiValentino:2021hoh,Capozzi:2021fjo,DiValentino:2021imh,Tanseri:2022zfe,diValentino:2022njd,DiValentino:2023fei,Gariazzo:2023joe,Craig:2024tky,Bertolez-Martinez:2024wez,Shao:2024mag}. In the Planck $\Lambda$CDM baseline model, a normal mass hierarchy is assumed, with the minimal mass of $\sum{m_\nu}=0.06$ eV. However, it is worth noting that the hierarchy and the neutrino oscillation experiments only determine the minimum value of $\sum{m_\nu}$, and $\sum{m_\nu}>0.06$ eV remains a plausible possibility - while an inverted hierarchy increases the lower bound to  $\sum{m_\nu}>0.1$ eV \cite{Esteban:2024eli}. Increasing the neutrino mass amplifies the Hubble tension, as it typically leads to lower values of $H_0$. Nevertheless, at the same time, it suppresses power on scales smaller than their free-streaming length. It is interesting then to use the mass parameter in combination with $N_{\rm eff}$ to study their interplay. Therefore, we consider the following two extensions of $\Lambda$CDM for the neutrino sector: $\Lambda$CDM+$N_{\rm eff}$ and $\Lambda$CDM+$N_{\rm eff}+\Sigma m_\nu$. The flat-prior ranges for $N_{\rm eff}$ and $\sum m_\nu$ are reported in Table \ref{tab:priors}. For the other models studied in this paper, we use the neutrino minimal setup, as in $\Lambda$CDM.

    \item \textit{Early Dark Energy.} Early dark energy (EDE) models account for a significant dark energy contribution in the early universe, and can be described for instance by inhomogeneous dark energy fluids or scalar fields (see e.g~\cite{Doran:2006kp,Hollenstein:2009ph,Calabrese:2010uf,Calabrese:2011hg,Pettorino:2013ia,Archidiacono:2014msa,Poulin:2018cxd,Poulin:2018dzj,Poulin:2018zxs,Niedermann:2019olb,Smith:2020rxx,Niedermann:2020dwg,Murgia:2020ryi,Klypin:2020tud,Hill:2020osr,Herold:2021ksg,Gomez-Valent:2021cbe,Reeves:2022aoi,Jiang:2022uyg,Simon:2022adh,Smith:2022hwi,Kamionkowski:2022pkx,Gomez-Valent:2022hkb,Niedermann:2023ssr,Poulin:2023lkg,Smith:2023oop,Cruz:2023lmn,Eskilt:2023nxm,Sharma:2023kzr,Efstathiou:2023fbn,Gsponer:2023wpm,Goldstein:2023gnw,Toda:2024ncp})\footnote{Early modified gravity models lead also to interesting results, see, e.g., \cite{SolaPeracaula:2019zsl,Braglia:2020auw,SolaPeracaula:2020vpg,Benevento:2022cql,Kable:2023bsg,FrancoAbellan:2023gec,Erdem:2024vsr}.}.  Our analysis focuses on the EDE implementation from~\cite{Poulin:2018cxd} where a light scalar field is initially frozen far from its potential minimum, making this component to behave like a cosmological constant under Hubble friction. At a critical redshift $z_c$ before recombination, the Hubble parameter drops below the field's mass, the scalar field rolls down its potential and oscillates around the minimum. The scalar field energy density must redshift faster than radiation so that the field becomes a subdominant component at the decoupling time \cite{Gomez-Valent:2021cbe,Vagnozzi:2021gjh}. A typical set of additional parameters used in this model is: the fractional contribution to the total energy density of the universe evaluated at the critical redshift $z_c$ at which it reaches the maximum value, $f_{\rm EDE}\equiv\rho_{\rm EDE}(z_c)/\rho_{tot}(z_c)$, and $\theta_i$, which is the parameter that usually describes the initial field displacement. This behavior increases the early universe’s energy density (prior to recombination), reduces the sound horizon, and raises the Hubble constant inferred from CMB observations, making EDE models a potential solution to the Hubble tension \cite{Poulin:2018cxd} and, consequently, a good candidate for our analysis\footnote{EDE could also help to explain the anomalous population of very massive galaxies at large redshifts observed by JWST \cite{Forconi:2023hsj,Shen:2024hpx}.}. To solve the coupled system of Einstein-Boltzmann equations in the EDE scenario, we used the publicly available software \texttt{CLASS\_EDE}\footnote{\url{https://github.com/mwt5345/class_ede}} \cite{Hill:2020osr}. In the Monte Carlo runs, we use the flat priors reported in Table \ref{tab:priors} for $f_{\rm EDE}$, $\log_{10}{z_c}$ and $\theta_i$.
\end{itemize}

Since we are dealing with multiple extensions of the standard model, we decided to perform a simple comparison with respect to $\Lambda$CDM. To achieve this, we utilize the Akaike information criterion (AIC)~\cite{Akaike}\footnote{There are many stronger (more robust) criteria for cosmological model selection which could be employed (see e.g. ~\cite{Verdineli:1995,Kass:1995loi,Marshall:2004zd,DIC,Trotta:2008qt,Keeley:2021dmx}), but this is out of the scope of our work. The use of AIC is more than enough for the purposes of this paper.}
\begin{equation}\label{eq:AIC}
    {\rm AIC}=\chi^2_{\rm min}+2n_p\,,
\end{equation}
with $\chi^2_{\rm min}$ the minimum value of $\chi^2$ and $n_p$ the number of fitting parameters. Apart from evaluating the quality of the fit through the first term in the right-hand side of  \eq{AIC}, this statistical tool duly penalizes the extra model parameters through the second term. It has been vastly used in the literature (e.g., in \cite{SolaPeracaula:2017esw,Vagnozzi:2018jhn,Gomez-Valent:2018hwc,Kreisch:2019yzn,Agrawal:2019lmo,Visinelli:2019qqu,RoyChoudhury:2019hls}). We compare the AIC values of the non-standard models to that of $\Lambda$CDM by computing the difference $\Delta {\rm AIC}={\rm AIC}_{\Lambda{\rm CDM}}-{\rm AIC}_i$, i.e., using the standard model as the benchmark. A positive difference of $\Delta$AIC  signals a preference for the non-standard model. More concretely, we quantify such a preference by means of the Jefferey's scale~\cite{Jeffreys:1939xee}; if $0 \leq \Delta\textrm{AIC}<2$ it is considered that there exists \textit{weak evidence} in favor of the new model $i$, compared to the standard model. If $2 \leq \Delta\textrm{AIC} < 6$, we speak instead of \textit{positive evidence}. If $6 \leq \Delta\textrm{AIC} < 10$, there is \textit{strong evidence}, and if $\Delta\textrm{AIC}>10$ we conclude that there is \textit{very strong evidence} supporting the  model $i$ against the $\Lambda$CDM.

\begin{figure*}[htp]
	\centering
    \includegraphics[width=\columnwidth]{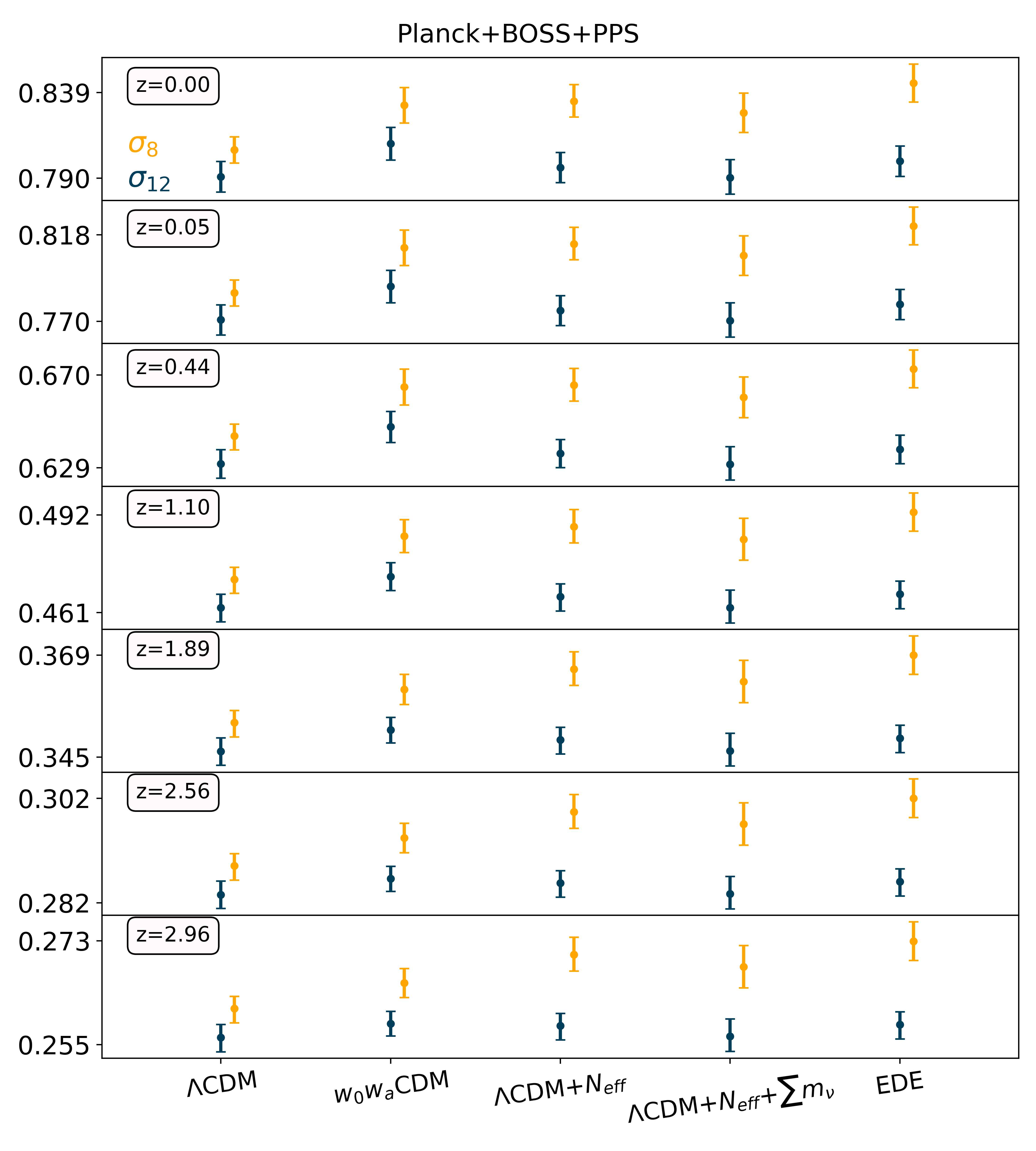}\includegraphics[width=\columnwidth]{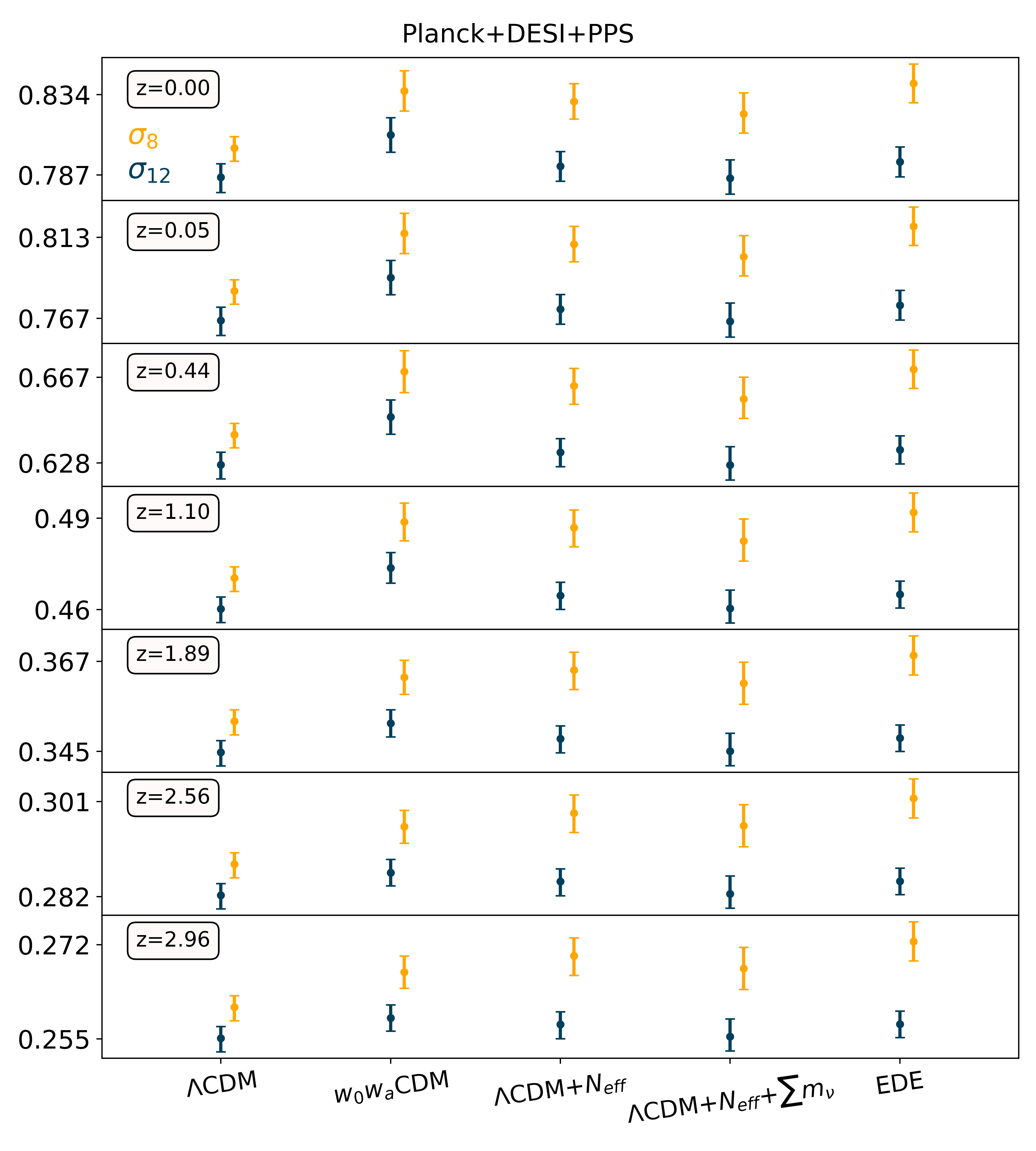}\caption{Constraints at 68\% C.L. on $\sigma_{12}$ (in green) and $\sigma_{8}$ (in orange) at different redshifts in the range $z\in[0,2.96]$, obtained from the analysis of Sec. \ref{sec:method} for each of the five cosmological models studied in this work. We show the results obtained by exploiting the full combination of data, i.e. Planck+BOSS+PPS (left panel) and Planck+DESI+PPS (right panel). Notice that the $\sigma_8$ error bars are slightly shifted on the $x$-axis in order to better distinguish them from the $\sigma_{12}$ results. This allows us to clearly visualize the relative shift (on the $y$-axis) between the two parameters for each model, which is present regardless of the redshift at which they are evaluated and reinforces the discussion made in Sec. \ref{sec:sec2}.}
	\label{fig:whisker}
\end{figure*}


\subsection{Results}\label{sec:results}

In Fig. \ref{fig:H0R8} we show the posterior distributions of $H_0$ obtained for the various models under study (cf. Sec. \ref{sec:models}), as well as the corresponding posteriors of $R_8$ using the datasets Planck+BOSS+PPS and Planck+DESI+PPS.  It is very well-known that anisotropic (3D) BAO data as those employed in this work do not give room for purely cosmological late-time solutions to the Hubble tension \cite{Knox:2019rjx,Gomez-Valent:2023uof}\footnote{Angular (or 2D) BAO data, in contrast, still allow for late-time solutions \cite{Bernui:2023byc,Akarsu:2023mfb,Gomez-Valent:2024ejh,Anchordoqui:2024gfa,Yadav:2024duq,Dwivedi:2024okk,Gomez-Valent:2024tdb}.}. In fact, late-time dynamical dark energy alone is completely unable to produce large enough values of $H_0$, even by taking into account the SH0ES data in the fitting analysis, as we do in the current study. Clearly, the $w_0w_a$CDM parameterization is no exception. On the other hand, as expected, regardless of the BAO dataset employed in the analysis (BOSS or DESI), we find values of the Hubble parameter that are much larger in those models that allow for some additional energy contribution before recombination compared to the $\Lambda$CDM. This additional energy induces a decrease of the sound horizon at the baryon-drag and decoupling epochs and forces the increase of $H(z)$ in the late universe in order to properly explain the BAO data and the location of the first peak of the CMB temperature angular power spectrum, which was measured very precisely by the Planck collaboration. This is the case of $\Lambda$CDM+$N_{\rm eff}$ (considering or not the variation of $\sum m_\nu$) and EDE, for which we obtain central values $H_0\gtrsim 71.2$ km/s/Mpc and $H_0\gtrsim 72$ km/s/Mpc, respectively. In Table \ref{tab:fit}, we provide for completeness the fitting results for all the model parameters. In the $\Lambda$CDM+$N_{\rm eff}$ model, an increase in $H_0$ is made possible by a rise in the number of ultra-relativistic species, which reaches $N_{\rm eff} = 3.5 - 3.6$. This value deviates from the standard one by approximately $4.2\sigma$ when the Planck+BOSS+PPS dataset is used for the fitting analysis, and by about $3.7\sigma$ when Planck+DESI+PPS is used instead. Similar results are found with a comparable dataset in, e.g., \cite{Yadav:2024duq}. In EDE, larger values of $H_0$ are possible thanks to an energy injection around redshift $z_c\sim 4200$, with a maximum EDE fraction $f_{\rm EDE}\sim 13\%$ (see Table \ref{tab:fit}). Our findings for $w_0w_a$CDM and EDE are aligned with those provided in \cite{Gomez-Valent:2024tdb} and \cite{Gsponer:2023wpm}, respectively. See also \cite{Poulin:2018cxd,Vagnozzi:2019ezj,Ballesteros:2020sik,Smith:2020rxx}. 

The right-hand side plot of Fig. \ref{fig:H0R8} reveals two key points for our discussion: (i) the differences in the values of $R_8$ obtained across the various models are significant, particularly between $\Lambda$CDM or $w_0w_a$CDM and the models that introduce new physics prior to the decoupling era; and (ii) in all cases, the values of $R_8$ are substantially smaller than $R_{12}$, even in $\Lambda$CDM and $w_0w_a$CDM, where $h$ remains far below the SH0ES measurement but is still larger than $h_*$. As a result, the comparison of the $\sigma_8(z)$ curves obtained from the Monte Carlo analyses for the various models becomes essentially meaningless, as they represent the amplitude of fluctuations at entirely different scales. In other words, comparing the values of $\sigma_8$ across the studied models does not yield definitive conclusions about their relative clustering performance. A more appropriate approach would be to compare the values of $\sigma_R$ at a fixed scale, such as $R_{12}$. Given the significant difference between $R_8$ and $R_{12}$, we can expect substantial changes in the conclusions derived from this more equitable comparison.

\begin{table*}[htp]
\centering
\begin{tabular}{c c c c c c}
\multicolumn{6}{c}{Planck+BOSS+PPS} \\
\hline\noalign{\smallskip}                             
Parameter & $\Lambda$CDM & $w_0w_a$CDM & $\Lambda$CDM + $N_{\rm eff}$ & $\Lambda$CDM + $N_{\rm eff}$ + $\Sigma m_{\nu}$  & EDE \\
 \hline\hline\noalign{\smallskip}    
$\omega_m$ & $0.1407\pm0.0008$ & $ 0.1430 \pm0.0010$  & $ 0.1511 _{- 0.0021 }^{+0.0028 }$ & $ 0.1514 \pm0.0026$ & $ 0.1570\pm0.0035$
 \\ [0.1cm]
$H_0$ [km/s/Mpc] & $ 68.57\pm0.39$ & $ 69.32\pm0.56$ & $ 71.28 _{- 0.67 }^{+ 0.78 }$ & $ 71.19 \pm 0.75$ & $ 72.06 _{- 0.83 }^{+ 0.84 }$ \\[0.1cm]
$\log(10^{10}A_s)$ & $ 3.051\pm 0.016$ & $ 3.042 _{- 0.014 }^{+ 0.015 }$ & $ 3.070_{- 0.017 }^{+ 0.015 }$ & $ 3.074 \pm 0.018$ & $ 3.073 _{- 0.017 }^{+ 0.016 }$ \\[0.1cm]
$n_s$ & $ 0.971 \pm 0.004$ & $ 0.965 \pm0.004$ & $ 0.986\pm 0.005$ & $ 0.987\pm 0.005$  & $ 0.992 \pm0.006$ \\[0.1cm]
$\sigma_8$ & $ 0.806\pm0.007$  & $ 0.832\pm0.010$ & $ 0.834\pm0.009$ & $ 0.828 \pm0.011$& $ 0.846\pm0.011$ \\[0.1cm]
$\sigma_{12}$ & $ 0.791\pm0.009$ & $ 0.810\pm0.009$ & $ 0.796\pm0.009$ & $ 0.791 _{- 0.009 }^{+ 0.010 }$  & $ 0.800 \pm0.008$ \\[0.1cm]
$S_{8}$ & $0.805\pm0.012$ & $0.829 \pm 0.012$  & $0.830 \pm0.013 $ & $0.826 \pm0.014$ & $0.848\pm 0.015$ \\[0.1cm]
$N_{\rm eff}$ & - & - & $ 3.58 _{- 0.11 }^{+ 0.14 }$ & $ 3.60\pm 0.13$ & - \\[0.1cm]
$\Sigma m_{\nu}$ [eV] & - & - & - & $<0.15$  & - \\[0.1cm]
$w_0$ & - & $ -0.859 _{- 0.054 }^{+ 0.050}$
 & - & -  & - \\[0.1cm]
$w_a$ & - & $ -0.82 _{- 0.21 }^{+ 0.24 }$
 & - & -  & - \\[0.1cm]
$f_{\rm EDE}$ & - & - & - & - &  $ 0.133\pm0.024$
\\[0.1cm]
$\log_{10} (z_c)$ & - & - & - & - & $ 3.619 _{- 0.133 }^{+ 0.082 }$

 \\[0.1cm]
$\theta_{i}$ & - & - & - & - &  $>2.13$
\\[0.1cm]
$\Omega_m^0$ &$0.299\pm0.005$ &$0.298\pm0.006$ &$0.297\pm0.005$& $0.299\pm0.005$&$0.302\pm0.005$ \\ [0.1cm]
$r_d$ [Mpc] &$ 147.53\pm 0.23$ &$ 147.06\pm 0.25$
&$ 142.026 _{- 1.36 }^{+ 0.97 }$
&$ 141.956 _{- 1.25 }^{+ 1.24 }$
&$ 139.64 _{- 1.51 }^{+ 1.53 }$
\\ [0.1cm]
$M$ & $-19.398\pm0.019$ & $-19.374\pm0.022$ & $-19.314\pm0.026$ & $-19.317\pm0.028$ & $-19.290\pm0.030$\\ [0.1cm]
\hline
$\chi^{2}_{\rm min}$ & 4279.90&4262.29& 4261.42&4258.96 &4246.83       \\ 
$\Delta{\rm AIC}$ & - &13.61 &16.48& 16.94&27.07        \\\hline                   
\vspace{0.1cm}
\end{tabular}

\begin{tabular}{c c c c c c}
\multicolumn{6}{c}{Planck+DESI+PPS} \\
\hline\noalign{\smallskip}                             
Parameter & $\Lambda$CDM & $w_0w_a$CDM & $\Lambda$CDM + $N_{\rm eff}$ & $\Lambda$CDM + $N_{\rm eff}$ + $\Sigma m_{\nu}$  & EDE \\
 \hline\hline\noalign{\smallskip}    
$\omega_m$ & $0.1402\pm0.0008$ & $ 0.1429\pm0.0010$ & $0.1494\pm0.0025$ & $ 0.1499\pm0.0027$ &$ 0.1556\pm0.0036$ \\ [0.1cm]
$H_0$ [km/s/Mpc] &$ 68.80\pm0.38$ & $ 69.78 _{- 0.60 }^{+ 0.59 }$
 &$ 71.30 _{- 0.72 }^{+ 0.73 }$
&$ 71.30 _{- 0.75}^{+ 0.76 }$ & 
$ 72.27 _{- 0.86 }^{+ 0.87 }$
\\[0.1cm]
$\log(10^{10}A_s)$ & $ 3.046 _{- 0.018 }^{+ 0.017 }$ &$ 3.043\pm0.016$
&$ 3.069 _{- 0.019 }^{+ 0.017 }$
&$ 3.070 _{- 0.018 }^{+ 0.016 }$
& $ 3.071\pm0.017$\\[0.1cm]
$n_s$ & $ 0.972\pm0.004$ &$ 0.965\pm0.004$  &$ 0.986\pm0.005$& $ 0.987\pm0.005$ & $ 0.993\pm0.006$\\[0.1cm]
$\sigma_8$ & $ 0.803 _{- 0.008 }^{+ 0.007 }$  &$ 0.836 _{- 0.011 }^{+ 0.012 }$
& $0.830\pm0.010$ &$ 0.824 _{- 0.011 }^{+ 0.012 }$
 & $ 0.842\pm0.011$ \\[0.1cm]
$\sigma_{12}$ & $ 0.785\pm0.008 $ &$ 0.810 \pm0.010$ & $0.792\pm0.009$ & $ 0.786 _{- 0.009 }^{+ 0.011 }$
 & $ 0.795 \pm 0.009$\\[0.1cm]
 $S_{8}$ & $0.798\pm0.011$& $0.827\pm0.013$ &   $0.822\pm0.013$&  $0.816\pm0.014$&  $0.838\pm0.015$\\[0.1cm]
$N_{\rm eff}$ & - & - & $ 3.53\pm 0.13$
 & $ 3.56\pm 0.14 $
 & - \\[0.1cm]
$\Sigma m_{\nu}$ [eV] & - & - & - & $<0.15$ & - \\[0.1cm]
$w_0$ & - & $ -0.808 _{- 0.071 }^{+ 0.064 }$ & - & - 
 & - \\[0.1cm]
$w_a$ & -& $ -1.10 _{- 0.29 }^{+ 0.34 }$
 & - & -  & - \\[0.1cm]
$f_{\rm EDE}$ & - & - & - & - & $ 0.129\pm0.025$
\\[0.1cm]
$\log_{10}(z_c)$ & - & - & - & - & $ 3.622 _{- 0.130 }^{+ 0.083 }$

 \\[0.1cm]
$\theta_{i}$ & - & - & - & - & $> 2.43$
 \\[0.1cm]
$\Omega_m^0$ & $ 0.296\pm0.005$ & $ 0.293\pm0.005$ & $ 0.294\pm0.005$&$ 0.295\pm0.005$& $ 0.298\pm0.005$\\ [0.1cm]
$r_d$ [Mpc] & $ 147.65 \pm 0.22$ &$ 147.11 _{- 0.26 }^{+ 0.25 }$ & $ 142.68 _{- 1.19 }^{+ 1.20 }$
&$ 142.47 _{- 1.30}^{+ 1.31 }$
&$ 140.08 _{- 1.58 }^{+ 1.56 }$
\\ [0.1cm]
$M$ & $-19.390_{-0.019}^{+0.018}$ & $-19.360\pm0.023$ & $-19.314_{-0.026}^{+0.027}$ & $-19.314\pm0.027$  & $-19.283\pm0.030$ \\ [0.1cm]
\hline
$\chi^2_{\rm min}$ & 4268.11 &4250.18&4256.69&4255.69&4235.97        \\ 
$\Delta{\rm AIC}$ & - &13.93 &9.42&8.42&26.14        \\\hline 

\end{tabular}\caption{Fitting results for the parameters of the various models described in Sec. \ref{sec:models} at 68\% C.L., obtained from the analysis with the datasets Planck+BOSS+PPS (upper half) and Planck+DESI+PPS (lower half), cf. Sec. \ref{sec:data}. Upper bounds are given at 95\% C.L. The parameter $\omega_m=\Omega_m^0 h^2$ is the sum of the baryon and CDM reduced density parameters. In the last rows of the two subtables, we provide the minimum values of $\chi^2$ as well as as the differences of AIC values (Eq. \eqref{eq:AIC}) with respect to $\Lambda$CDM.}  \label{tab:fit}
\end{table*}

\begin{figure*}[htp]
	\centering
	\includegraphics[width=\textwidth]{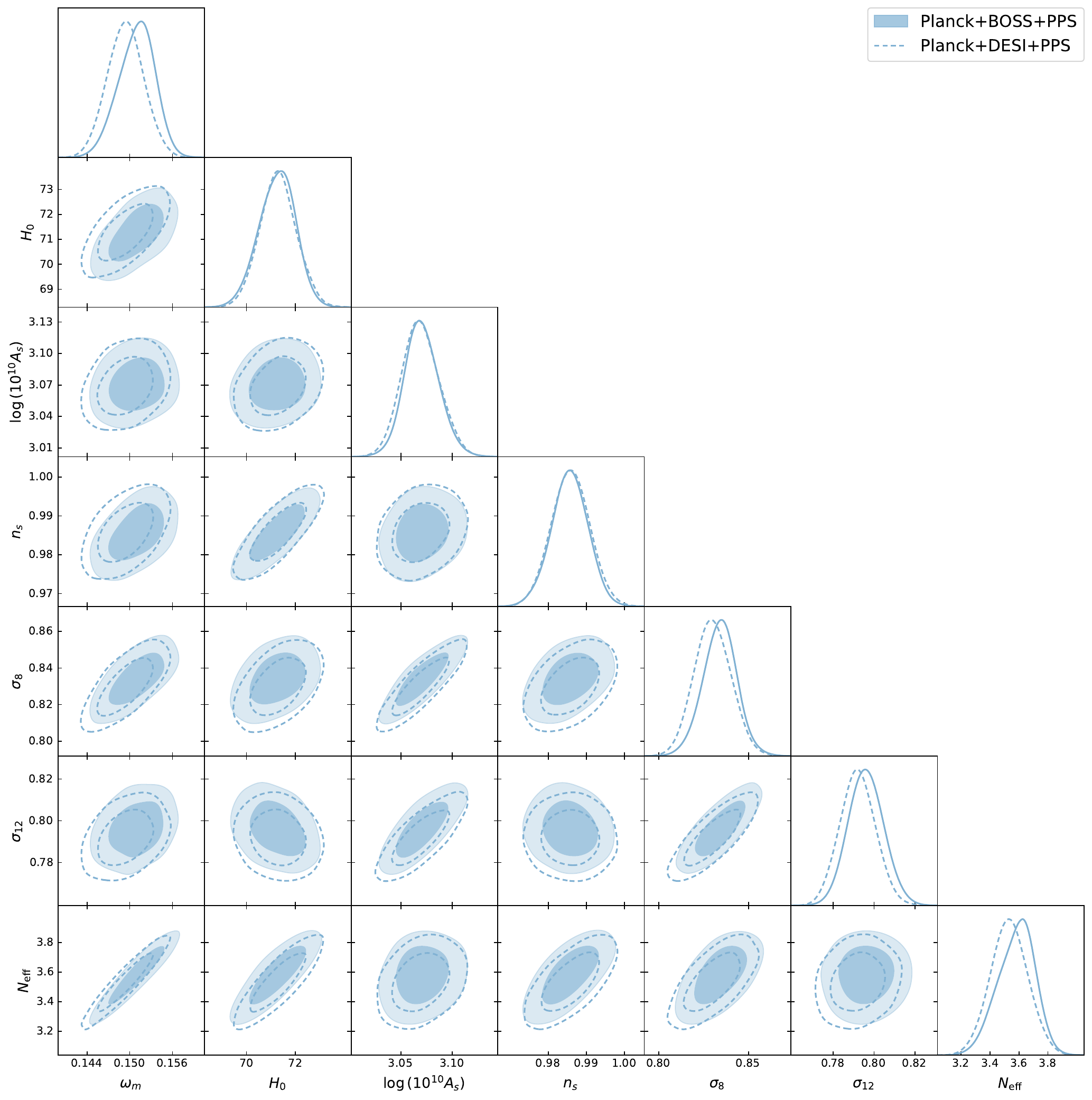}
	\caption{Triangular plot for $\Lambda$CDM$+N_{\rm eff}$ obtained from the fitting analysis with Planck+BOSS+PPS (solid contours) and Planck+DESI+PPS (dashed contours), cf. Sec. \ref{sec:data}.  The parameter $\omega_m=\Omega_m^0 h^2$ is the sum of the baryon and CDM reduced density parameters, $\omega_b$ and $\omega_{cdm}$ respectively. The results for $\sigma_8$ and $\sigma_{12}$ are those computed at $z=0$. }
	\label{fig:nnu}
\end{figure*}

\begin{figure*}[htp]
	\centering
	\includegraphics[width=\textwidth]{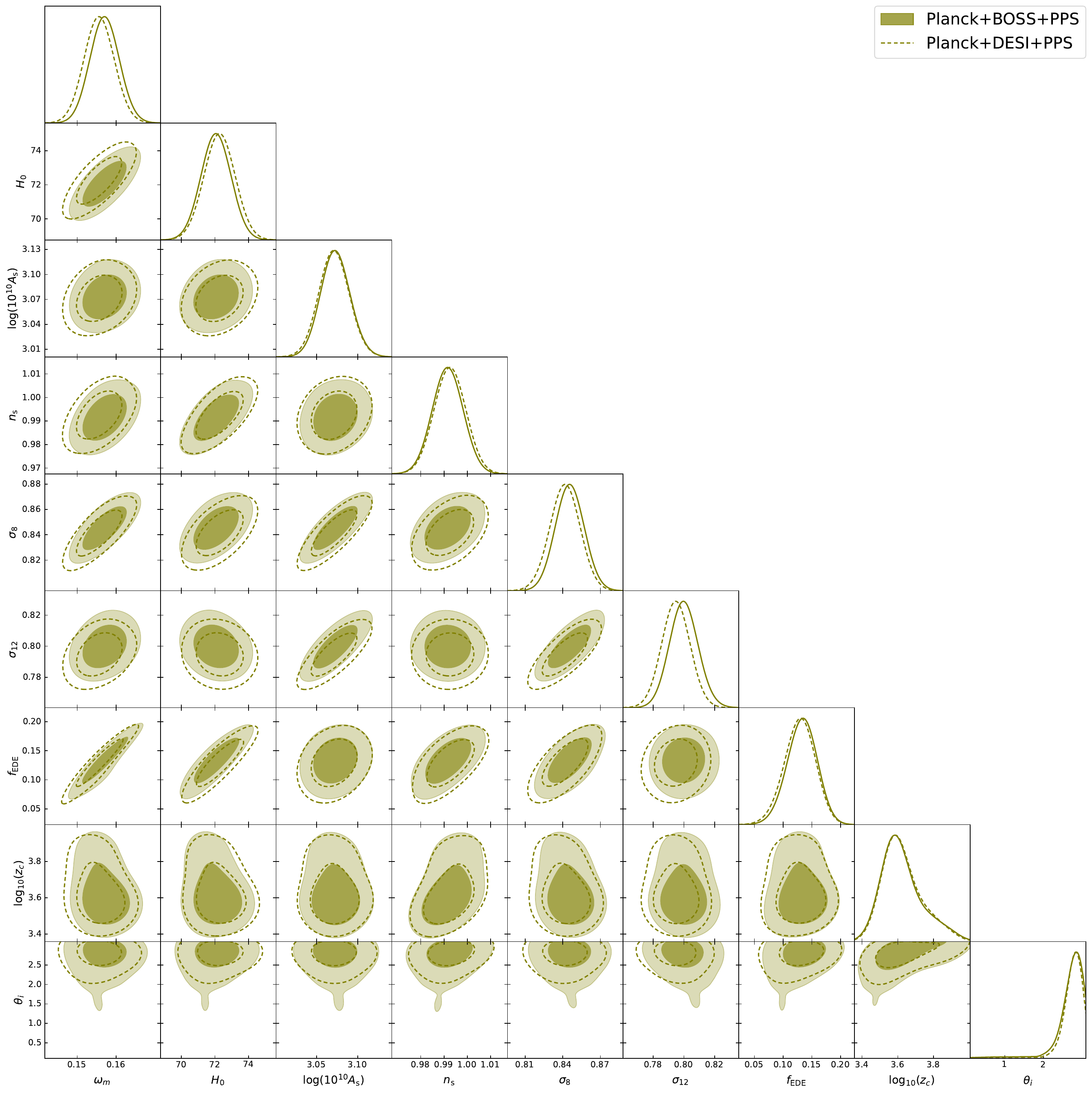}
	\caption{Triangular plot for EDE obtained from the fitting analysis with Planck+BOSS+PPS (solid contours) and Planck+DESI+PPS (dashed contours), cf. Sec. \ref{sec:data}. The parameter $\omega_m=\Omega_m^0 h^2$ is the sum of the baryon and CDM reduced density parameters, $\omega_b$ and $\omega_{cdm}$ respectively. The results for $\sigma_8$ and $\sigma_{12}$ are those computed at $z=0$.} 
	\label{fig:EDE}
\end{figure*}

We illustrate this in Fig.~\ref{fig:whisker}, which shows the constraints on $\sigma_8(z)$ and $\sigma_{12}(z)$ obtained for all the models at seven different redshifts, ranging from $z=0$ to $z=2.96$. Interestingly, no statistically significant differences are found in the values of $\sigma_{12}(z)$ across the models, irrespective of the redshift. At any given $z$, the $\sigma_{12}$ values in all models are consistent within $<1\sigma$ confidence level. This indicates that the data constrain the shape of $\sigma_{12}(z)$ in a manner that is nearly independent of the very different underlying physics in the models under study. In contrast, the values of $\sigma_8(z)$ deviate substantially from the stable values of $\sigma_{12}(z)$, with the largest differences being found in those models that favor higher values of $H_0$. Let us consider, for instance, the constraints on these two functions for EDE. They are displayed in Table \ref{tab:fit}. The constraints obtained with Planck+BOSS+PPS and Planck+DESI+PPS are very similar, so let us focus on the case of Planck+DESI+PPS for concreteness. In particular, let us consider the redshift $z=0$. We find $\sigma_{12}=0.795\pm 0.009$, which is fully compatible with the $\Lambda$CDM constraint, $\sigma_{12,\Lambda}=0.785\pm 0.008$. However, $\sigma_8=0.842\pm 0.011$ is $3.3\sigma$ higher than $\sigma_{12}$ in EDE and $2.9\sigma$ higher than $\sigma_{8,\Lambda}=0.803^{+0.007}_{-0.008}$. Hence, comparing the values of $\sigma_{8}$ obtained in $\Lambda$CDM and EDE, as is commonly done in the literature, would lead to the conclusion that EDE produces significantly greater clustering in the universe than the standard model. This, in turn, would suggest that EDE is only capable of resolving the Hubble tension at the expense of worsening substantially the growth tension. This conclusion is clearly biased, and is caused by the incorrect use of $\sigma_8$. The values of $\sigma_{12}$ and $\sigma_{12,\Lambda}$ indicate that the increase in the amplitude of fluctuations at linear scales in EDE is only mild and does not lead to a significant worsening of the growth tension.  A similar discussion applies to the other models.  Although in EDE and $\Lambda$CDM+$N_{\rm eff}$ there is an obvious increase of the reduced matter density parameter $\omega_m$ required to not enhance the early integrated Sachs-Wolfe effect, the data also allows for an increase of the cosmological constant (or $H_0$) that helps to keep the value of the matter density parameter $\Omega_m^0$ close to the one found in $\Lambda$CDM, and this contributes to have the growth of clustering under control in the late universe \cite{Gomez-Valent:2022hkb}. The contour plots obtained for EDE, shown in Fig. \ref{fig:EDE}, reinforce our arguments\footnote{We display the triangle plots of all the models under study in the last pages of the manuscript, after the bibliography list.}. One can check that the correlation between $f_{\rm EDE}$ and $\sigma_{12}$ is derisory, whereas for $\sigma_8$ the correlation is positive and non-negligible. Therefore, the parameter responsible for the increase of $H_0$, i.e., $f_{\rm EDE}$, does not lead to a significant increase in $\sigma_{12}$ because the model can adjust other parameters to compensate for its effects at the perturbation level. Similar compensations are also encountered in the other models. For instance, in Fig. \ref{fig:nnu} we observe only a positive correlation between $\sigma_8$ and the parameter that helps to increase the size of the sound horizon at recombination in $\Lambda$CDM+$N_{\rm eff}$ and $\Lambda$CDM+$N_{\rm eff}$+$\sum m_\nu$, $N_{\rm eff}$. This correlation is again absent in the $\sigma_{12}-N_{\rm eff}$ plane. 

 When quantified in terms of the parameter $S_8$ measured by weak lensing surveys, the growth tension remains small in all the models under study according to the latest data release from the Kilo-Degree Survey (KiDS) \cite{Wright:2025xka}, $S_8=0.815^{+0.016}_{-0.021}$. In fact, the tension is non-existent, since the differences between the posterior values of $S_8$ and the KiDS measurement are always below the $1\sigma$ level, with the only exception of EDE with Planck+DES+PPS, in which the difference grows to the still mild 1.4$\sigma$ CL. This conclusion is also reached when the tension is quantified in the plane $(\Omega_m,S_8)$, since the constraints on $\Omega_m$ peak close to 0.3 for all the models (cf. Table \ref{tab:fit}), in full agreement with KiDS \cite{Wright:2025xka}.

A comment on the results obtained for the sum of neutrino masses in $\Lambda$CDM+$N_{\rm eff}$+$\sum m_\nu$ is in order. We find an upper bound $\sum m_\nu<0.15$ eV at 95\% C.L. using Planck+BOSS+PPS and Planck+DESI+PPS. This constraint is a bit weaker than other constraints in the literature obtained with CMB and BAO data when $N_{\rm eff}$ is fixed to the standard value (see, e.g., \cite{Planck:2018vyg,Jiang:2024viw,Escudero:2024uea,Herold:2024nvk,Loverde:2024nfi,Allali:2024aiv,DESI:2024mwx}). The reason is that larger neutrino masses can suppress power at small scales, making the amplitude of fluctuations to be even closer to $\Lambda$CDM. Notice that this conclusion can be reached again only by looking at the value of $\sigma_{12}$. We find in both, $\Lambda$CDM and $\Lambda$CDM+$N_{\rm eff}$+$\sum m_\nu$, central values of $\sigma_{12}=0.785-0.791$. The values of $\sigma_8$ obtained in the analysis of Planck+BOSS+PPS and Planck+DESI+PPS, instead, are $1.7\sigma$ and $1.5\sigma$ larger in $\Lambda$CDM+$N_{\rm eff}$+$\sum m_\nu$ than in $\Lambda$CDM. Again, we can understand better the physics behind our results by looking at the values of $\sigma_{12}$.

We can ask now what error in $\chi^2$ would be committed if we considered data from RSD and peculiar velocities \cite{Guzzo:2008ac,Song:2008qt,Blake:2011rj,Blake:2013nif,Simpson:2015yfa,Gil-Marin:2016wya,eBOSS:2020gbb,Said:2020epb,Avila:2021dqv,Mohammad:2018mdy,Okumura:2015lvp}, treating the latter as data on $f(z)\sigma_8(z)$ or $f(z)\sigma_{12}(z)$. These data cover the redshift range $z \in (0.01, 1.50)$. We perform a very simple exercise to obtain a rough estimate. Let us take the best-fit values for the various models (using, e.g., the results obtained with Planck+DESI+PPS) and compute the following two $\chi^2$:  

\begin{equation}
\chi^2_{f\sigma_8} = \sum_{i=1}^{15}\left(\frac{f(z_i)\sigma_8(z_i)-f(z_i)\sigma_8(z_i)|_{\rm obs}}{\sigma_i}\right)^2\,,
\end{equation}
and

\begin{equation}
\chi^2_{f\sigma_{12}} = \sum_{i=1}^{15}\left(\frac{f(z_i)\sigma_{12}(z_i)-f(z_i)\sigma_{12}(z_i)|_{\rm obs}}{\sigma_i}\right)^2\,.
\end{equation}
Here, $f(z_i)\sigma_{12}(z_i)|_{\rm obs}$ are just the same observational values of $f(z_i)\sigma_8(z_i)|_{\rm obs}$, but treated as data points on $f(z_i)\sigma_{12}(z_i)$ (see discussion in Sec. \ref{sec:sec2}). With this exercise we will show that by incorrectly treating the data as data on $f(z)\sigma_8(z)$ the models with larger values of $H_0$ are unfairly penalized, since the clustering in these models is overestimated when it is evaluated in terms of the $\sigma_8$ parameter. Notice that the theoretical quantity of interest can be computed as follows, 

\begin{equation}
f(z)\sigma_R(z) = -(1+z)\frac{d\sigma_R}{dz}\,,
\end{equation}
and the derivative on the right-hand side can be calculated numerically with \texttt{CLASS}, by simply evaluating $d\sigma_R/dz\simeq (\sigma_R(z+\delta z)-\sigma_R(z))/\delta z$. Here, $\delta z$ is just a small variation in redshift, which we take to be equal to $10^{-3}$. For the $\Lambda$CDM model, we obtain $\chi^2_{f\sigma_{8}}\approx35$, whereas $\chi^2_{f\sigma_{12}}\approx10$. Models favoring a Hubble parameter around $\sim 71-72$ km/s/Mpc show an even clearer indication of the problem at hand. For instance, in the $\Lambda$CDM+$N_{\rm eff}$+$\sum m_\nu$ scenario, we find $\chi^2_{f\sigma_{8}}\approx78$ and $\chi^2_{f\sigma_{12}}\approx11$. The disparity is even more pronounced in the EDE scenario, where $\chi^2_{f\sigma_{8}}$ increases dramatically to $\approx118$. Remarkably, when evaluated in terms of $\sigma_{12}$, the chi-square remains stable since also in the latter case we obtain a much smaller value than $\chi^2_{f\sigma_8}$, $\chi^2_{f\sigma_{12}}\approx13$. Thus, all the best-fit models are able to accommodate well the data if the latter are treated as data on $f(z)\sigma_{12}(z)$, while are strongly (and artificially!) penalized if the data are treated as data on $f(z)\sigma_{8}(z)$.

Before concluding this section, we want to discuss more in detail the fitting results displayed in Table \ref{tab:fit} for the various models.  As explained in Sec. \ref{sec:models}, our choice of models have been motivated by their ability to change the large-scale structure formation processes in the universe compared to the standard model, and, in some cases, also by their potential ability to raise the value of $H_0$.  Interestingly, by means of the AIC in \eq{AIC} and simply comparing the $\chi^2$ values, we find a very substantial improvement in the description of the cosmological data employed in this study in all the non-standard cosmologies, compared to $\Lambda$CDM. 
Using Planck+BOSS+PPS we find values $\Delta {\rm AIC}>10$ for all models, which allows us to conclude that there exists very strong Bayesian evidence supporting the non-standard models against the $\Lambda$CDM. For $w_0w_a$CDM with the BOSS combination, $\Delta{\rm AIC}=13.61$ and it remains stable when the BAO data are taken from DESI; for $\Lambda$CDM+$N_{\rm eff}$ and $\Lambda$CDM+$N_{\rm eff}+\sum m_\nu$, $\Delta{\rm AIC}\sim 16-17$ using BOSS and such value is lowered to $\Delta{\rm AIC}\sim 9$ when we use the Planck+DESI+PPS combination; however, it is still above the \textit{strong evidence} threshold. For EDE we obtain the highest $\Delta{\rm AIC}$ value for both data combinations, which suggests very strong evidence in favor of a non-zero fraction of energy injected before recombination ($\Delta{\rm AIC}\sim 27$ for BOSS and $\Delta{\rm AIC}\sim 26$ for DESI). Hence, EDE is very strongly preferred over the other models due to its efficiency in alleviating the Hubble tension, based on the data analyzed in this study\footnote{If the SH0ES data were not employed, considerably lower values of $H_0$ would be obtained, see, e.g., \cite{Ivanov:2020ril,Gomez-Valent:2022hkb}. Nevertheless, this decrease has been shown to be partially due to volume (or marginalization) effects, which can be avoided by the use of profile likelihoods instead of posterior distributions \cite{Smith:2020rxx,Herold:2021ksg,Gomez-Valent:2022hkb}. See however \cite{Efstathiou:2023fbn}.}. Despite the fact that, to the best of our knowledge, not all of these combinations of models and data have been previously explored, our results aligns with previous analysis that employ observations of CMB, BAO and Type Ia Supernovae, see e.g.~\cite{Gomez-Valent:2024ejh,Giare:2024ocw,Giare:2024gpk,Yadav:2024duq,Khalife:2023qbu,Gsponer:2023wpm,Simon:2022adh,Herold:2022iib}.
   
Rather than focusing on these very interesting evidences, which are admittedly a subject of heated discussion in the literature, in this work we want to emphasize that any potential theoretical resolution of the Hubble tension will obviously entail large values of $H_0$ and, consequently, strong biases will plague our analysis of the LSS if it is performed in terms of the quantity $\sigma_8$.
The above discussion, along with the examples provided in Sec. \ref{sec:sec2}, underscores the urgent need to abandon the use of $\sigma_8$ in galaxy clustering analyses and in the comparison of the clustering properties of cosmological models. Our arguments build upon the work of A.G. Sánchez \cite{Sanchez:2020vvb} and consolidate insights previously presented in a scattered form across the literature \cite{Gomez-Valent:2021cbe,eBOSS:2021poy,Sanchez:2021plj,Gomez-Valent:2022hkb,Gomez-Valent:2022bku,Secco:2022kqg,Semenaite:2022unt,Garcia-Garcia:2024gzy,Gomez-Valent:2023hov,Gomez-Valent:2024tdb,Esposito:2024qlo,Gomez-Valent:2024ejh}.


\section{Conclusions}\label{sec:conclusions}

The widespread use of $\sigma_8$ in cosmology is no longer justified. Five years ago, A.G. Sánchez already provided significant arguments against its use \cite{Sanchez:2020vvb}. The dependence of $\sigma_8$ on $h$ arises both from its cosmological impact on the linear matter power spectrum and also from the scale $R_8 = 8h^{-1}$ Mpc in the window function used to calculate the root-mean-square mass fluctuations. This second dependence makes comparisons of measured $\sigma_8$ values reliant on the priors assumed for $h$ in the analyses. Consequently, such comparisons are prone to significant biases if the assumed priors for $h$ differ in the various galaxy clustering experiments. A problem also arises from the comparison of values of $\sigma_8$ derived in fitting analyses of models with different posterior distributions for $H_0$, or between these values and the measured ones. These biases can distort the statistical significance of the growth tension and lead to incorrect evaluations of the models' impact on large-scale structure formation in the universe. The solution to this issue is both simple and costless: replace $\sigma_8$ with a new parameter $\sigma_R$ associated to a scale $R$ that is unaffected by $h$, such as $\sigma_{12}$. The latter represents the root-mean-square mass fluctuations at a fixed scale of $R_{12}=12$ Mpc \cite{Sanchez:2020vvb}. Since Sánchez's proposal \cite{Sanchez:2020vvb}, some authors have already employed $\sigma_{12}$ to improve the characterization of the amplitude of perturbations in several works and in the context of different models \cite{Gomez-Valent:2021cbe,eBOSS:2021poy,Sanchez:2021plj,Gomez-Valent:2022hkb,Gomez-Valent:2022bku,Secco:2022kqg,Semenaite:2022unt,Garcia-Garcia:2024gzy,Gomez-Valent:2023hov,Gomez-Valent:2024tdb,Esposito:2024qlo,Gomez-Valent:2024ejh}. Nevertheless, the necessity of transitioning from $\sigma_8$ to $\sigma_{12}$ has apparently gone unnoticed by the majority of the community. 

In this paper, which builds upon those previous works, we aimed to further motivate this transition with some novel and crystal-clear examples that bring to light the impact of the bias. In addition, we have obtained an approximate analytical formula of the difference $\sigma_8-\sigma_{12}$ as a function of $h$, Eq. \eqref{eq:bias}, and illustrated how the bias impacts our interpretation of the RSD and peculiar velocity measurements of $f(z)\sigma_R(z)$. Finally, we have performed a detailed full Monte Carlo analysis considering five alternative cosmological models, namely, the $\Lambda$CDM, the $w_0w_a$CDM parameterization of dynamical dark energy, an extension of the standard model considering a larger amount of ultra-relativistic species with or without a non-minimal sum of the neutrino masses, and early dark energy. This selection of models encompasses scenarios featuring both late-time and early-time new physics, with different abilities to alleviate the Hubble tension. We have constrained them using state-of-the-art CMB data from Planck, BAO data from BOSS and DESI, and SNIa data from the Pantheon+ compilation, including the SH0ES calibration of the absolute magnitude of supernovae. Our study demonstrates that the shapes of $\sigma_{12}(z)$ obtained across the five models are stable and compatible with each other within $1\sigma$ C.L., while the shapes of $\sigma_8(z)$ suffer from spurious very large positive shifts, more conspicuously in the models that alleviate the $H_0$ tension. These shifts do not point to a real enhancement of clustering power, but are mainly induced by the shift in the scale $R_8$, which makes $\sigma_8$ to characterize the amplitude of fluctuations at completely different physical scales. In terms of $\sigma_{12}$, we do not observe any statistically significant increase of the amplitude of fluctuations at linear scales in models that are able to soften or even erase the Hubble tension, like EDE. This is in contrast to $\sigma_8$, which points exactly in the opposite direction. It is therefore clear that accounting for the existing bias by replacing $\sigma_8$ with $\sigma_{12}$ allows us to get a more reliable interpretation of the results obtained from fitting analyses and evaluate better the performance of the various models regarding the LSS growth. Moreover, our results underscore the importance of analyzing the ability of the models to simultaneously assess the Hubble and growth tension in a more robust plane like $H_0-\sigma_{12}$ when RSD data are considered. Weak lensing analyses can be still safely performed in terms of the quantity $S_8$ \cite{Secco:2022kqg,Garcia-Garcia:2024gzy}. In addition, objections previously raised against EDE and other models that introduce new physics at early times should be duly revisited, given that these models do not appear to significantly increase the amount of structure in the universe with respect to the $\Lambda$CDM, at least at linear scales.

Removing the bias introduced by $\sigma_8$ in current cosmological analyses is already urgent, and future data from ongoing surveys like Euclid and DESI will make the issue even more pressing. In view of this, we encourage the community to make that simple step towards the abandonment of $\sigma_8$ and the adoption of $\sigma_{12}$, particularly in the conflictive cases analyzed in this work.


\section*{Acknowledgments}

MF is funded by the PRIN (Progetti di ricerca di Rilevante Interesse Nazionale) number 2022WJ9J33. AF acknowledges support from the INFN project “InDark”. AGV is funded by “la Caixa” Foundation (ID 100010434) and the European Union's Horizon 2020 research and innovation programme under the Marie Sklodowska-Curie grant agreement No 847648, with fellowship code LCF/BQ/PI23/11970027. MF, AF and AGV also acknowledge the participation in the COST Action CA21136 “Addressing observational tensions in cosmology with systematics and fundamental physics” (CosmoVerse). The authors thank the anonymous referee for their insightful comments and suggestions as well as Ariel Sánchez for his very valuable feedback.

\bibliography{Bibliography}

\begin{thebibliography}{264}%
\makeatletter
\providecommand \@ifxundefined [1]{%
 \@ifx{#1\undefined}
}%
\providecommand \@ifnum [1]{%
 \ifnum #1\expandafter \@firstoftwo
 \else \expandafter \@secondoftwo
 \fi
}%
\providecommand \@ifx [1]{%
 \ifx #1\expandafter \@firstoftwo
 \else \expandafter \@secondoftwo
 \fi
}%
\providecommand \natexlab [1]{#1}%
\providecommand \enquote  [1]{``#1''}%
\providecommand \bibnamefont  [1]{#1}%
\providecommand \bibfnamefont [1]{#1}%
\providecommand \citenamefont [1]{#1}%
\providecommand \href@noop [0]{\@secondoftwo}%
\providecommand \href [0]{\begingroup \@sanitize@url \@href}%
\providecommand \@href[1]{\@@startlink{#1}\@@href}%
\providecommand \@@href[1]{\endgroup#1\@@endlink}%
\providecommand \@sanitize@url [0]{\catcode `\\12\catcode `\$12\catcode
  `\&12\catcode `\#12\catcode `\^12\catcode `\_12\catcode `\%12\relax}%
\providecommand \@@startlink[1]{}%
\providecommand \@@endlink[0]{}%
\providecommand \url  [0]{\begingroup\@sanitize@url \@url }%
\providecommand \@url [1]{\endgroup\@href {#1}{\urlprefix }}%
\providecommand \urlprefix  [0]{URL }%
\providecommand \Eprint [0]{\href }%
\providecommand \doibase [0]{http://dx.doi.org/}%
\providecommand \selectlanguage [0]{\@gobble}%
\providecommand \bibinfo  [0]{\@secondoftwo}%
\providecommand \bibfield  [0]{\@secondoftwo}%
\providecommand \translation [1]{[#1]}%
\providecommand \BibitemOpen [0]{}%
\providecommand \bibitemStop [0]{}%
\providecommand \bibitemNoStop [0]{.\EOS\space}%
\providecommand \EOS [0]{\spacefactor3000\relax}%
\providecommand \BibitemShut  [1]{\csname bibitem#1\endcsname}%
\let\auto@bib@innerbib\@empty
\bibitem [{\citenamefont {Bennett}\ \emph {et~al.}(2013)\citenamefont {Bennett}
  \emph {et~al.}}]{WMAP:2012fli}%
  \BibitemOpen
  \bibfield  {author} {\bibinfo {author} {\bibfnamefont {C.~L.}\ \bibnamefont
  {Bennett}} \emph {et~al.} (\bibinfo {collaboration} {WMAP}),\ }\href
  {\doibase 10.1088/0067-0049/208/2/20} {\bibfield  {journal} {\bibinfo
  {journal} {Astrophys. J. Suppl.}\ }\textbf {\bibinfo {volume} {208}},\
  \bibinfo {pages} {20} (\bibinfo {year} {2013})},\ \Eprint
  {http://arxiv.org/abs/1212.5225} {arXiv:1212.5225 [astro-ph.CO]} \BibitemShut
  {NoStop}%
\bibitem [{\citenamefont {Aghanim}\ \emph
  {et~al.}(2020{\natexlab{a}})\citenamefont {Aghanim} \emph
  {et~al.}}]{Planck:2018vyg}%
  \BibitemOpen
  \bibfield  {author} {\bibinfo {author} {\bibfnamefont {N.}~\bibnamefont
  {Aghanim}} \emph {et~al.} (\bibinfo {collaboration} {Planck}),\ }\href
  {\doibase 10.1051/0004-6361/201833910} {\bibfield  {journal} {\bibinfo
  {journal} {Astron. Astrophys.}\ }\textbf {\bibinfo {volume} {641}},\ \bibinfo
  {pages} {A6} (\bibinfo {year} {2020}{\natexlab{a}})},\ \bibinfo {note}
  {[Erratum: Astron.Astrophys. 652, C4 (2021)]},\ \Eprint
  {http://arxiv.org/abs/1807.06209} {arXiv:1807.06209 [astro-ph.CO]}
  \BibitemShut {NoStop}%
\bibitem [{\citenamefont {Aiola}\ \emph {et~al.}(2020)\citenamefont {Aiola}
  \emph {et~al.}}]{ACT:2020gnv}%
  \BibitemOpen
  \bibfield  {author} {\bibinfo {author} {\bibfnamefont {S.}~\bibnamefont
  {Aiola}} \emph {et~al.} (\bibinfo {collaboration} {ACT}),\ }\href {\doibase
  10.1088/1475-7516/2020/12/047} {\bibfield  {journal} {\bibinfo  {journal}
  {JCAP}\ }\textbf {\bibinfo {volume} {12}},\ \bibinfo {pages} {047} (\bibinfo
  {year} {2020})},\ \Eprint {http://arxiv.org/abs/2007.07288} {arXiv:2007.07288
  [astro-ph.CO]} \BibitemShut {NoStop}%
\bibitem [{\citenamefont {Chou}\ \emph {et~al.}(2025)\citenamefont {Chou} \emph
  {et~al.}}]{SPTpol:2025kpo}%
  \BibitemOpen
  \bibfield  {author} {\bibinfo {author} {\bibfnamefont {T.~L.}\ \bibnamefont
  {Chou}} \emph {et~al.} (\bibinfo {collaboration} {SPTpol}),\ }\href {\doibase
  10.1103/PhysRevD.111.123513} {\bibfield  {journal} {\bibinfo  {journal}
  {Phys. Rev. D}\ }\textbf {\bibinfo {volume} {111}},\ \bibinfo {pages}
  {123513} (\bibinfo {year} {2025})},\ \Eprint
  {http://arxiv.org/abs/2501.06890} {arXiv:2501.06890 [astro-ph.CO]}
  \BibitemShut {NoStop}%
\bibitem [{\citenamefont {Alam}\ \emph
  {et~al.}(2017{\natexlab{a}})\citenamefont {Alam} \emph
  {et~al.}}]{BOSS:2016wmc}%
  \BibitemOpen
  \bibfield  {author} {\bibinfo {author} {\bibfnamefont {S.}~\bibnamefont
  {Alam}} \emph {et~al.} (\bibinfo {collaboration} {BOSS}),\ }\href {\doibase
  10.1093/mnras/stx721} {\bibfield  {journal} {\bibinfo  {journal} {Mon. Not.
  Roy. Astron. Soc.}\ }\textbf {\bibinfo {volume} {470}},\ \bibinfo {pages}
  {2617} (\bibinfo {year} {2017}{\natexlab{a}})},\ \Eprint
  {http://arxiv.org/abs/1607.03155} {arXiv:1607.03155 [astro-ph.CO]}
  \BibitemShut {NoStop}%
\bibitem [{\citenamefont {Heymans}\ \emph {et~al.}(2021)\citenamefont {Heymans}
  \emph {et~al.}}]{Heymans:2020gsg}%
  \BibitemOpen
  \bibfield  {author} {\bibinfo {author} {\bibfnamefont {C.}~\bibnamefont
  {Heymans}} \emph {et~al.},\ }\href {\doibase 10.1051/0004-6361/202039063}
  {\bibfield  {journal} {\bibinfo  {journal} {Astron. Astrophys.}\ }\textbf
  {\bibinfo {volume} {646}},\ \bibinfo {pages} {A140} (\bibinfo {year}
  {2021})},\ \Eprint {http://arxiv.org/abs/2007.15632} {arXiv:2007.15632
  [astro-ph.CO]} \BibitemShut {NoStop}%
\bibitem [{\citenamefont {Abbott}\ \emph {et~al.}(2022)\citenamefont {Abbott}
  \emph {et~al.}}]{DES:2021wwk}%
  \BibitemOpen
  \bibfield  {author} {\bibinfo {author} {\bibfnamefont {T.~M.~C.}\
  \bibnamefont {Abbott}} \emph {et~al.} (\bibinfo {collaboration} {DES}),\
  }\href {\doibase 10.1103/PhysRevD.105.023520} {\bibfield  {journal} {\bibinfo
   {journal} {Phys. Rev. D}\ }\textbf {\bibinfo {volume} {105}},\ \bibinfo
  {pages} {023520} (\bibinfo {year} {2022})},\ \Eprint
  {http://arxiv.org/abs/2105.13549} {arXiv:2105.13549 [astro-ph.CO]}
  \BibitemShut {NoStop}%
\bibitem [{\citenamefont {Adame}\ \emph
  {et~al.}(2024{\natexlab{a}})\citenamefont {Adame} \emph
  {et~al.}}]{DESI:2024hhd}%
  \BibitemOpen
  \bibfield  {author} {\bibinfo {author} {\bibfnamefont {A.~G.}\ \bibnamefont
  {Adame}} \emph {et~al.} (\bibinfo {collaboration} {DESI}),\ }\href@noop {} {\
   (\bibinfo {year} {2024}{\natexlab{a}})},\ \Eprint
  {http://arxiv.org/abs/2411.12022} {arXiv:2411.12022 [astro-ph.CO]}
  \BibitemShut {NoStop}%
\bibitem [{\citenamefont {Huterer}(2023)}]{Huterer:2022dds}%
  \BibitemOpen
  \bibfield  {author} {\bibinfo {author} {\bibfnamefont {D.}~\bibnamefont
  {Huterer}},\ }\href {\doibase 10.1007/s00159-023-00147-4} {\bibfield
  {journal} {\bibinfo  {journal} {Astron. Astrophys. Rev.}\ }\textbf {\bibinfo
  {volume} {31}},\ \bibinfo {pages} {2} (\bibinfo {year} {2023})},\ \Eprint
  {http://arxiv.org/abs/2212.05003} {arXiv:2212.05003 [astro-ph.CO]}
  \BibitemShut {NoStop}%
\bibitem [{\citenamefont {Kolb}(2019)}]{Kolb:1990vq}%
  \BibitemOpen
  \bibfield  {author} {\bibinfo {author} {\bibfnamefont {E.~W.}\ \bibnamefont
  {Kolb}},\ }\href {\doibase 10.1201/9780429492860} {\emph {\bibinfo {title}
  {{The Early Universe}}}},\ Vol.~\bibinfo {volume} {69}\ (\bibinfo
  {publisher} {Taylor and Francis},\ \bibinfo {year} {2019})\BibitemShut
  {NoStop}%
\bibitem [{\citenamefont {Liddle}\ and\ \citenamefont
  {Lyth}(2000)}]{Liddle:2000cg}%
  \BibitemOpen
  \bibfield  {author} {\bibinfo {author} {\bibfnamefont {A.~R.}\ \bibnamefont
  {Liddle}}\ and\ \bibinfo {author} {\bibfnamefont {D.~H.}\ \bibnamefont
  {Lyth}},\ }\href {\doibase 10.1017/CBO9781139175180} {\emph {\bibinfo {title}
  {{Cosmological inflation and large scale structure}}}}\ (\bibinfo {year}
  {2000})\BibitemShut {NoStop}%
\bibitem [{\citenamefont {Sánchez}(2020)}]{Sanchez:2020vvb}%
  \BibitemOpen
  \bibfield  {author} {\bibinfo {author} {\bibfnamefont {A.~G.}\ \bibnamefont
  {Sánchez}},\ }\href {\doibase 10.1103/PhysRevD.102.123511} {\bibfield
  {journal} {\bibinfo  {journal} {Phys. Rev. D}\ }\textbf {\bibinfo {volume}
  {102}},\ \bibinfo {pages} {123511} (\bibinfo {year} {2020})},\ \Eprint
  {http://arxiv.org/abs/2002.07829} {arXiv:2002.07829 [astro-ph.CO]}
  \BibitemShut {NoStop}%
\bibitem [{\citenamefont {G\'omez-Valent}\ and\ \citenamefont
  {Sol\`a~Peracaula}(2018)}]{Gomez-Valent:2018nib}%
  \BibitemOpen
  \bibfield  {author} {\bibinfo {author} {\bibfnamefont {A.}~\bibnamefont
  {G\'omez-Valent}}\ and\ \bibinfo {author} {\bibfnamefont {J.}~\bibnamefont
  {Sol\`a~Peracaula}},\ }\href {\doibase 10.1093/mnras/sty1028} {\bibfield
  {journal} {\bibinfo  {journal} {Mon. Not. Roy. Astron. Soc.}\ }\textbf
  {\bibinfo {volume} {478}},\ \bibinfo {pages} {126} (\bibinfo {year}
  {2018})},\ \Eprint {http://arxiv.org/abs/1801.08501} {arXiv:1801.08501
  [astro-ph.CO]} \BibitemShut {NoStop}%
\bibitem [{\citenamefont {Sol\`a~Peracaula}\ \emph {et~al.}(2020)\citenamefont
  {Sol\`a~Peracaula}, \citenamefont {G\'omez-Valent}, \citenamefont
  {de~Cruz~P\'erez},\ and\ \citenamefont
  {Moreno-Pulido}}]{SolaPeracaula:2020vpg}%
  \BibitemOpen
  \bibfield  {author} {\bibinfo {author} {\bibfnamefont {J.}~\bibnamefont
  {Sol\`a~Peracaula}}, \bibinfo {author} {\bibfnamefont {A.}~\bibnamefont
  {G\'omez-Valent}}, \bibinfo {author} {\bibfnamefont {J.}~\bibnamefont
  {de~Cruz~P\'erez}}, \ and\ \bibinfo {author} {\bibfnamefont {C.}~\bibnamefont
  {Moreno-Pulido}},\ }\href {\doibase 10.1088/1361-6382/abbc43} {\bibfield
  {journal} {\bibinfo  {journal} {Class. Quant. Grav.}\ }\textbf {\bibinfo
  {volume} {37}},\ \bibinfo {pages} {245003} (\bibinfo {year} {2020})},\
  \Eprint {http://arxiv.org/abs/2006.04273} {arXiv:2006.04273 [astro-ph.CO]}
  \BibitemShut {NoStop}%
\bibitem [{\citenamefont {Di~Valentino}\ \emph
  {et~al.}(2021{\natexlab{a}})\citenamefont {Di~Valentino} \emph
  {et~al.}}]{DiValentino:2020vvd}%
  \BibitemOpen
  \bibfield  {author} {\bibinfo {author} {\bibfnamefont {E.}~\bibnamefont
  {Di~Valentino}} \emph {et~al.},\ }\href {\doibase
  10.1016/j.astropartphys.2021.102604} {\bibfield  {journal} {\bibinfo
  {journal} {Astropart. Phys.}\ }\textbf {\bibinfo {volume} {131}},\ \bibinfo
  {pages} {102604} (\bibinfo {year} {2021}{\natexlab{a}})},\ \Eprint
  {http://arxiv.org/abs/2008.11285} {arXiv:2008.11285 [astro-ph.CO]}
  \BibitemShut {NoStop}%
\bibitem [{\citenamefont {Marra}\ and\ \citenamefont
  {Perivolaropoulos}(2021)}]{Marra:2021fvf}%
  \BibitemOpen
  \bibfield  {author} {\bibinfo {author} {\bibfnamefont {V.}~\bibnamefont
  {Marra}}\ and\ \bibinfo {author} {\bibfnamefont {L.}~\bibnamefont
  {Perivolaropoulos}},\ }\href {\doibase 10.1103/PhysRevD.104.L021303}
  {\bibfield  {journal} {\bibinfo  {journal} {Phys. Rev. D}\ }\textbf {\bibinfo
  {volume} {104}},\ \bibinfo {pages} {L021303} (\bibinfo {year} {2021})},\
  \Eprint {http://arxiv.org/abs/2102.06012} {arXiv:2102.06012 [astro-ph.CO]}
  \BibitemShut {NoStop}%
\bibitem [{\citenamefont {Hall}(2021)}]{Hall:2021qjk}%
  \BibitemOpen
  \bibfield  {author} {\bibinfo {author} {\bibfnamefont {A.}~\bibnamefont
  {Hall}},\ }\href {\doibase 10.1093/mnras/stab1563} {\bibfield  {journal}
  {\bibinfo  {journal} {Mon. Not. Roy. Astron. Soc.}\ }\textbf {\bibinfo
  {volume} {505}},\ \bibinfo {pages} {4935} (\bibinfo {year} {2021})},\ \Eprint
  {http://arxiv.org/abs/2104.12880} {arXiv:2104.12880 [astro-ph.CO]}
  \BibitemShut {NoStop}%
\bibitem [{\citenamefont {Abdalla}\ \emph {et~al.}(2022)\citenamefont {Abdalla}
  \emph {et~al.}}]{Abdalla:2022yfr}%
  \BibitemOpen
  \bibfield  {author} {\bibinfo {author} {\bibfnamefont {E.}~\bibnamefont
  {Abdalla}} \emph {et~al.},\ }\href {\doibase 10.1016/j.jheap.2022.04.002}
  {\bibfield  {journal} {\bibinfo  {journal} {JHEAp}\ }\textbf {\bibinfo
  {volume} {34}},\ \bibinfo {pages} {49} (\bibinfo {year} {2022})},\ \Eprint
  {http://arxiv.org/abs/2203.06142} {arXiv:2203.06142 [astro-ph.CO]}
  \BibitemShut {NoStop}%
\bibitem [{\citenamefont {Pedreira}\ \emph {et~al.}(2024)\citenamefont
  {Pedreira}, \citenamefont {Benetti}, \citenamefont {Ferreira}, \citenamefont
  {Graef},\ and\ \citenamefont {Herold}}]{Pedreira:2023qqt}%
  \BibitemOpen
  \bibfield  {author} {\bibinfo {author} {\bibfnamefont {I.~d. O.~C.}\
  \bibnamefont {Pedreira}}, \bibinfo {author} {\bibfnamefont {M.}~\bibnamefont
  {Benetti}}, \bibinfo {author} {\bibfnamefont {E.~G.~M.}\ \bibnamefont
  {Ferreira}}, \bibinfo {author} {\bibfnamefont {L.~L.}\ \bibnamefont {Graef}},
  \ and\ \bibinfo {author} {\bibfnamefont {L.}~\bibnamefont {Herold}},\ }\href
  {\doibase 10.1103/PhysRevD.109.103525} {\bibfield  {journal} {\bibinfo
  {journal} {Phys. Rev. D}\ }\textbf {\bibinfo {volume} {109}},\ \bibinfo
  {pages} {103525} (\bibinfo {year} {2024})},\ \Eprint
  {http://arxiv.org/abs/2311.04977} {arXiv:2311.04977 [astro-ph.CO]}
  \BibitemShut {NoStop}%
\bibitem [{\citenamefont {Vagnozzi}(2023)}]{Vagnozzi:2023nrq}%
  \BibitemOpen
  \bibfield  {author} {\bibinfo {author} {\bibfnamefont {S.}~\bibnamefont
  {Vagnozzi}},\ }\href {\doibase 10.3390/universe9090393} {\bibfield  {journal}
  {\bibinfo  {journal} {Universe}\ }\textbf {\bibinfo {volume} {9}},\ \bibinfo
  {pages} {393} (\bibinfo {year} {2023})},\ \Eprint
  {http://arxiv.org/abs/2308.16628} {arXiv:2308.16628 [astro-ph.CO]}
  \BibitemShut {NoStop}%
\bibitem [{\citenamefont {Sakr}(2023)}]{Sakr:2023bms}%
  \BibitemOpen
  \bibfield  {author} {\bibinfo {author} {\bibfnamefont {Z.}~\bibnamefont
  {Sakr}},\ }\href {\doibase 10.3390/universe9080366} {\bibfield  {journal}
  {\bibinfo  {journal} {Universe}\ }\textbf {\bibinfo {volume} {9}},\ \bibinfo
  {pages} {366} (\bibinfo {year} {2023})},\ \Eprint
  {http://arxiv.org/abs/2305.02863} {arXiv:2305.02863 [astro-ph.CO]}
  \BibitemShut {NoStop}%
\bibitem [{\citenamefont {Secco}\ \emph {et~al.}(2023)\citenamefont {Secco},
  \citenamefont {Karwal}, \citenamefont {Hu},\ and\ \citenamefont
  {Krause}}]{Secco:2022kqg}%
  \BibitemOpen
  \bibfield  {author} {\bibinfo {author} {\bibfnamefont {L.~F.}\ \bibnamefont
  {Secco}}, \bibinfo {author} {\bibfnamefont {T.}~\bibnamefont {Karwal}},
  \bibinfo {author} {\bibfnamefont {W.}~\bibnamefont {Hu}}, \ and\ \bibinfo
  {author} {\bibfnamefont {E.}~\bibnamefont {Krause}},\ }\href {\doibase
  10.1103/PhysRevD.107.083532} {\bibfield  {journal} {\bibinfo  {journal}
  {Phys. Rev. D}\ }\textbf {\bibinfo {volume} {107}},\ \bibinfo {pages}
  {083532} (\bibinfo {year} {2023})},\ \Eprint
  {http://arxiv.org/abs/2209.12997} {arXiv:2209.12997 [astro-ph.CO]}
  \BibitemShut {NoStop}%
\bibitem [{\citenamefont {Simon}\ \emph {et~al.}(2025)\citenamefont {Simon},
  \citenamefont {Adi}, \citenamefont {Bernal}, \citenamefont {Kovetz},
  \citenamefont {Poulin},\ and\ \citenamefont {Smith}}]{Simon:2024jmu}%
  \BibitemOpen
  \bibfield  {author} {\bibinfo {author} {\bibfnamefont {T.}~\bibnamefont
  {Simon}}, \bibinfo {author} {\bibfnamefont {T.}~\bibnamefont {Adi}}, \bibinfo
  {author} {\bibfnamefont {J.~L.}\ \bibnamefont {Bernal}}, \bibinfo {author}
  {\bibfnamefont {E.~D.}\ \bibnamefont {Kovetz}}, \bibinfo {author}
  {\bibfnamefont {V.}~\bibnamefont {Poulin}}, \ and\ \bibinfo {author}
  {\bibfnamefont {T.~L.}\ \bibnamefont {Smith}},\ }\href {\doibase
  10.1103/PhysRevD.111.023523} {\bibfield  {journal} {\bibinfo  {journal}
  {Phys. Rev. D}\ }\textbf {\bibinfo {volume} {111}},\ \bibinfo {pages}
  {023523} (\bibinfo {year} {2025})},\ \Eprint
  {http://arxiv.org/abs/2410.21459} {arXiv:2410.21459 [astro-ph.CO]}
  \BibitemShut {NoStop}%
\bibitem [{\citenamefont {Roy~Choudhury}\ and\ \citenamefont
  {Okumura}(2024)}]{RoyChoudhury:2024wri}%
  \BibitemOpen
  \bibfield  {author} {\bibinfo {author} {\bibfnamefont {S.}~\bibnamefont
  {Roy~Choudhury}}\ and\ \bibinfo {author} {\bibfnamefont {T.}~\bibnamefont
  {Okumura}},\ }\href {\doibase 10.3847/2041-8213/ad8c26} {\bibfield  {journal}
  {\bibinfo  {journal} {Astrophys. J. Lett.}\ }\textbf {\bibinfo {volume}
  {976}},\ \bibinfo {pages} {L11} (\bibinfo {year} {2024})},\ \Eprint
  {http://arxiv.org/abs/2409.13022} {arXiv:2409.13022 [astro-ph.CO]}
  \BibitemShut {NoStop}%
\bibitem [{\citenamefont {Liu}\ \emph {et~al.}(2024{\natexlab{a}})\citenamefont
  {Liu}, \citenamefont {Yu},\ and\ \citenamefont {Wu}}]{Liu:2024vlt}%
  \BibitemOpen
  \bibfield  {author} {\bibinfo {author} {\bibfnamefont {Y.}~\bibnamefont
  {Liu}}, \bibinfo {author} {\bibfnamefont {H.}~\bibnamefont {Yu}}, \ and\
  \bibinfo {author} {\bibfnamefont {P.}~\bibnamefont {Wu}},\ }\href {\doibase
  10.1103/PhysRevD.110.L021304} {\bibfield  {journal} {\bibinfo  {journal}
  {Phys. Rev. D}\ }\textbf {\bibinfo {volume} {110}},\ \bibinfo {pages}
  {L021304} (\bibinfo {year} {2024}{\natexlab{a}})},\ \Eprint
  {http://arxiv.org/abs/2406.02956} {arXiv:2406.02956 [astro-ph.CO]}
  \BibitemShut {NoStop}%
\bibitem [{\citenamefont {Toda}\ and\ \citenamefont
  {Seto}(2025)}]{Toda:2024uff}%
  \BibitemOpen
  \bibfield  {author} {\bibinfo {author} {\bibfnamefont {Y.}~\bibnamefont
  {Toda}}\ and\ \bibinfo {author} {\bibfnamefont {O.}~\bibnamefont {Seto}},\
  }\href {\doibase 10.1103/PhysRevD.111.083551} {\bibfield  {journal} {\bibinfo
   {journal} {Phys. Rev. D}\ }\textbf {\bibinfo {volume} {111}},\ \bibinfo
  {pages} {083551} (\bibinfo {year} {2025})},\ \Eprint
  {http://arxiv.org/abs/2410.21925} {arXiv:2410.21925 [astro-ph.CO]}
  \BibitemShut {NoStop}%
\bibitem [{\citenamefont {G\'omez-Valent}\ \emph {et~al.}(2021)\citenamefont
  {G\'omez-Valent}, \citenamefont {Zheng}, \citenamefont {Amendola},
  \citenamefont {Pettorino},\ and\ \citenamefont
  {Wetterich}}]{Gomez-Valent:2021cbe}%
  \BibitemOpen
  \bibfield  {author} {\bibinfo {author} {\bibfnamefont {A.}~\bibnamefont
  {G\'omez-Valent}}, \bibinfo {author} {\bibfnamefont {Z.}~\bibnamefont
  {Zheng}}, \bibinfo {author} {\bibfnamefont {L.}~\bibnamefont {Amendola}},
  \bibinfo {author} {\bibfnamefont {V.}~\bibnamefont {Pettorino}}, \ and\
  \bibinfo {author} {\bibfnamefont {C.}~\bibnamefont {Wetterich}},\ }\href
  {\doibase 10.1103/PhysRevD.104.083536} {\bibfield  {journal} {\bibinfo
  {journal} {Phys. Rev. D}\ }\textbf {\bibinfo {volume} {104}},\ \bibinfo
  {pages} {083536} (\bibinfo {year} {2021})},\ \Eprint
  {http://arxiv.org/abs/2107.11065} {arXiv:2107.11065 [astro-ph.CO]}
  \BibitemShut {NoStop}%
\bibitem [{\citenamefont {Semenaite}\ \emph {et~al.}(2022)\citenamefont
  {Semenaite} \emph {et~al.}}]{eBOSS:2021poy}%
  \BibitemOpen
  \bibfield  {author} {\bibinfo {author} {\bibfnamefont {A.}~\bibnamefont
  {Semenaite}} \emph {et~al.} (\bibinfo {collaboration} {eBOSS}),\ }\href
  {\doibase 10.1093/mnras/stac829} {\bibfield  {journal} {\bibinfo  {journal}
  {Mon. Not. Roy. Astron. Soc.}\ }\textbf {\bibinfo {volume} {512}},\ \bibinfo
  {pages} {5657} (\bibinfo {year} {2022})},\ \Eprint
  {http://arxiv.org/abs/2111.03156} {arXiv:2111.03156 [astro-ph.CO]}
  \BibitemShut {NoStop}%
\bibitem [{\citenamefont {S\'anchez}\ \emph {et~al.}(2022)\citenamefont
  {S\'anchez}, \citenamefont {Ruiz}, \citenamefont {Jara},\ and\ \citenamefont
  {Padilla}}]{Sanchez:2021plj}%
  \BibitemOpen
  \bibfield  {author} {\bibinfo {author} {\bibfnamefont {A.~G.}\ \bibnamefont
  {S\'anchez}}, \bibinfo {author} {\bibfnamefont {A.~N.}\ \bibnamefont {Ruiz}},
  \bibinfo {author} {\bibfnamefont {J.~G.}\ \bibnamefont {Jara}}, \ and\
  \bibinfo {author} {\bibfnamefont {N.~D.}\ \bibnamefont {Padilla}},\ }\href
  {\doibase 10.1093/mnras/stac1656} {\bibfield  {journal} {\bibinfo  {journal}
  {Mon. Not. Roy. Astron. Soc.}\ }\textbf {\bibinfo {volume} {514}},\ \bibinfo
  {pages} {5673} (\bibinfo {year} {2022})},\ \Eprint
  {http://arxiv.org/abs/2108.12710} {arXiv:2108.12710 [astro-ph.CO]}
  \BibitemShut {NoStop}%
\bibitem [{\citenamefont {G\'omez-Valent}(2022)}]{Gomez-Valent:2022hkb}%
  \BibitemOpen
  \bibfield  {author} {\bibinfo {author} {\bibfnamefont {A.}~\bibnamefont
  {G\'omez-Valent}},\ }\href {\doibase 10.1103/PhysRevD.106.063506} {\bibfield
  {journal} {\bibinfo  {journal} {Phys. Rev. D}\ }\textbf {\bibinfo {volume}
  {106}},\ \bibinfo {pages} {063506} (\bibinfo {year} {2022})},\ \Eprint
  {http://arxiv.org/abs/2203.16285} {arXiv:2203.16285 [astro-ph.CO]}
  \BibitemShut {NoStop}%
\bibitem [{\citenamefont {G\'omez-Valent}\ \emph {et~al.}(2022)\citenamefont
  {G\'omez-Valent}, \citenamefont {Zheng}, \citenamefont {Amendola},
  \citenamefont {Wetterich},\ and\ \citenamefont
  {Pettorino}}]{Gomez-Valent:2022bku}%
  \BibitemOpen
  \bibfield  {author} {\bibinfo {author} {\bibfnamefont {A.}~\bibnamefont
  {G\'omez-Valent}}, \bibinfo {author} {\bibfnamefont {Z.}~\bibnamefont
  {Zheng}}, \bibinfo {author} {\bibfnamefont {L.}~\bibnamefont {Amendola}},
  \bibinfo {author} {\bibfnamefont {C.}~\bibnamefont {Wetterich}}, \ and\
  \bibinfo {author} {\bibfnamefont {V.}~\bibnamefont {Pettorino}},\ }\href
  {\doibase 10.1103/PhysRevD.106.103522} {\bibfield  {journal} {\bibinfo
  {journal} {Phys. Rev. D}\ }\textbf {\bibinfo {volume} {106}},\ \bibinfo
  {pages} {103522} (\bibinfo {year} {2022})},\ \Eprint
  {http://arxiv.org/abs/2207.14487} {arXiv:2207.14487 [astro-ph.CO]}
  \BibitemShut {NoStop}%
\bibitem [{\citenamefont {Semenaite}\ \emph {et~al.}(2023)\citenamefont
  {Semenaite}, \citenamefont {S\'anchez}, \citenamefont {Pezzotta},
  \citenamefont {Hou}, \citenamefont {Eggemeier}, \citenamefont {Crocce},
  \citenamefont {Zhao}, \citenamefont {Brownstein}, \citenamefont {Rossi},\
  and\ \citenamefont {Schneider}}]{Semenaite:2022unt}%
  \BibitemOpen
  \bibfield  {author} {\bibinfo {author} {\bibfnamefont {A.}~\bibnamefont
  {Semenaite}}, \bibinfo {author} {\bibfnamefont {A.~G.}\ \bibnamefont
  {S\'anchez}}, \bibinfo {author} {\bibfnamefont {A.}~\bibnamefont {Pezzotta}},
  \bibinfo {author} {\bibfnamefont {J.}~\bibnamefont {Hou}}, \bibinfo {author}
  {\bibfnamefont {A.}~\bibnamefont {Eggemeier}}, \bibinfo {author}
  {\bibfnamefont {M.}~\bibnamefont {Crocce}}, \bibinfo {author} {\bibfnamefont
  {C.}~\bibnamefont {Zhao}}, \bibinfo {author} {\bibfnamefont {J.~R.}\
  \bibnamefont {Brownstein}}, \bibinfo {author} {\bibfnamefont
  {G.}~\bibnamefont {Rossi}}, \ and\ \bibinfo {author} {\bibfnamefont {D.~P.}\
  \bibnamefont {Schneider}},\ }\href {\doibase 10.1093/mnras/stad849}
  {\bibfield  {journal} {\bibinfo  {journal} {Mon. Not. Roy. Astron. Soc.}\
  }\textbf {\bibinfo {volume} {521}},\ \bibinfo {pages} {5013} (\bibinfo {year}
  {2023})},\ \Eprint {http://arxiv.org/abs/2210.07304} {arXiv:2210.07304
  [astro-ph.CO]} \BibitemShut {NoStop}%
\bibitem [{\citenamefont {Garc\'\i{}a-Garc\'\i{}a}\ \emph
  {et~al.}(2024)\citenamefont {Garc\'\i{}a-Garc\'\i{}a}, \citenamefont
  {Zennaro}, \citenamefont {Aric\`o}, \citenamefont {Alonso},\ and\
  \citenamefont {Angulo}}]{Garcia-Garcia:2024gzy}%
  \BibitemOpen
  \bibfield  {author} {\bibinfo {author} {\bibfnamefont {C.}~\bibnamefont
  {Garc\'\i{}a-Garc\'\i{}a}}, \bibinfo {author} {\bibfnamefont
  {M.}~\bibnamefont {Zennaro}}, \bibinfo {author} {\bibfnamefont
  {G.}~\bibnamefont {Aric\`o}}, \bibinfo {author} {\bibfnamefont
  {D.}~\bibnamefont {Alonso}}, \ and\ \bibinfo {author} {\bibfnamefont {R.~E.}\
  \bibnamefont {Angulo}},\ }\href {\doibase 10.1088/1475-7516/2024/08/024}
  {\bibfield  {journal} {\bibinfo  {journal} {JCAP}\ }\textbf {\bibinfo
  {volume} {08}},\ \bibinfo {pages} {024} (\bibinfo {year} {2024})},\ \Eprint
  {http://arxiv.org/abs/2403.13794} {arXiv:2403.13794 [astro-ph.CO]}
  \BibitemShut {NoStop}%
\bibitem [{\citenamefont {G\'omez-Valent}\ \emph
  {et~al.}(2024{\natexlab{a}})\citenamefont {G\'omez-Valent}, \citenamefont
  {Mavromatos},\ and\ \citenamefont {Sol\`a~Peracaula}}]{Gomez-Valent:2023hov}%
  \BibitemOpen
  \bibfield  {author} {\bibinfo {author} {\bibfnamefont {A.}~\bibnamefont
  {G\'omez-Valent}}, \bibinfo {author} {\bibfnamefont {N.~E.}\ \bibnamefont
  {Mavromatos}}, \ and\ \bibinfo {author} {\bibfnamefont {J.}~\bibnamefont
  {Sol\`a~Peracaula}},\ }\href {\doibase 10.1088/1361-6382/ad0fb8} {\bibfield
  {journal} {\bibinfo  {journal} {Class. Quant. Grav.}\ }\textbf {\bibinfo
  {volume} {41}},\ \bibinfo {pages} {015026} (\bibinfo {year}
  {2024}{\natexlab{a}})},\ \Eprint {http://arxiv.org/abs/2305.15774}
  {arXiv:2305.15774 [gr-qc]} \BibitemShut {NoStop}%
\bibitem [{\citenamefont {G\'omez-Valent}\ and\ \citenamefont
  {Sol\`a~Peracaula}(2024)}]{Gomez-Valent:2024tdb}%
  \BibitemOpen
  \bibfield  {author} {\bibinfo {author} {\bibfnamefont {A.}~\bibnamefont
  {G\'omez-Valent}}\ and\ \bibinfo {author} {\bibfnamefont {J.}~\bibnamefont
  {Sol\`a~Peracaula}},\ }\href {\doibase 10.3847/1538-4357/ad7a62} {\bibfield
  {journal} {\bibinfo  {journal} {Astrophys. J.}\ }\textbf {\bibinfo {volume}
  {975}},\ \bibinfo {pages} {64} (\bibinfo {year} {2024})},\ \Eprint
  {http://arxiv.org/abs/2404.18845} {arXiv:2404.18845 [astro-ph.CO]}
  \BibitemShut {NoStop}%
\bibitem [{\citenamefont {Esposito}\ \emph {et~al.}(2024)\citenamefont
  {Esposito}, \citenamefont {S\'anchez}, \citenamefont {Bel},\ and\
  \citenamefont {Ruiz}}]{Esposito:2024qlo}%
  \BibitemOpen
  \bibfield  {author} {\bibinfo {author} {\bibfnamefont {M.}~\bibnamefont
  {Esposito}}, \bibinfo {author} {\bibfnamefont {A.~G.}\ \bibnamefont
  {S\'anchez}}, \bibinfo {author} {\bibfnamefont {J.}~\bibnamefont {Bel}}, \
  and\ \bibinfo {author} {\bibfnamefont {A.~N.}\ \bibnamefont {Ruiz}},\ }\href
  {\doibase 10.1093/mnras/stae2351} {\bibfield  {journal} {\bibinfo  {journal}
  {Mon. Not. Roy. Astron. Soc.}\ }\textbf {\bibinfo {volume} {534}},\ \bibinfo
  {pages} {3906} (\bibinfo {year} {2024})},\ \Eprint
  {http://arxiv.org/abs/2406.08539} {arXiv:2406.08539 [astro-ph.CO]}
  \BibitemShut {NoStop}%
\bibitem [{\citenamefont {G\'omez-Valent}\ and\ \citenamefont
  {Sol\`a~Peracaula}(2025)}]{Gomez-Valent:2024ejh}%
  \BibitemOpen
  \bibfield  {author} {\bibinfo {author} {\bibfnamefont {A.}~\bibnamefont
  {G\'omez-Valent}}\ and\ \bibinfo {author} {\bibfnamefont {J.}~\bibnamefont
  {Sol\`a~Peracaula}},\ }\href {\doibase 10.1016/j.physletb.2025.139391}
  {\bibfield  {journal} {\bibinfo  {journal} {Phys. Lett. B}\ }\textbf
  {\bibinfo {volume} {864}},\ \bibinfo {pages} {139391} (\bibinfo {year}
  {2025})},\ \Eprint {http://arxiv.org/abs/2412.15124} {arXiv:2412.15124
  [astro-ph.CO]} \BibitemShut {NoStop}%
\bibitem [{\citenamefont {Perivolaropoulos}\ and\ \citenamefont
  {Skara}(2022{\natexlab{a}})}]{Perivolaropoulos:2021jda}%
  \BibitemOpen
  \bibfield  {author} {\bibinfo {author} {\bibfnamefont {L.}~\bibnamefont
  {Perivolaropoulos}}\ and\ \bibinfo {author} {\bibfnamefont {F.}~\bibnamefont
  {Skara}},\ }\href {\doibase 10.1016/j.newar.2022.101659} {\bibfield
  {journal} {\bibinfo  {journal} {New Astron. Rev.}\ }\textbf {\bibinfo
  {volume} {95}},\ \bibinfo {pages} {101659} (\bibinfo {year}
  {2022}{\natexlab{a}})},\ \Eprint {http://arxiv.org/abs/2105.05208}
  {arXiv:2105.05208 [astro-ph.CO]} \BibitemShut {NoStop}%
\bibitem [{\citenamefont {Aluri}\ \emph {et~al.}(2023)\citenamefont {Aluri}
  \emph {et~al.}}]{Aluri:2022hzs}%
  \BibitemOpen
  \bibfield  {author} {\bibinfo {author} {\bibfnamefont {P.~K.}\ \bibnamefont
  {Aluri}} \emph {et~al.},\ }\href {\doibase 10.1088/1361-6382/acbefc}
  {\bibfield  {journal} {\bibinfo  {journal} {Class. Quant. Grav.}\ }\textbf
  {\bibinfo {volume} {40}},\ \bibinfo {pages} {094001} (\bibinfo {year}
  {2023})},\ \Eprint {http://arxiv.org/abs/2207.05765} {arXiv:2207.05765
  [astro-ph.CO]} \BibitemShut {NoStop}%
\bibitem [{\citenamefont {Riess}\ \emph {et~al.}(2022)\citenamefont {Riess}
  \emph {et~al.}}]{Riess:2021jrx}%
  \BibitemOpen
  \bibfield  {author} {\bibinfo {author} {\bibfnamefont {A.~G.}\ \bibnamefont
  {Riess}} \emph {et~al.},\ }\href {\doibase 10.3847/2041-8213/ac5c5b}
  {\bibfield  {journal} {\bibinfo  {journal} {Astrophys. J. Lett.}\ }\textbf
  {\bibinfo {volume} {934}},\ \bibinfo {pages} {L7} (\bibinfo {year} {2022})},\
  \Eprint {http://arxiv.org/abs/2112.04510} {arXiv:2112.04510 [astro-ph.CO]}
  \BibitemShut {NoStop}%
\bibitem [{\citenamefont {Freedman}(2021)}]{Freedman:2021ahq}%
  \BibitemOpen
  \bibfield  {author} {\bibinfo {author} {\bibfnamefont {W.~L.}\ \bibnamefont
  {Freedman}},\ }\href {\doibase 10.3847/1538-4357/ac0e95} {\bibfield
  {journal} {\bibinfo  {journal} {Astrophys. J.}\ }\textbf {\bibinfo {volume}
  {919}},\ \bibinfo {pages} {16} (\bibinfo {year} {2021})},\ \Eprint
  {http://arxiv.org/abs/2106.15656} {arXiv:2106.15656 [astro-ph.CO]}
  \BibitemShut {NoStop}%
\bibitem [{\citenamefont {Perivolaropoulos}\ and\ \citenamefont
  {Skara}(2022{\natexlab{b}})}]{Perivolaropoulos:2022khd}%
  \BibitemOpen
  \bibfield  {author} {\bibinfo {author} {\bibfnamefont {L.}~\bibnamefont
  {Perivolaropoulos}}\ and\ \bibinfo {author} {\bibfnamefont {F.}~\bibnamefont
  {Skara}},\ }\href {\doibase 10.3390/universe8100502} {\bibfield  {journal}
  {\bibinfo  {journal} {Universe}\ }\textbf {\bibinfo {volume} {8}},\ \bibinfo
  {pages} {502} (\bibinfo {year} {2022}{\natexlab{b}})},\ \Eprint
  {http://arxiv.org/abs/2208.11169} {arXiv:2208.11169 [astro-ph.CO]}
  \BibitemShut {NoStop}%
\bibitem [{\citenamefont {Freedman}\ and\ \citenamefont
  {Madore}(2023)}]{Freedman:2023jcz}%
  \BibitemOpen
  \bibfield  {author} {\bibinfo {author} {\bibfnamefont {W.~L.}\ \bibnamefont
  {Freedman}}\ and\ \bibinfo {author} {\bibfnamefont {B.~F.}\ \bibnamefont
  {Madore}},\ }\href {\doibase 10.1088/1475-7516/2023/11/050} {\bibfield
  {journal} {\bibinfo  {journal} {JCAP}\ }\textbf {\bibinfo {volume} {11}},\
  \bibinfo {pages} {050} (\bibinfo {year} {2023})},\ \Eprint
  {http://arxiv.org/abs/2309.05618} {arXiv:2309.05618 [astro-ph.CO]}
  \BibitemShut {NoStop}%
\bibitem [{\citenamefont {Perivolaropoulos}\ and\ \citenamefont
  {Skara}(2023)}]{Perivolaropoulos:2023iqj}%
  \BibitemOpen
  \bibfield  {author} {\bibinfo {author} {\bibfnamefont {L.}~\bibnamefont
  {Perivolaropoulos}}\ and\ \bibinfo {author} {\bibfnamefont {F.}~\bibnamefont
  {Skara}},\ }\href {\doibase 10.1093/mnras/stad451} {\bibfield  {journal}
  {\bibinfo  {journal} {Mon. Not. Roy. Astron. Soc.}\ }\textbf {\bibinfo
  {volume} {520}},\ \bibinfo {pages} {5110} (\bibinfo {year} {2023})},\ \Eprint
  {http://arxiv.org/abs/2301.01024} {arXiv:2301.01024 [astro-ph.CO]}
  \BibitemShut {NoStop}%
\bibitem [{\citenamefont {Freedman}\ \emph {et~al.}(2024)\citenamefont
  {Freedman}, \citenamefont {Madore}, \citenamefont {Jang}, \citenamefont
  {Hoyt}, \citenamefont {Lee},\ and\ \citenamefont {Owens}}]{Freedman:2024eph}%
  \BibitemOpen
  \bibfield  {author} {\bibinfo {author} {\bibfnamefont {W.~L.}\ \bibnamefont
  {Freedman}}, \bibinfo {author} {\bibfnamefont {B.~F.}\ \bibnamefont
  {Madore}}, \bibinfo {author} {\bibfnamefont {I.~S.}\ \bibnamefont {Jang}},
  \bibinfo {author} {\bibfnamefont {T.~J.}\ \bibnamefont {Hoyt}}, \bibinfo
  {author} {\bibfnamefont {A.~J.}\ \bibnamefont {Lee}}, \ and\ \bibinfo
  {author} {\bibfnamefont {K.~A.}\ \bibnamefont {Owens}},\ }\href@noop {} {\
  (\bibinfo {year} {2024})},\ \Eprint {http://arxiv.org/abs/2408.06153}
  {arXiv:2408.06153 [astro-ph.CO]} \BibitemShut {NoStop}%
\bibitem [{\citenamefont {Wojtak}\ and\ \citenamefont
  {Hjorth}(2024)}]{Wojtak:2024mgg}%
  \BibitemOpen
  \bibfield  {author} {\bibinfo {author} {\bibfnamefont {R.}~\bibnamefont
  {Wojtak}}\ and\ \bibinfo {author} {\bibfnamefont {J.}~\bibnamefont
  {Hjorth}},\ }\href {\doibase 10.1093/mnras/stae1977} {\bibfield  {journal}
  {\bibinfo  {journal} {Mon. Not. Roy. Astron. Soc.}\ }\textbf {\bibinfo
  {volume} {533}},\ \bibinfo {pages} {2319} (\bibinfo {year} {2024})},\ \Eprint
  {http://arxiv.org/abs/2403.10388} {arXiv:2403.10388 [astro-ph.CO]}
  \BibitemShut {NoStop}%
\bibitem [{\citenamefont {Gall}\ \emph {et~al.}(2024)\citenamefont {Gall},
  \citenamefont {Izzo}, \citenamefont {Wojtak},\ and\ \citenamefont
  {Hjorth}}]{Gall:2024oyl}%
  \BibitemOpen
  \bibfield  {author} {\bibinfo {author} {\bibfnamefont {C.}~\bibnamefont
  {Gall}}, \bibinfo {author} {\bibfnamefont {L.}~\bibnamefont {Izzo}}, \bibinfo
  {author} {\bibfnamefont {R.}~\bibnamefont {Wojtak}}, \ and\ \bibinfo {author}
  {\bibfnamefont {J.}~\bibnamefont {Hjorth}},\ }\href@noop {} {\  (\bibinfo
  {year} {2024})},\ \Eprint {http://arxiv.org/abs/2411.05642} {arXiv:2411.05642
  [astro-ph.CO]} \BibitemShut {NoStop}%
\bibitem [{\citenamefont {Sharma}\ \emph {et~al.}(2025)\citenamefont {Sharma},
  \citenamefont {Perivolaropoulos},\ and\ \citenamefont
  {Sami}}]{Sharma:2025ucg}%
  \BibitemOpen
  \bibfield  {author} {\bibinfo {author} {\bibfnamefont {M.~K.}\ \bibnamefont
  {Sharma}}, \bibinfo {author} {\bibfnamefont {L.}~\bibnamefont
  {Perivolaropoulos}}, \ and\ \bibinfo {author} {\bibfnamefont
  {M.}~\bibnamefont {Sami}},\ }\href@noop {} {\  (\bibinfo {year} {2025})},\
  \Eprint {http://arxiv.org/abs/2504.19807} {arXiv:2504.19807 [astro-ph.CO]}
  \BibitemShut {NoStop}%
\bibitem [{\citenamefont {Efstathiou}(2020)}]{Efstathiou:2020wxn}%
  \BibitemOpen
  \bibfield  {author} {\bibinfo {author} {\bibfnamefont {G.}~\bibnamefont
  {Efstathiou}},\ }\href@noop {} {\  (\bibinfo {year} {2020})},\ \Eprint
  {http://arxiv.org/abs/2007.10716} {arXiv:2007.10716 [astro-ph.CO]}
  \BibitemShut {NoStop}%
\bibitem [{\citenamefont {Perivolaropoulos}(2024)}]{Perivolaropoulos:2024yxv}%
  \BibitemOpen
  \bibfield  {author} {\bibinfo {author} {\bibfnamefont {L.}~\bibnamefont
  {Perivolaropoulos}},\ }\href {\doibase 10.1103/PhysRevD.110.123518}
  {\bibfield  {journal} {\bibinfo  {journal} {Phys. Rev. D}\ }\textbf {\bibinfo
  {volume} {110}},\ \bibinfo {pages} {123518} (\bibinfo {year} {2024})},\
  \Eprint {http://arxiv.org/abs/2408.11031} {arXiv:2408.11031 [astro-ph.CO]}
  \BibitemShut {NoStop}%
\bibitem [{\citenamefont {Schwarz}\ \emph {et~al.}(2016)\citenamefont
  {Schwarz}, \citenamefont {Copi}, \citenamefont {Huterer},\ and\ \citenamefont
  {Starkman}}]{Schwarz:2015cma}%
  \BibitemOpen
  \bibfield  {author} {\bibinfo {author} {\bibfnamefont {D.~J.}\ \bibnamefont
  {Schwarz}}, \bibinfo {author} {\bibfnamefont {C.~J.}\ \bibnamefont {Copi}},
  \bibinfo {author} {\bibfnamefont {D.}~\bibnamefont {Huterer}}, \ and\
  \bibinfo {author} {\bibfnamefont {G.~D.}\ \bibnamefont {Starkman}},\ }\href
  {\doibase 10.1088/0264-9381/33/18/184001} {\bibfield  {journal} {\bibinfo
  {journal} {Class. Quant. Grav.}\ }\textbf {\bibinfo {volume} {33}},\ \bibinfo
  {pages} {184001} (\bibinfo {year} {2016})},\ \Eprint
  {http://arxiv.org/abs/1510.07929} {arXiv:1510.07929 [astro-ph.CO]}
  \BibitemShut {NoStop}%
\bibitem [{\citenamefont {Handley}(2021)}]{Handley:2019tkm}%
  \BibitemOpen
  \bibfield  {author} {\bibinfo {author} {\bibfnamefont {W.}~\bibnamefont
  {Handley}},\ }\href {\doibase 10.1103/PhysRevD.103.L041301} {\bibfield
  {journal} {\bibinfo  {journal} {Phys. Rev. D}\ }\textbf {\bibinfo {volume}
  {103}},\ \bibinfo {pages} {L041301} (\bibinfo {year} {2021})},\ \Eprint
  {http://arxiv.org/abs/1908.09139} {arXiv:1908.09139 [astro-ph.CO]}
  \BibitemShut {NoStop}%
\bibitem [{\citenamefont {Di~Valentino}\ \emph {et~al.}(2019)\citenamefont
  {Di~Valentino}, \citenamefont {Melchiorri},\ and\ \citenamefont
  {Silk}}]{DiValentino:2019qzk}%
  \BibitemOpen
  \bibfield  {author} {\bibinfo {author} {\bibfnamefont {E.}~\bibnamefont
  {Di~Valentino}}, \bibinfo {author} {\bibfnamefont {A.}~\bibnamefont
  {Melchiorri}}, \ and\ \bibinfo {author} {\bibfnamefont {J.}~\bibnamefont
  {Silk}},\ }\href {\doibase 10.1038/s41550-019-0906-9} {\bibfield  {journal}
  {\bibinfo  {journal} {Nature Astron.}\ }\textbf {\bibinfo {volume} {4}},\
  \bibinfo {pages} {196} (\bibinfo {year} {2019})},\ \Eprint
  {http://arxiv.org/abs/1911.02087} {arXiv:1911.02087 [astro-ph.CO]}
  \BibitemShut {NoStop}%
\bibitem [{\citenamefont {Minami}\ and\ \citenamefont
  {Komatsu}(2020)}]{Minami:2020odp}%
  \BibitemOpen
  \bibfield  {author} {\bibinfo {author} {\bibfnamefont {Y.}~\bibnamefont
  {Minami}}\ and\ \bibinfo {author} {\bibfnamefont {E.}~\bibnamefont
  {Komatsu}},\ }\href {\doibase 10.1103/PhysRevLett.125.221301} {\bibfield
  {journal} {\bibinfo  {journal} {Phys. Rev. Lett.}\ }\textbf {\bibinfo
  {volume} {125}},\ \bibinfo {pages} {221301} (\bibinfo {year} {2020})},\
  \Eprint {http://arxiv.org/abs/2011.11254} {arXiv:2011.11254 [astro-ph.CO]}
  \BibitemShut {NoStop}%
\bibitem [{\citenamefont {Komatsu}(2022)}]{Komatsu:2022nvu}%
  \BibitemOpen
  \bibfield  {author} {\bibinfo {author} {\bibfnamefont {E.}~\bibnamefont
  {Komatsu}},\ }\href {\doibase 10.1038/s42254-022-00452-4} {\bibfield
  {journal} {\bibinfo  {journal} {Nature Rev. Phys.}\ }\textbf {\bibinfo
  {volume} {4}},\ \bibinfo {pages} {452} (\bibinfo {year} {2022})},\ \Eprint
  {http://arxiv.org/abs/2202.13919} {arXiv:2202.13919 [astro-ph.CO]}
  \BibitemShut {NoStop}%
\bibitem [{\citenamefont {Galloni}\ \emph {et~al.}(2022)\citenamefont
  {Galloni}, \citenamefont {Bartolo}, \citenamefont {Matarrese}, \citenamefont
  {Migliaccio}, \citenamefont {Ricciardone},\ and\ \citenamefont
  {Vittorio}}]{Galloni:2022rgg}%
  \BibitemOpen
  \bibfield  {author} {\bibinfo {author} {\bibfnamefont {G.}~\bibnamefont
  {Galloni}}, \bibinfo {author} {\bibfnamefont {N.}~\bibnamefont {Bartolo}},
  \bibinfo {author} {\bibfnamefont {S.}~\bibnamefont {Matarrese}}, \bibinfo
  {author} {\bibfnamefont {M.}~\bibnamefont {Migliaccio}}, \bibinfo {author}
  {\bibfnamefont {A.}~\bibnamefont {Ricciardone}}, \ and\ \bibinfo {author}
  {\bibfnamefont {N.}~\bibnamefont {Vittorio}},\ }\href {\doibase
  10.1088/1475-7516/2022/09/046} {\bibfield  {journal} {\bibinfo  {journal}
  {JCAP}\ }\textbf {\bibinfo {volume} {09}},\ \bibinfo {pages} {046} (\bibinfo
  {year} {2022})},\ \Eprint {http://arxiv.org/abs/2202.12858} {arXiv:2202.12858
  [astro-ph.CO]} \BibitemShut {NoStop}%
\bibitem [{\citenamefont {Diego-Palazuelos}\ \emph {et~al.}(2022)\citenamefont
  {Diego-Palazuelos} \emph {et~al.}}]{Diego-Palazuelos:2022dsq}%
  \BibitemOpen
  \bibfield  {author} {\bibinfo {author} {\bibfnamefont {P.}~\bibnamefont
  {Diego-Palazuelos}} \emph {et~al.},\ }\href {\doibase
  10.1103/PhysRevLett.128.091302} {\bibfield  {journal} {\bibinfo  {journal}
  {Phys. Rev. Lett.}\ }\textbf {\bibinfo {volume} {128}},\ \bibinfo {pages}
  {091302} (\bibinfo {year} {2022})},\ \Eprint
  {http://arxiv.org/abs/2201.07682} {arXiv:2201.07682 [astro-ph.CO]}
  \BibitemShut {NoStop}%
\bibitem [{\citenamefont {Yeung}\ and\ \citenamefont
  {Chu}(2022)}]{Yeung:2022smn}%
  \BibitemOpen
  \bibfield  {author} {\bibinfo {author} {\bibfnamefont {S.}~\bibnamefont
  {Yeung}}\ and\ \bibinfo {author} {\bibfnamefont {M.-C.}\ \bibnamefont
  {Chu}},\ }\href {\doibase 10.1103/PhysRevD.105.083508} {\bibfield  {journal}
  {\bibinfo  {journal} {Phys. Rev. D}\ }\textbf {\bibinfo {volume} {105}},\
  \bibinfo {pages} {083508} (\bibinfo {year} {2022})},\ \Eprint
  {http://arxiv.org/abs/2201.03799} {arXiv:2201.03799 [astro-ph.CO]}
  \BibitemShut {NoStop}%
\bibitem [{\citenamefont {Giar\`e}(2023)}]{Giare:2023xoc}%
  \BibitemOpen
  \bibfield  {author} {\bibinfo {author} {\bibfnamefont {W.}~\bibnamefont
  {Giar\`e}},\ }\href {\doibase 10.1007/978-981-99-0177-7\_36} {\  (\bibinfo
  {year} {2023}),\ 10.1007/978-981-99-0177-7\_36},\ \Eprint
  {http://arxiv.org/abs/2305.16919} {arXiv:2305.16919 [astro-ph.CO]}
  \BibitemShut {NoStop}%
\bibitem [{\citenamefont {Galloni}\ \emph {et~al.}(2023)\citenamefont
  {Galloni}, \citenamefont {Ballardini}, \citenamefont {Bartolo}, \citenamefont
  {Gruppuso}, \citenamefont {Pagano},\ and\ \citenamefont
  {Ricciardone}}]{Galloni:2023pie}%
  \BibitemOpen
  \bibfield  {author} {\bibinfo {author} {\bibfnamefont {G.}~\bibnamefont
  {Galloni}}, \bibinfo {author} {\bibfnamefont {M.}~\bibnamefont {Ballardini}},
  \bibinfo {author} {\bibfnamefont {N.}~\bibnamefont {Bartolo}}, \bibinfo
  {author} {\bibfnamefont {A.}~\bibnamefont {Gruppuso}}, \bibinfo {author}
  {\bibfnamefont {L.}~\bibnamefont {Pagano}}, \ and\ \bibinfo {author}
  {\bibfnamefont {A.}~\bibnamefont {Ricciardone}},\ }\href {\doibase
  10.1088/1475-7516/2023/10/013} {\bibfield  {journal} {\bibinfo  {journal}
  {JCAP}\ }\textbf {\bibinfo {volume} {10}},\ \bibinfo {pages} {013} (\bibinfo
  {year} {2023})},\ \Eprint {http://arxiv.org/abs/2305.18184} {arXiv:2305.18184
  [astro-ph.CO]} \BibitemShut {NoStop}%
\bibitem [{\citenamefont {Jung}\ \emph {et~al.}(2024)\citenamefont {Jung},
  \citenamefont {Aghanim}, \citenamefont {Sorce}, \citenamefont {Seidel},
  \citenamefont {Dolag},\ and\ \citenamefont {Douspis}}]{Jung:2024slj}%
  \BibitemOpen
  \bibfield  {author} {\bibinfo {author} {\bibfnamefont {G.}~\bibnamefont
  {Jung}}, \bibinfo {author} {\bibfnamefont {N.}~\bibnamefont {Aghanim}},
  \bibinfo {author} {\bibfnamefont {J.}~\bibnamefont {Sorce}}, \bibinfo
  {author} {\bibfnamefont {B.}~\bibnamefont {Seidel}}, \bibinfo {author}
  {\bibfnamefont {K.}~\bibnamefont {Dolag}}, \ and\ \bibinfo {author}
  {\bibfnamefont {M.}~\bibnamefont {Douspis}},\ }\href {\doibase
  10.1051/0004-6361/202451238} {\bibfield  {journal} {\bibinfo  {journal}
  {Astron. Astrophys.}\ }\textbf {\bibinfo {volume} {692}},\ \bibinfo {pages}
  {A180} (\bibinfo {year} {2024})},\ \Eprint {http://arxiv.org/abs/2406.11543}
  {arXiv:2406.11543 [astro-ph.CO]} \BibitemShut {NoStop}%
\bibitem [{\citenamefont {Samandar}\ \emph {et~al.}(2024)\citenamefont
  {Samandar} \emph {et~al.}}]{COMPACT:2024cud}%
  \BibitemOpen
  \bibfield  {author} {\bibinfo {author} {\bibfnamefont {A.}~\bibnamefont
  {Samandar}} \emph {et~al.} (\bibinfo {collaboration} {COMPACT}),\ }\href
  {\doibase 10.1088/1475-7516/2024/11/020} {\bibfield  {journal} {\bibinfo
  {journal} {JCAP}\ }\textbf {\bibinfo {volume} {11}},\ \bibinfo {pages} {020}
  (\bibinfo {year} {2024})},\ \Eprint {http://arxiv.org/abs/2407.09400}
  {arXiv:2407.09400 [astro-ph.CO]} \BibitemShut {NoStop}%
\bibitem [{\citenamefont {Secrest}\ \emph {et~al.}(2021)\citenamefont
  {Secrest}, \citenamefont {von Hausegger}, \citenamefont {Rameez},
  \citenamefont {Mohayaee}, \citenamefont {Sarkar},\ and\ \citenamefont
  {Colin}}]{Secrest:2020has}%
  \BibitemOpen
  \bibfield  {author} {\bibinfo {author} {\bibfnamefont {N.~J.}\ \bibnamefont
  {Secrest}}, \bibinfo {author} {\bibfnamefont {S.}~\bibnamefont {von
  Hausegger}}, \bibinfo {author} {\bibfnamefont {M.}~\bibnamefont {Rameez}},
  \bibinfo {author} {\bibfnamefont {R.}~\bibnamefont {Mohayaee}}, \bibinfo
  {author} {\bibfnamefont {S.}~\bibnamefont {Sarkar}}, \ and\ \bibinfo {author}
  {\bibfnamefont {J.}~\bibnamefont {Colin}},\ }\href {\doibase
  10.3847/2041-8213/abdd40} {\bibfield  {journal} {\bibinfo  {journal}
  {Astrophys. J. Lett.}\ }\textbf {\bibinfo {volume} {908}},\ \bibinfo {pages}
  {L51} (\bibinfo {year} {2021})},\ \Eprint {http://arxiv.org/abs/2009.14826}
  {arXiv:2009.14826 [astro-ph.CO]} \BibitemShut {NoStop}%
\bibitem [{\citenamefont {Dom\`enech}\ \emph {et~al.}(2022)\citenamefont
  {Dom\`enech}, \citenamefont {Mohayaee}, \citenamefont {Patil},\ and\
  \citenamefont {Sarkar}}]{Domenech:2022mvt}%
  \BibitemOpen
  \bibfield  {author} {\bibinfo {author} {\bibfnamefont {G.}~\bibnamefont
  {Dom\`enech}}, \bibinfo {author} {\bibfnamefont {R.}~\bibnamefont
  {Mohayaee}}, \bibinfo {author} {\bibfnamefont {S.~P.}\ \bibnamefont {Patil}},
  \ and\ \bibinfo {author} {\bibfnamefont {S.}~\bibnamefont {Sarkar}},\ }\href
  {\doibase 10.1088/1475-7516/2022/10/019} {\bibfield  {journal} {\bibinfo
  {journal} {JCAP}\ }\textbf {\bibinfo {volume} {10}},\ \bibinfo {pages} {019}
  (\bibinfo {year} {2022})},\ \Eprint {http://arxiv.org/abs/2207.01569}
  {arXiv:2207.01569 [astro-ph.CO]} \BibitemShut {NoStop}%
\bibitem [{\citenamefont {Mittal}\ \emph {et~al.}(2024)\citenamefont {Mittal},
  \citenamefont {Oayda},\ and\ \citenamefont {Lewis}}]{Mittal:2023xub}%
  \BibitemOpen
  \bibfield  {author} {\bibinfo {author} {\bibfnamefont {V.}~\bibnamefont
  {Mittal}}, \bibinfo {author} {\bibfnamefont {O.~T.}\ \bibnamefont {Oayda}}, \
  and\ \bibinfo {author} {\bibfnamefont {G.~F.}\ \bibnamefont {Lewis}},\ }\href
  {\doibase 10.1093/mnras/stae1057} {\bibfield  {journal} {\bibinfo  {journal}
  {Mon. Not. Roy. Astron. Soc.}\ }\textbf {\bibinfo {volume} {527}},\ \bibinfo
  {pages} {8497} (\bibinfo {year} {2024})},\ \bibinfo {note} {[Erratum:
  Mon.Not.Roy.Astron.Soc. 530, 4763--4764 (2024)]},\ \Eprint
  {http://arxiv.org/abs/2311.14938} {arXiv:2311.14938 [astro-ph.CO]}
  \BibitemShut {NoStop}%
\bibitem [{\citenamefont {Wagenveld}\ \emph {et~al.}(2023)\citenamefont
  {Wagenveld}, \citenamefont {Kl\"ockner},\ and\ \citenamefont
  {Schwarz}}]{Wagenveld:2023kvi}%
  \BibitemOpen
  \bibfield  {author} {\bibinfo {author} {\bibfnamefont {J.~D.}\ \bibnamefont
  {Wagenveld}}, \bibinfo {author} {\bibfnamefont {H.-R.}\ \bibnamefont
  {Kl\"ockner}}, \ and\ \bibinfo {author} {\bibfnamefont {D.~J.}\ \bibnamefont
  {Schwarz}},\ }\href {\doibase 10.1051/0004-6361/202346210} {\bibfield
  {journal} {\bibinfo  {journal} {Astron. Astrophys.}\ }\textbf {\bibinfo
  {volume} {675}},\ \bibinfo {pages} {A72} (\bibinfo {year} {2023})},\ \Eprint
  {http://arxiv.org/abs/2305.15335} {arXiv:2305.15335 [astro-ph.CO]}
  \BibitemShut {NoStop}%
\bibitem [{\citenamefont {Oayda}\ \emph {et~al.}(2024)\citenamefont {Oayda},
  \citenamefont {Mittal}, \citenamefont {Lewis},\ and\ \citenamefont
  {Murphy}}]{Oayda:2024hnu}%
  \BibitemOpen
  \bibfield  {author} {\bibinfo {author} {\bibfnamefont {O.~T.}\ \bibnamefont
  {Oayda}}, \bibinfo {author} {\bibfnamefont {V.}~\bibnamefont {Mittal}},
  \bibinfo {author} {\bibfnamefont {G.~F.}\ \bibnamefont {Lewis}}, \ and\
  \bibinfo {author} {\bibfnamefont {T.}~\bibnamefont {Murphy}},\ }\href
  {\doibase 10.1093/mnras/stae1399} {\bibfield  {journal} {\bibinfo  {journal}
  {Mon. Not. Roy. Astron. Soc.}\ }\textbf {\bibinfo {volume} {531}},\ \bibinfo
  {pages} {4545} (\bibinfo {year} {2024})},\ \Eprint
  {http://arxiv.org/abs/2406.01871} {arXiv:2406.01871 [astro-ph.CO]}
  \BibitemShut {NoStop}%
\bibitem [{\citenamefont {Labb\'e}\ \emph {et~al.}(2023)\citenamefont {Labb\'e}
  \emph {et~al.}}]{Labbe:2022ahb}%
  \BibitemOpen
  \bibfield  {author} {\bibinfo {author} {\bibfnamefont {I.}~\bibnamefont
  {Labb\'e}} \emph {et~al.},\ }\href {\doibase 10.1038/s41586-023-05786-2}
  {\bibfield  {journal} {\bibinfo  {journal} {Nature}\ }\textbf {\bibinfo
  {volume} {616}},\ \bibinfo {pages} {266} (\bibinfo {year} {2023})},\ \bibinfo
  {note} {arXiv:2207.12446.},\ \Eprint {http://arxiv.org/abs/2207.12446}
  {arXiv:2207.12446 [astro-ph.GA]} \BibitemShut {NoStop}%
\bibitem [{\citenamefont {Menci}\ \emph {et~al.}(2020)\citenamefont {Menci}
  \emph {et~al.}}]{Menci:2020ybl}%
  \BibitemOpen
  \bibfield  {author} {\bibinfo {author} {\bibfnamefont {N.}~\bibnamefont
  {Menci}} \emph {et~al.},\ }\href {\doibase 10.3847/1538-4357/aba9d2}
  {\bibfield  {journal} {\bibinfo  {journal} {Astrophys. J.}\ }\textbf
  {\bibinfo {volume} {900}},\ \bibinfo {pages} {108} (\bibinfo {year}
  {2020})},\ \Eprint {http://arxiv.org/abs/2007.12453} {arXiv:2007.12453
  [astro-ph.CO]} \BibitemShut {NoStop}%
\bibitem [{\citenamefont {Menci}\ \emph {et~al.}(2022)\citenamefont {Menci},
  \citenamefont {Castellano}, \citenamefont {Santini}, \citenamefont {Merlin},
  \citenamefont {Fontana},\ and\ \citenamefont {Shankar}}]{Menci:2022wia}%
  \BibitemOpen
  \bibfield  {author} {\bibinfo {author} {\bibfnamefont {N.}~\bibnamefont
  {Menci}}, \bibinfo {author} {\bibfnamefont {M.}~\bibnamefont {Castellano}},
  \bibinfo {author} {\bibfnamefont {P.}~\bibnamefont {Santini}}, \bibinfo
  {author} {\bibfnamefont {E.}~\bibnamefont {Merlin}}, \bibinfo {author}
  {\bibfnamefont {A.}~\bibnamefont {Fontana}}, \ and\ \bibinfo {author}
  {\bibfnamefont {F.}~\bibnamefont {Shankar}},\ }\href {\doibase
  10.3847/2041-8213/ac96e9} {\bibfield  {journal} {\bibinfo  {journal}
  {Astrophys. J. Lett.}\ }\textbf {\bibinfo {volume} {938}},\ \bibinfo {pages}
  {L5} (\bibinfo {year} {2022})},\ \Eprint {http://arxiv.org/abs/2208.11471}
  {arXiv:2208.11471 [astro-ph.CO]} \BibitemShut {NoStop}%
\bibitem [{\citenamefont {Forconi}\ \emph {et~al.}(2023)\citenamefont
  {Forconi}, \citenamefont {Ruchika}, \citenamefont {Melchiorri}, \citenamefont
  {Mena},\ and\ \citenamefont {Menci}}]{Forconi:2023izg}%
  \BibitemOpen
  \bibfield  {author} {\bibinfo {author} {\bibfnamefont {M.}~\bibnamefont
  {Forconi}}, \bibinfo {author} {\bibnamefont {Ruchika}}, \bibinfo {author}
  {\bibfnamefont {A.}~\bibnamefont {Melchiorri}}, \bibinfo {author}
  {\bibfnamefont {O.}~\bibnamefont {Mena}}, \ and\ \bibinfo {author}
  {\bibfnamefont {N.}~\bibnamefont {Menci}},\ }\href {\doibase
  10.1088/1475-7516/2023/10/012} {\bibfield  {journal} {\bibinfo  {journal}
  {JCAP}\ }\textbf {\bibinfo {volume} {10}},\ \bibinfo {pages} {012} (\bibinfo
  {year} {2023})},\ \Eprint {http://arxiv.org/abs/2306.07781} {arXiv:2306.07781
  [astro-ph.CO]} \BibitemShut {NoStop}%
\bibitem [{\citenamefont {Forconi}\ \emph {et~al.}(2024)\citenamefont
  {Forconi}, \citenamefont {Giar\`e}, \citenamefont {Mena}, \citenamefont
  {Ruchika}, \citenamefont {Di~Valentino}, \citenamefont {Melchiorri},\ and\
  \citenamefont {Nunes}}]{Forconi:2023hsj}%
  \BibitemOpen
  \bibfield  {author} {\bibinfo {author} {\bibfnamefont {M.}~\bibnamefont
  {Forconi}}, \bibinfo {author} {\bibfnamefont {W.}~\bibnamefont {Giar\`e}},
  \bibinfo {author} {\bibfnamefont {O.}~\bibnamefont {Mena}}, \bibinfo {author}
  {\bibnamefont {Ruchika}}, \bibinfo {author} {\bibfnamefont {E.}~\bibnamefont
  {Di~Valentino}}, \bibinfo {author} {\bibfnamefont {A.}~\bibnamefont
  {Melchiorri}}, \ and\ \bibinfo {author} {\bibfnamefont {R.~C.}\ \bibnamefont
  {Nunes}},\ }\href {\doibase 10.1088/1475-7516/2024/05/097} {\bibfield
  {journal} {\bibinfo  {journal} {JCAP}\ }\textbf {\bibinfo {volume} {05}},\
  \bibinfo {pages} {097} (\bibinfo {year} {2024})},\ \Eprint
  {http://arxiv.org/abs/2312.11074} {arXiv:2312.11074 [astro-ph.CO]}
  \BibitemShut {NoStop}%
\bibitem [{\citenamefont {Menci}\ \emph
  {et~al.}(2024{\natexlab{a}})\citenamefont {Menci}, \citenamefont {Adil},
  \citenamefont {Mukhopadhyay}, \citenamefont {Sen},\ and\ \citenamefont
  {Vagnozzi}}]{Menci:2024rbq}%
  \BibitemOpen
  \bibfield  {author} {\bibinfo {author} {\bibfnamefont {N.}~\bibnamefont
  {Menci}}, \bibinfo {author} {\bibfnamefont {S.~A.}\ \bibnamefont {Adil}},
  \bibinfo {author} {\bibfnamefont {U.}~\bibnamefont {Mukhopadhyay}}, \bibinfo
  {author} {\bibfnamefont {A.~A.}\ \bibnamefont {Sen}}, \ and\ \bibinfo
  {author} {\bibfnamefont {S.}~\bibnamefont {Vagnozzi}},\ }\href {\doibase
  10.1088/1475-7516/2024/07/072} {\bibfield  {journal} {\bibinfo  {journal}
  {JCAP}\ }\textbf {\bibinfo {volume} {07}},\ \bibinfo {pages} {072} (\bibinfo
  {year} {2024}{\natexlab{a}})},\ \Eprint {http://arxiv.org/abs/2401.12659}
  {arXiv:2401.12659 [astro-ph.CO]} \BibitemShut {NoStop}%
\bibitem [{\citenamefont {Menci}\ \emph
  {et~al.}(2024{\natexlab{b}})\citenamefont {Menci}, \citenamefont {Sen},\ and\
  \citenamefont {Castellano}}]{Menci:2024hop}%
  \BibitemOpen
  \bibfield  {author} {\bibinfo {author} {\bibfnamefont {N.}~\bibnamefont
  {Menci}}, \bibinfo {author} {\bibfnamefont {A.~A.}\ \bibnamefont {Sen}}, \
  and\ \bibinfo {author} {\bibfnamefont {M.}~\bibnamefont {Castellano}},\
  }\href {\doibase 10.3847/1538-4357/ad8d5b} {\bibfield  {journal} {\bibinfo
  {journal} {Astrophys. J.}\ }\textbf {\bibinfo {volume} {976}},\ \bibinfo
  {pages} {227} (\bibinfo {year} {2024}{\natexlab{b}})},\ \Eprint
  {http://arxiv.org/abs/2410.22940} {arXiv:2410.22940 [astro-ph.CO]}
  \BibitemShut {NoStop}%
\bibitem [{\citenamefont {Macaulay}\ \emph {et~al.}(2013)\citenamefont
  {Macaulay}, \citenamefont {Wehus},\ and\ \citenamefont
  {Eriksen}}]{Macaulay:2013swa}%
  \BibitemOpen
  \bibfield  {author} {\bibinfo {author} {\bibfnamefont {E.}~\bibnamefont
  {Macaulay}}, \bibinfo {author} {\bibfnamefont {I.~K.}\ \bibnamefont {Wehus}},
  \ and\ \bibinfo {author} {\bibfnamefont {H.~K.}\ \bibnamefont {Eriksen}},\
  }\href {\doibase 10.1103/PhysRevLett.111.161301} {\bibfield  {journal}
  {\bibinfo  {journal} {Phys. Rev. Lett.}\ }\textbf {\bibinfo {volume} {111}},\
  \bibinfo {pages} {161301} (\bibinfo {year} {2013})},\ \Eprint
  {http://arxiv.org/abs/1303.6583} {arXiv:1303.6583 [astro-ph.CO]} \BibitemShut
  {NoStop}%
\bibitem [{\citenamefont {Joudaki}\ \emph {et~al.}(2018)\citenamefont {Joudaki}
  \emph {et~al.}}]{Joudaki:2017zdt}%
  \BibitemOpen
  \bibfield  {author} {\bibinfo {author} {\bibfnamefont {S.}~\bibnamefont
  {Joudaki}} \emph {et~al.},\ }\href {\doibase 10.1093/mnras/stx2820}
  {\bibfield  {journal} {\bibinfo  {journal} {Mon. Not. Roy. Astron. Soc.}\
  }\textbf {\bibinfo {volume} {474}},\ \bibinfo {pages} {4894} (\bibinfo {year}
  {2018})},\ \Eprint {http://arxiv.org/abs/1707.06627} {arXiv:1707.06627
  [astro-ph.CO]} \BibitemShut {NoStop}%
\bibitem [{\citenamefont {G\'omez-Valent}\ and\ \citenamefont
  {Sol\`a}(2017)}]{Gomez-Valent:2017idt}%
  \BibitemOpen
  \bibfield  {author} {\bibinfo {author} {\bibfnamefont {A.}~\bibnamefont
  {G\'omez-Valent}}\ and\ \bibinfo {author} {\bibfnamefont {J.}~\bibnamefont
  {Sol\`a}},\ }\href {\doibase 10.1209/0295-5075/120/39001} {\bibfield
  {journal} {\bibinfo  {journal} {EPL}\ }\textbf {\bibinfo {volume} {120}},\
  \bibinfo {pages} {39001} (\bibinfo {year} {2017})},\ \Eprint
  {http://arxiv.org/abs/1711.00692} {arXiv:1711.00692 [astro-ph.CO]}
  \BibitemShut {NoStop}%
\bibitem [{\citenamefont {Nesseris}\ \emph {et~al.}(2017)\citenamefont
  {Nesseris}, \citenamefont {Pantazis},\ and\ \citenamefont
  {Perivolaropoulos}}]{Nesseris:2017vor}%
  \BibitemOpen
  \bibfield  {author} {\bibinfo {author} {\bibfnamefont {S.}~\bibnamefont
  {Nesseris}}, \bibinfo {author} {\bibfnamefont {G.}~\bibnamefont {Pantazis}},
  \ and\ \bibinfo {author} {\bibfnamefont {L.}~\bibnamefont
  {Perivolaropoulos}},\ }\href {\doibase 10.1103/PhysRevD.96.023542} {\bibfield
   {journal} {\bibinfo  {journal} {Phys. Rev. D}\ }\textbf {\bibinfo {volume}
  {96}},\ \bibinfo {pages} {023542} (\bibinfo {year} {2017})},\ \Eprint
  {http://arxiv.org/abs/1703.10538} {arXiv:1703.10538 [astro-ph.CO]}
  \BibitemShut {NoStop}%
\bibitem [{\citenamefont {Benisty}(2021)}]{Benisty:2020kdt}%
  \BibitemOpen
  \bibfield  {author} {\bibinfo {author} {\bibfnamefont {D.}~\bibnamefont
  {Benisty}},\ }\href {\doibase 10.1016/j.dark.2020.100766} {\bibfield
  {journal} {\bibinfo  {journal} {Phys. Dark Univ.}\ }\textbf {\bibinfo
  {volume} {31}},\ \bibinfo {pages} {100766} (\bibinfo {year} {2021})},\
  \Eprint {http://arxiv.org/abs/2005.03751} {arXiv:2005.03751 [astro-ph.CO]}
  \BibitemShut {NoStop}%
\bibitem [{\citenamefont {Wright}\ \emph {et~al.}(2020)\citenamefont {Wright},
  \citenamefont {Hildebrandt}, \citenamefont {van~den Busch}, \citenamefont
  {Heymans}, \citenamefont {Joachimi}, \citenamefont {Kannawadi},\ and\
  \citenamefont {Kuijken}}]{Wright:2020ppw}%
  \BibitemOpen
  \bibfield  {author} {\bibinfo {author} {\bibfnamefont {A.~H.}\ \bibnamefont
  {Wright}}, \bibinfo {author} {\bibfnamefont {H.}~\bibnamefont {Hildebrandt}},
  \bibinfo {author} {\bibfnamefont {J.~L.}\ \bibnamefont {van~den Busch}},
  \bibinfo {author} {\bibfnamefont {C.}~\bibnamefont {Heymans}}, \bibinfo
  {author} {\bibfnamefont {B.}~\bibnamefont {Joachimi}}, \bibinfo {author}
  {\bibfnamefont {A.}~\bibnamefont {Kannawadi}}, \ and\ \bibinfo {author}
  {\bibfnamefont {K.}~\bibnamefont {Kuijken}},\ }\href {\doibase
  10.1051/0004-6361/202038389} {\bibfield  {journal} {\bibinfo  {journal}
  {Astron. Astrophys.}\ }\textbf {\bibinfo {volume} {640}},\ \bibinfo {pages}
  {L14} (\bibinfo {year} {2020})},\ \Eprint {http://arxiv.org/abs/2005.04207}
  {arXiv:2005.04207 [astro-ph.CO]} \BibitemShut {NoStop}%
\bibitem [{\citenamefont {Nunes}\ and\ \citenamefont
  {Vagnozzi}(2021)}]{Nunes:2021ipq}%
  \BibitemOpen
  \bibfield  {author} {\bibinfo {author} {\bibfnamefont {R.~C.}\ \bibnamefont
  {Nunes}}\ and\ \bibinfo {author} {\bibfnamefont {S.}~\bibnamefont
  {Vagnozzi}},\ }\href {\doibase 10.1093/mnras/stab1613} {\bibfield  {journal}
  {\bibinfo  {journal} {Mon. Not. Roy. Astron. Soc.}\ }\textbf {\bibinfo
  {volume} {505}},\ \bibinfo {pages} {5427} (\bibinfo {year} {2021})},\ \Eprint
  {http://arxiv.org/abs/2106.01208} {arXiv:2106.01208 [astro-ph.CO]}
  \BibitemShut {NoStop}%
\bibitem [{\citenamefont {Nguyen}\ \emph {et~al.}(2023)\citenamefont {Nguyen},
  \citenamefont {Huterer},\ and\ \citenamefont {Wen}}]{Nguyen:2023fip}%
  \BibitemOpen
  \bibfield  {author} {\bibinfo {author} {\bibfnamefont {N.-M.}\ \bibnamefont
  {Nguyen}}, \bibinfo {author} {\bibfnamefont {D.}~\bibnamefont {Huterer}}, \
  and\ \bibinfo {author} {\bibfnamefont {Y.}~\bibnamefont {Wen}},\ }\href
  {\doibase 10.1103/PhysRevLett.131.111001} {\bibfield  {journal} {\bibinfo
  {journal} {Phys. Rev. Lett.}\ }\textbf {\bibinfo {volume} {131}},\ \bibinfo
  {pages} {111001} (\bibinfo {year} {2023})},\ \Eprint
  {http://arxiv.org/abs/2302.01331} {arXiv:2302.01331 [astro-ph.CO]}
  \BibitemShut {NoStop}%
\bibitem [{\citenamefont {Adil}\ \emph {et~al.}(2023)\citenamefont {Adil},
  \citenamefont {Akarsu}, \citenamefont {Malekjani}, \citenamefont {Colg\'ain},
  \citenamefont {Pourojaghi}, \citenamefont {Sen},\ and\ \citenamefont
  {Sheikh-Jabbari}}]{Adil:2023jtu}%
  \BibitemOpen
  \bibfield  {author} {\bibinfo {author} {\bibfnamefont {S.~A.}\ \bibnamefont
  {Adil}}, \bibinfo {author} {\bibfnamefont {O.}~\bibnamefont {Akarsu}},
  \bibinfo {author} {\bibfnamefont {M.}~\bibnamefont {Malekjani}}, \bibinfo
  {author} {\bibfnamefont {E.~O.}\ \bibnamefont {Colg\'ain}}, \bibinfo {author}
  {\bibfnamefont {S.}~\bibnamefont {Pourojaghi}}, \bibinfo {author}
  {\bibfnamefont {A.~A.}\ \bibnamefont {Sen}}, \ and\ \bibinfo {author}
  {\bibfnamefont {M.~M.}\ \bibnamefont {Sheikh-Jabbari}},\ }\href {\doibase
  10.1093/mnrasl/slad165} {\bibfield  {journal} {\bibinfo  {journal} {Mon. Not.
  Roy. Astron. Soc.}\ }\textbf {\bibinfo {volume} {528}},\ \bibinfo {pages}
  {L20} (\bibinfo {year} {2023})},\ \Eprint {http://arxiv.org/abs/2303.06928}
  {arXiv:2303.06928 [astro-ph.CO]} \BibitemShut {NoStop}%
\bibitem [{\citenamefont {Sailer}\ \emph {et~al.}(2025)\citenamefont {Sailer}
  \emph {et~al.}}]{Sailer:2024coh}%
  \BibitemOpen
  \bibfield  {author} {\bibinfo {author} {\bibfnamefont {N.}~\bibnamefont
  {Sailer}} \emph {et~al.},\ }\href {\doibase 10.1088/1475-7516/2025/06/008}
  {\bibfield  {journal} {\bibinfo  {journal} {JCAP}\ }\textbf {\bibinfo
  {volume} {06}},\ \bibinfo {pages} {008} (\bibinfo {year} {2025})},\ \Eprint
  {http://arxiv.org/abs/2407.04607} {arXiv:2407.04607 [astro-ph.CO]}
  \BibitemShut {NoStop}%
\bibitem [{\citenamefont {Toda}\ \emph
  {et~al.}(2024{\natexlab{a}})\citenamefont {Toda}, \citenamefont
  {G\'omez-Valent},\ and\ \citenamefont {Koyama}}]{Toda:2024fgv}%
  \BibitemOpen
  \bibfield  {author} {\bibinfo {author} {\bibfnamefont {Y.}~\bibnamefont
  {Toda}}, \bibinfo {author} {\bibfnamefont {A.}~\bibnamefont
  {G\'omez-Valent}}, \ and\ \bibinfo {author} {\bibfnamefont {K.}~\bibnamefont
  {Koyama}},\ }\href {\doibase 10.1088/1475-7516/2024/12/033} {\bibfield
  {journal} {\bibinfo  {journal} {JCAP}\ }\textbf {\bibinfo {volume} {12}},\
  \bibinfo {pages} {033} (\bibinfo {year} {2024}{\natexlab{a}})},\ \Eprint
  {http://arxiv.org/abs/2408.16388} {arXiv:2408.16388 [astro-ph.CO]}
  \BibitemShut {NoStop}%
\bibitem [{\citenamefont {Akarsu}\ \emph {et~al.}(2024)\citenamefont {Akarsu},
  \citenamefont {Colg\'ain}, \citenamefont {Sen},\ and\ \citenamefont
  {Sheikh-Jabbari}}]{Akarsu:2024hsu}%
  \BibitemOpen
  \bibfield  {author} {\bibinfo {author} {\bibfnamefont {O.}~\bibnamefont
  {Akarsu}}, \bibinfo {author} {\bibfnamefont {E.~O.}\ \bibnamefont
  {Colg\'ain}}, \bibinfo {author} {\bibfnamefont {A.~A.}\ \bibnamefont {Sen}},
  \ and\ \bibinfo {author} {\bibfnamefont {M.~M.}\ \bibnamefont
  {Sheikh-Jabbari}},\ }\href@noop {} {\  (\bibinfo {year} {2024})},\ \Eprint
  {http://arxiv.org/abs/2410.23134} {arXiv:2410.23134 [astro-ph.CO]}
  \BibitemShut {NoStop}%
\bibitem [{\citenamefont {Artis}\ \emph {et~al.}(2025)\citenamefont {Artis}
  \emph {et~al.}}]{Artis:2024zag}%
  \BibitemOpen
  \bibfield  {author} {\bibinfo {author} {\bibfnamefont {E.}~\bibnamefont
  {Artis}} \emph {et~al.},\ }\href {\doibase 10.1051/0004-6361/202452584}
  {\bibfield  {journal} {\bibinfo  {journal} {Astron. Astrophys.}\ }\textbf
  {\bibinfo {volume} {696}},\ \bibinfo {pages} {A5} (\bibinfo {year} {2025})},\
  \Eprint {http://arxiv.org/abs/2410.09499} {arXiv:2410.09499 [astro-ph.CO]}
  \BibitemShut {NoStop}%
\bibitem [{\citenamefont {Kim}\ \emph {et~al.}(2024)\citenamefont {Kim} \emph
  {et~al.}}]{ACT:2024okh}%
  \BibitemOpen
  \bibfield  {author} {\bibinfo {author} {\bibfnamefont {J.}~\bibnamefont
  {Kim}} \emph {et~al.},\ }\href {\doibase 10.1088/1475-7516/2024/12/022}
  {\bibfield  {journal} {\bibinfo  {journal} {JCAP}\ }\textbf {\bibinfo
  {volume} {12}},\ \bibinfo {pages} {022} (\bibinfo {year} {2024})},\ \Eprint
  {http://arxiv.org/abs/2407.04606} {arXiv:2407.04606 [astro-ph.CO]}
  \BibitemShut {NoStop}%
\bibitem [{\citenamefont {Qu}\ \emph {et~al.}(2025)\citenamefont {Qu} \emph
  {et~al.}}]{ACT:2024nrz}%
  \BibitemOpen
  \bibfield  {author} {\bibinfo {author} {\bibfnamefont {F.~J.}\ \bibnamefont
  {Qu}} \emph {et~al.},\ }\href {\doibase 10.1103/PhysRevD.111.103503}
  {\bibfield  {journal} {\bibinfo  {journal} {Phys. Rev. D}\ }\textbf {\bibinfo
  {volume} {111}},\ \bibinfo {pages} {103503} (\bibinfo {year} {2025})},\
  \Eprint {http://arxiv.org/abs/2410.10808} {arXiv:2410.10808 [astro-ph.CO]}
  \BibitemShut {NoStop}%
\bibitem [{\citenamefont {Wright}\ \emph {et~al.}(2025)\citenamefont {Wright}
  \emph {et~al.}}]{Wright:2025xka}%
  \BibitemOpen
  \bibfield  {author} {\bibinfo {author} {\bibfnamefont {A.~H.}\ \bibnamefont
  {Wright}} \emph {et~al.},\ }\href@noop {} {\  (\bibinfo {year} {2025})},\
  \Eprint {http://arxiv.org/abs/2503.19441} {arXiv:2503.19441 [astro-ph.CO]}
  \BibitemShut {NoStop}%
\bibitem [{\citenamefont {Avila}\ \emph {et~al.}(2021)\citenamefont {Avila},
  \citenamefont {Bernui}, \citenamefont {de~Carvalho},\ and\ \citenamefont
  {Novaes}}]{Avila:2021dqv}%
  \BibitemOpen
  \bibfield  {author} {\bibinfo {author} {\bibfnamefont {F.}~\bibnamefont
  {Avila}}, \bibinfo {author} {\bibfnamefont {A.}~\bibnamefont {Bernui}},
  \bibinfo {author} {\bibfnamefont {E.}~\bibnamefont {de~Carvalho}}, \ and\
  \bibinfo {author} {\bibfnamefont {C.~P.}\ \bibnamefont {Novaes}},\ }\href
  {\doibase 10.1093/mnras/stab1488} {\bibfield  {journal} {\bibinfo  {journal}
  {Mon. Not. Roy. Astron. Soc.}\ }\textbf {\bibinfo {volume} {505}},\ \bibinfo
  {pages} {3404} (\bibinfo {year} {2021})},\ \Eprint
  {http://arxiv.org/abs/2105.10583} {arXiv:2105.10583 [astro-ph.CO]}
  \BibitemShut {NoStop}%
\bibitem [{\citenamefont {Said}\ \emph {et~al.}(2020)\citenamefont {Said},
  \citenamefont {Colless}, \citenamefont {Magoulas}, \citenamefont {Lucey},\
  and\ \citenamefont {Hudson}}]{Said:2020epb}%
  \BibitemOpen
  \bibfield  {author} {\bibinfo {author} {\bibfnamefont {K.}~\bibnamefont
  {Said}}, \bibinfo {author} {\bibfnamefont {M.}~\bibnamefont {Colless}},
  \bibinfo {author} {\bibfnamefont {C.}~\bibnamefont {Magoulas}}, \bibinfo
  {author} {\bibfnamefont {J.~R.}\ \bibnamefont {Lucey}}, \ and\ \bibinfo
  {author} {\bibfnamefont {M.~J.}\ \bibnamefont {Hudson}},\ }\href {\doibase
  10.1093/mnras/staa2032} {\bibfield  {journal} {\bibinfo  {journal} {Mon. Not.
  Roy. Astron. Soc.}\ }\textbf {\bibinfo {volume} {497}},\ \bibinfo {pages}
  {1275} (\bibinfo {year} {2020})},\ \Eprint {http://arxiv.org/abs/2007.04993}
  {arXiv:2007.04993 [astro-ph.CO]} \BibitemShut {NoStop}%
\bibitem [{\citenamefont {Simpson}\ \emph {et~al.}(2016)\citenamefont
  {Simpson}, \citenamefont {Blake}, \citenamefont {Peacock}, \citenamefont
  {Baldry}, \citenamefont {Bland-Hawthorn}, \citenamefont {Heavens},
  \citenamefont {Heymans}, \citenamefont {Loveday},\ and\ \citenamefont
  {Norberg}}]{Simpson:2015yfa}%
  \BibitemOpen
  \bibfield  {author} {\bibinfo {author} {\bibfnamefont {F.}~\bibnamefont
  {Simpson}}, \bibinfo {author} {\bibfnamefont {C.}~\bibnamefont {Blake}},
  \bibinfo {author} {\bibfnamefont {J.~A.}\ \bibnamefont {Peacock}}, \bibinfo
  {author} {\bibfnamefont {I.}~\bibnamefont {Baldry}}, \bibinfo {author}
  {\bibfnamefont {J.}~\bibnamefont {Bland-Hawthorn}}, \bibinfo {author}
  {\bibfnamefont {A.}~\bibnamefont {Heavens}}, \bibinfo {author} {\bibfnamefont
  {C.}~\bibnamefont {Heymans}}, \bibinfo {author} {\bibfnamefont
  {J.}~\bibnamefont {Loveday}}, \ and\ \bibinfo {author} {\bibfnamefont
  {P.}~\bibnamefont {Norberg}},\ }\href {\doibase 10.1103/PhysRevD.93.023525}
  {\bibfield  {journal} {\bibinfo  {journal} {Phys. Rev. D}\ }\textbf {\bibinfo
  {volume} {93}},\ \bibinfo {pages} {023525} (\bibinfo {year} {2016})},\
  \Eprint {http://arxiv.org/abs/1505.03865} {arXiv:1505.03865 [astro-ph.CO]}
  \BibitemShut {NoStop}%
\bibitem [{\citenamefont {Blake}\ \emph {et~al.}(2011)\citenamefont {Blake}
  \emph {et~al.}}]{Blake:2011rj}%
  \BibitemOpen
  \bibfield  {author} {\bibinfo {author} {\bibfnamefont {C.}~\bibnamefont
  {Blake}} \emph {et~al.},\ }\href {\doibase 10.1111/j.1365-2966.2011.18903.x}
  {\bibfield  {journal} {\bibinfo  {journal} {Mon. Not. Roy. Astron. Soc.}\
  }\textbf {\bibinfo {volume} {415}},\ \bibinfo {pages} {2876} (\bibinfo {year}
  {2011})},\ \Eprint {http://arxiv.org/abs/1104.2948} {arXiv:1104.2948
  [astro-ph.CO]} \BibitemShut {NoStop}%
\bibitem [{\citenamefont {Blake}\ \emph {et~al.}(2013)\citenamefont {Blake}
  \emph {et~al.}}]{Blake:2013nif}%
  \BibitemOpen
  \bibfield  {author} {\bibinfo {author} {\bibfnamefont {C.}~\bibnamefont
  {Blake}} \emph {et~al.},\ }\href {\doibase 10.1093/mnras/stt1791} {\bibfield
  {journal} {\bibinfo  {journal} {Mon. Not. Roy. Astron. Soc.}\ }\textbf
  {\bibinfo {volume} {436}},\ \bibinfo {pages} {3089} (\bibinfo {year}
  {2013})},\ \Eprint {http://arxiv.org/abs/1309.5556} {arXiv:1309.5556
  [astro-ph.CO]} \BibitemShut {NoStop}%
\bibitem [{\citenamefont {Gil-Mar\'\i{}n}\ \emph {et~al.}(2017)\citenamefont
  {Gil-Mar\'\i{}n}, \citenamefont {Percival}, \citenamefont {Verde},
  \citenamefont {Brownstein}, \citenamefont {Chuang}, \citenamefont {Kitaura},
  \citenamefont {Rodr\'\i{}guez-Torres},\ and\ \citenamefont
  {Olmstead}}]{Gil-Marin:2016wya}%
  \BibitemOpen
  \bibfield  {author} {\bibinfo {author} {\bibfnamefont {H.}~\bibnamefont
  {Gil-Mar\'\i{}n}}, \bibinfo {author} {\bibfnamefont {W.~J.}\ \bibnamefont
  {Percival}}, \bibinfo {author} {\bibfnamefont {L.}~\bibnamefont {Verde}},
  \bibinfo {author} {\bibfnamefont {J.~R.}\ \bibnamefont {Brownstein}},
  \bibinfo {author} {\bibfnamefont {C.-H.}\ \bibnamefont {Chuang}}, \bibinfo
  {author} {\bibfnamefont {F.-S.}\ \bibnamefont {Kitaura}}, \bibinfo {author}
  {\bibfnamefont {S.~A.}\ \bibnamefont {Rodr\'\i{}guez-Torres}}, \ and\
  \bibinfo {author} {\bibfnamefont {M.~D.}\ \bibnamefont {Olmstead}},\ }\href
  {\doibase 10.1093/mnras/stw2679} {\bibfield  {journal} {\bibinfo  {journal}
  {Mon. Not. Roy. Astron. Soc.}\ }\textbf {\bibinfo {volume} {465}},\ \bibinfo
  {pages} {1757} (\bibinfo {year} {2017})},\ \Eprint
  {http://arxiv.org/abs/1606.00439} {arXiv:1606.00439 [astro-ph.CO]}
  \BibitemShut {NoStop}%
\bibitem [{\citenamefont {Mohammad}\ \emph {et~al.}(2018)\citenamefont
  {Mohammad} \emph {et~al.}}]{Mohammad:2018mdy}%
  \BibitemOpen
  \bibfield  {author} {\bibinfo {author} {\bibfnamefont {F.~G.}\ \bibnamefont
  {Mohammad}} \emph {et~al.},\ }\href {\doibase 10.1051/0004-6361/201833853}
  {\bibfield  {journal} {\bibinfo  {journal} {Astron. Astrophys.}\ }\textbf
  {\bibinfo {volume} {619}},\ \bibinfo {pages} {A17} (\bibinfo {year}
  {2018})},\ \Eprint {http://arxiv.org/abs/1807.05999} {arXiv:1807.05999
  [astro-ph.CO]} \BibitemShut {NoStop}%
\bibitem [{\citenamefont {Song}\ and\ \citenamefont
  {Percival}(2009)}]{Song:2008qt}%
  \BibitemOpen
  \bibfield  {author} {\bibinfo {author} {\bibfnamefont {Y.-S.}\ \bibnamefont
  {Song}}\ and\ \bibinfo {author} {\bibfnamefont {W.~J.}\ \bibnamefont
  {Percival}},\ }\href {\doibase 10.1088/1475-7516/2009/10/004} {\bibfield
  {journal} {\bibinfo  {journal} {JCAP}\ }\textbf {\bibinfo {volume} {10}},\
  \bibinfo {pages} {004} (\bibinfo {year} {2009})},\ \Eprint
  {http://arxiv.org/abs/0807.0810} {arXiv:0807.0810 [astro-ph]} \BibitemShut
  {NoStop}%
\bibitem [{\citenamefont {Guzzo}\ \emph {et~al.}(2008)\citenamefont {Guzzo}
  \emph {et~al.}}]{Guzzo:2008ac}%
  \BibitemOpen
  \bibfield  {author} {\bibinfo {author} {\bibfnamefont {L.}~\bibnamefont
  {Guzzo}} \emph {et~al.},\ }\href {\doibase 10.1038/nature06555} {\bibfield
  {journal} {\bibinfo  {journal} {Nature}\ }\textbf {\bibinfo {volume} {451}},\
  \bibinfo {pages} {541} (\bibinfo {year} {2008})},\ \Eprint
  {http://arxiv.org/abs/0802.1944} {arXiv:0802.1944 [astro-ph]} \BibitemShut
  {NoStop}%
\bibitem [{\citenamefont {Okumura}\ \emph {et~al.}(2016)\citenamefont {Okumura}
  \emph {et~al.}}]{Okumura:2015lvp}%
  \BibitemOpen
  \bibfield  {author} {\bibinfo {author} {\bibfnamefont {T.}~\bibnamefont
  {Okumura}} \emph {et~al.},\ }\href {\doibase 10.1093/pasj/psw029} {\bibfield
  {journal} {\bibinfo  {journal} {Publ. Astron. Soc. Jap.}\ }\textbf {\bibinfo
  {volume} {68}},\ \bibinfo {pages} {38} (\bibinfo {year} {2016})},\ \Eprint
  {http://arxiv.org/abs/1511.08083} {arXiv:1511.08083 [astro-ph.CO]}
  \BibitemShut {NoStop}%
\bibitem [{\citenamefont {Hou}\ \emph {et~al.}(2020)\citenamefont {Hou} \emph
  {et~al.}}]{eBOSS:2020gbb}%
  \BibitemOpen
  \bibfield  {author} {\bibinfo {author} {\bibfnamefont {J.}~\bibnamefont
  {Hou}} \emph {et~al.} (\bibinfo {collaboration} {eBOSS}),\ }\href {\doibase
  10.1093/mnras/staa3234} {\bibfield  {journal} {\bibinfo  {journal} {Mon. Not.
  Roy. Astron. Soc.}\ }\textbf {\bibinfo {volume} {500}},\ \bibinfo {pages}
  {1201} (\bibinfo {year} {2020})},\ \Eprint {http://arxiv.org/abs/2007.08998}
  {arXiv:2007.08998 [astro-ph.CO]} \BibitemShut {NoStop}%
\bibitem [{\citenamefont {Adame}\ \emph
  {et~al.}(2024{\natexlab{b}})\citenamefont {Adame} \emph
  {et~al.}}]{DESI:2023dwi}%
  \BibitemOpen
  \bibfield  {author} {\bibinfo {author} {\bibfnamefont {A.~G.}\ \bibnamefont
  {Adame}} \emph {et~al.} (\bibinfo {collaboration} {DESI}),\ }\href {\doibase
  10.3847/1538-3881/ad0b08} {\bibfield  {journal} {\bibinfo  {journal} {Astron.
  J.}\ }\textbf {\bibinfo {volume} {167}},\ \bibinfo {pages} {62} (\bibinfo
  {year} {2024}{\natexlab{b}})},\ \Eprint {http://arxiv.org/abs/2306.06307}
  {arXiv:2306.06307 [astro-ph.CO]} \BibitemShut {NoStop}%
\bibitem [{\citenamefont {Laureijs}\ \emph {et~al.}(2011)\citenamefont
  {Laureijs} \emph {et~al.}}]{EUCLID:2011zbd}%
  \BibitemOpen
  \bibfield  {author} {\bibinfo {author} {\bibfnamefont {R.}~\bibnamefont
  {Laureijs}} \emph {et~al.} (\bibinfo {collaboration} {EUCLID}),\ }\href@noop
  {} {\  (\bibinfo {year} {2011})},\ \Eprint {http://arxiv.org/abs/1110.3193}
  {arXiv:1110.3193 [astro-ph.CO]} \BibitemShut {NoStop}%
\bibitem [{\citenamefont {Amendola}\ \emph {et~al.}(2013)\citenamefont
  {Amendola} \emph {et~al.}}]{EuclidTheoryWorkingGroup:2012gxx}%
  \BibitemOpen
  \bibfield  {author} {\bibinfo {author} {\bibfnamefont {L.}~\bibnamefont
  {Amendola}} \emph {et~al.} (\bibinfo {collaboration} {Euclid Theory Working
  Group}),\ }\href {\doibase 10.12942/lrr-2013-6} {\bibfield  {journal}
  {\bibinfo  {journal} {Living Rev. Rel.}\ }\textbf {\bibinfo {volume} {16}},\
  \bibinfo {pages} {6} (\bibinfo {year} {2013})},\ \Eprint
  {http://arxiv.org/abs/1206.1225} {arXiv:1206.1225 [astro-ph.CO]} \BibitemShut
  {NoStop}%
\bibitem [{\citenamefont {Jackson}(1972)}]{Jackson:1972}%
  \BibitemOpen
  \bibfield  {author} {\bibinfo {author} {\bibfnamefont {J.~C.}\ \bibnamefont
  {Jackson}},\ }\href {\doibase 10.1093/mnras/156.1.1P} {\bibfield  {journal}
  {\bibinfo  {journal} {Mon. Not. Roy. Astron. Soc.}\ }\textbf {\bibinfo
  {volume} {156}},\ \bibinfo {pages} {1P} (\bibinfo {year} {1972})}\BibitemShut
  {NoStop}%
\bibitem [{\citenamefont {Kaiser}(1987)}]{Kaiser:1987qv}%
  \BibitemOpen
  \bibfield  {author} {\bibinfo {author} {\bibfnamefont {N.}~\bibnamefont
  {Kaiser}},\ }\href {\doibase 10.1093/mnras/227.1.1} {\bibfield  {journal}
  {\bibinfo  {journal} {Mon. Not. Roy. Astron. Soc.}\ }\textbf {\bibinfo
  {volume} {227}},\ \bibinfo {pages} {1} (\bibinfo {year} {1987})}\BibitemShut
  {NoStop}%
\bibitem [{\citenamefont {Amendola}\ \emph {et~al.}(2022)\citenamefont
  {Amendola}, \citenamefont {Pietroni},\ and\ \citenamefont
  {Quartin}}]{Amendola:2022vte}%
  \BibitemOpen
  \bibfield  {author} {\bibinfo {author} {\bibfnamefont {L.}~\bibnamefont
  {Amendola}}, \bibinfo {author} {\bibfnamefont {M.}~\bibnamefont {Pietroni}},
  \ and\ \bibinfo {author} {\bibfnamefont {M.}~\bibnamefont {Quartin}},\ }\href
  {\doibase 10.1088/1475-7516/2022/11/023} {\bibfield  {journal} {\bibinfo
  {journal} {JCAP}\ }\textbf {\bibinfo {volume} {11}},\ \bibinfo {pages} {023}
  (\bibinfo {year} {2022})},\ \Eprint {http://arxiv.org/abs/2205.00569}
  {arXiv:2205.00569 [astro-ph.CO]} \BibitemShut {NoStop}%
\bibitem [{\citenamefont {Schirra}\ \emph {et~al.}(2024)\citenamefont
  {Schirra}, \citenamefont {Quartin},\ and\ \citenamefont
  {Amendola}}]{Schirra:2024rjq}%
  \BibitemOpen
  \bibfield  {author} {\bibinfo {author} {\bibfnamefont {A.~P.}\ \bibnamefont
  {Schirra}}, \bibinfo {author} {\bibfnamefont {M.}~\bibnamefont {Quartin}}, \
  and\ \bibinfo {author} {\bibfnamefont {L.}~\bibnamefont {Amendola}},\
  }\href@noop {} {\  (\bibinfo {year} {2024})},\ \Eprint
  {http://arxiv.org/abs/2406.15347} {arXiv:2406.15347 [astro-ph.CO]}
  \BibitemShut {NoStop}%
\bibitem [{\citenamefont {Linder}(2005)}]{Linder:2005in}%
  \BibitemOpen
  \bibfield  {author} {\bibinfo {author} {\bibfnamefont {E.~V.}\ \bibnamefont
  {Linder}},\ }\href {\doibase 10.1103/PhysRevD.72.043529} {\bibfield
  {journal} {\bibinfo  {journal} {Phys. Rev. D}\ }\textbf {\bibinfo {volume}
  {72}},\ \bibinfo {pages} {043529} (\bibinfo {year} {2005})},\ \Eprint
  {http://arxiv.org/abs/astro-ph/0507263} {arXiv:astro-ph/0507263} \BibitemShut
  {NoStop}%
\bibitem [{\citenamefont {Basilakos}\ and\ \citenamefont
  {Sol\`a}(2015)}]{Basilakos:2015vra}%
  \BibitemOpen
  \bibfield  {author} {\bibinfo {author} {\bibfnamefont {S.}~\bibnamefont
  {Basilakos}}\ and\ \bibinfo {author} {\bibfnamefont {J.}~\bibnamefont
  {Sol\`a}},\ }\href {\doibase 10.1103/PhysRevD.92.123501} {\bibfield
  {journal} {\bibinfo  {journal} {Phys. Rev. D}\ }\textbf {\bibinfo {volume}
  {92}},\ \bibinfo {pages} {123501} (\bibinfo {year} {2015})},\ \Eprint
  {http://arxiv.org/abs/1509.06732} {arXiv:1509.06732 [astro-ph.CO]}
  \BibitemShut {NoStop}%
\bibitem [{\citenamefont {Cort\^es}\ and\ \citenamefont
  {Batista}(2024)}]{Cortes:2024yon}%
  \BibitemOpen
  \bibfield  {author} {\bibinfo {author} {\bibfnamefont {I.~B.~S.}\
  \bibnamefont {Cort\^es}}\ and\ \bibinfo {author} {\bibfnamefont {R.~C.}\
  \bibnamefont {Batista}},\ }\href@noop {} {\  (\bibinfo {year} {2024})},\
  \Eprint {http://arxiv.org/abs/2411.00963} {arXiv:2411.00963 [astro-ph.CO]}
  \BibitemShut {NoStop}%
\bibitem [{\citenamefont {Torrado}\ and\ \citenamefont
  {Lewis}(2021)}]{Torrado:2020dgo}%
  \BibitemOpen
  \bibfield  {author} {\bibinfo {author} {\bibfnamefont {J.}~\bibnamefont
  {Torrado}}\ and\ \bibinfo {author} {\bibfnamefont {A.}~\bibnamefont
  {Lewis}},\ }\href {\doibase 10.1088/1475-7516/2021/05/057} {\bibfield
  {journal} {\bibinfo  {journal} {JCAP}\ }\textbf {\bibinfo {volume} {05}},\
  \bibinfo {pages} {057} (\bibinfo {year} {2021})},\ \Eprint
  {http://arxiv.org/abs/2005.05290} {arXiv:2005.05290 [astro-ph.IM]}
  \BibitemShut {NoStop}%
\bibitem [{\citenamefont {Lewis}\ and\ \citenamefont
  {Bridle}(2002)}]{Lewis:2002ah}%
  \BibitemOpen
  \bibfield  {author} {\bibinfo {author} {\bibfnamefont {A.}~\bibnamefont
  {Lewis}}\ and\ \bibinfo {author} {\bibfnamefont {S.}~\bibnamefont {Bridle}},\
  }\href {\doibase 10.1103/PhysRevD.66.103511} {\bibfield  {journal} {\bibinfo
  {journal} {Phys. Rev. D}\ }\textbf {\bibinfo {volume} {66}},\ \bibinfo
  {pages} {103511} (\bibinfo {year} {2002})},\ \Eprint
  {http://arxiv.org/abs/astro-ph/0205436} {arXiv:astro-ph/0205436} \BibitemShut
  {NoStop}%
\bibitem [{\citenamefont {Lewis}(2013)}]{Lewis:2013hha}%
  \BibitemOpen
  \bibfield  {author} {\bibinfo {author} {\bibfnamefont {A.}~\bibnamefont
  {Lewis}},\ }\href {\doibase 10.1103/PhysRevD.87.103529} {\bibfield  {journal}
  {\bibinfo  {journal} {Phys. Rev. D}\ }\textbf {\bibinfo {volume} {87}},\
  \bibinfo {pages} {103529} (\bibinfo {year} {2013})},\ \Eprint
  {http://arxiv.org/abs/1304.4473} {arXiv:1304.4473 [astro-ph.CO]} \BibitemShut
  {NoStop}%
\bibitem [{\citenamefont {Neal}(2005)}]{neal2005taking}%
  \BibitemOpen
  \bibfield  {author} {\bibinfo {author} {\bibfnamefont {R.~M.}\ \bibnamefont
  {Neal}},\ }\href@noop {} {\enquote {\bibinfo {title} {Taking bigger
  metropolis steps by dragging fast variables},}\ } (\bibinfo {year} {2005}),\
  \Eprint {http://arxiv.org/abs/math/0502099} {arXiv:math/0502099 [math.ST]}
  \BibitemShut {NoStop}%
\bibitem [{\citenamefont {Gelman}\ and\ \citenamefont
  {Rubin}(1992)}]{Gelman:1992zz}%
  \BibitemOpen
  \bibfield  {author} {\bibinfo {author} {\bibfnamefont {A.}~\bibnamefont
  {Gelman}}\ and\ \bibinfo {author} {\bibfnamefont {D.~B.}\ \bibnamefont
  {Rubin}},\ }\href {\doibase 10.1214/ss/1177011136} {\bibfield  {journal}
  {\bibinfo  {journal} {Statist. Sci.}\ }\textbf {\bibinfo {volume} {7}},\
  \bibinfo {pages} {457} (\bibinfo {year} {1992})}\BibitemShut {NoStop}%
\bibitem [{\citenamefont {Lewis}(2019)}]{Lewis:2019xzd}%
  \BibitemOpen
  \bibfield  {author} {\bibinfo {author} {\bibfnamefont {A.}~\bibnamefont
  {Lewis}},\ }\href@noop {} {\  (\bibinfo {year} {2019})},\ \Eprint
  {http://arxiv.org/abs/1910.13970} {arXiv:1910.13970 [astro-ph.IM]}
  \BibitemShut {NoStop}%
\bibitem [{\citenamefont {Lesgourgues}(2011)}]{Lesgourgues:2011re}%
  \BibitemOpen
  \bibfield  {author} {\bibinfo {author} {\bibfnamefont {J.}~\bibnamefont
  {Lesgourgues}},\ }\href@noop {} {\  (\bibinfo {year} {2011})},\ \bibinfo
  {note} {arXiv:1104.2932},\ \Eprint {http://arxiv.org/abs/1104.2932}
  {arXiv:1104.2932 [astro-ph.IM]} \BibitemShut {NoStop}%
\bibitem [{\citenamefont {Blas}\ \emph {et~al.}(2011)\citenamefont {Blas},
  \citenamefont {Lesgourgues},\ and\ \citenamefont {Tram}}]{Blas:2011rf}%
  \BibitemOpen
  \bibfield  {author} {\bibinfo {author} {\bibfnamefont {D.}~\bibnamefont
  {Blas}}, \bibinfo {author} {\bibfnamefont {J.}~\bibnamefont {Lesgourgues}}, \
  and\ \bibinfo {author} {\bibfnamefont {T.}~\bibnamefont {Tram}},\ }\href
  {\doibase 10.1088/1475-7516/2011/07/034} {\bibfield  {journal} {\bibinfo
  {journal} {JCAP}\ }\textbf {\bibinfo {volume} {07}},\ \bibinfo {pages} {034}
  (\bibinfo {year} {2011})},\ \Eprint {http://arxiv.org/abs/1104.2933}
  {arXiv:1104.2933 [astro-ph.CO]} \BibitemShut {NoStop}%
\bibitem [{\citenamefont {Aghanim}\ \emph
  {et~al.}(2020{\natexlab{b}})\citenamefont {Aghanim} \emph
  {et~al.}}]{Planck:2019nip}%
  \BibitemOpen
  \bibfield  {author} {\bibinfo {author} {\bibfnamefont {N.}~\bibnamefont
  {Aghanim}} \emph {et~al.} (\bibinfo {collaboration} {Planck}),\ }\href
  {\doibase 10.1051/0004-6361/201936386} {\bibfield  {journal} {\bibinfo
  {journal} {Astron. Astrophys.}\ }\textbf {\bibinfo {volume} {641}},\ \bibinfo
  {pages} {A5} (\bibinfo {year} {2020}{\natexlab{b}})},\ \Eprint
  {http://arxiv.org/abs/1907.12875} {arXiv:1907.12875 [astro-ph.CO]}
  \BibitemShut {NoStop}%
\bibitem [{\citenamefont {Brout}\ \emph {et~al.}(2022)\citenamefont {Brout}
  \emph {et~al.}}]{Brout:2022vxf}%
  \BibitemOpen
  \bibfield  {author} {\bibinfo {author} {\bibfnamefont {D.}~\bibnamefont
  {Brout}} \emph {et~al.},\ }\href {\doibase 10.3847/1538-4357/ac8e04}
  {\bibfield  {journal} {\bibinfo  {journal} {Astrophys. J.}\ }\textbf
  {\bibinfo {volume} {938}},\ \bibinfo {pages} {110} (\bibinfo {year}
  {2022})},\ \Eprint {http://arxiv.org/abs/2202.04077} {arXiv:2202.04077
  [astro-ph.CO]} \BibitemShut {NoStop}%
\bibitem [{\citenamefont {Scolnic}\ \emph {et~al.}(2022)\citenamefont {Scolnic}
  \emph {et~al.}}]{Scolnic:2021amr}%
  \BibitemOpen
  \bibfield  {author} {\bibinfo {author} {\bibfnamefont {D.}~\bibnamefont
  {Scolnic}} \emph {et~al.},\ }\href {\doibase 10.3847/1538-4357/ac8b7a}
  {\bibfield  {journal} {\bibinfo  {journal} {Astrophys. J.}\ }\textbf
  {\bibinfo {volume} {938}},\ \bibinfo {pages} {113} (\bibinfo {year}
  {2022})},\ \Eprint {http://arxiv.org/abs/2112.03863} {arXiv:2112.03863
  [astro-ph.CO]} \BibitemShut {NoStop}%
\bibitem [{\citenamefont {Adame}\ \emph
  {et~al.}(2025{\natexlab{a}})\citenamefont {Adame} \emph
  {et~al.}}]{DESI:2024mwx}%
  \BibitemOpen
  \bibfield  {author} {\bibinfo {author} {\bibfnamefont {A.~G.}\ \bibnamefont
  {Adame}} \emph {et~al.} (\bibinfo {collaboration} {DESI}),\ }\href {\doibase
  10.1088/1475-7516/2025/02/021} {\bibfield  {journal} {\bibinfo  {journal}
  {JCAP}\ }\textbf {\bibinfo {volume} {02}},\ \bibinfo {pages} {021} (\bibinfo
  {year} {2025}{\natexlab{a}})},\ \Eprint {http://arxiv.org/abs/2404.03002}
  {arXiv:2404.03002 [astro-ph.CO]} \BibitemShut {NoStop}%
\bibitem [{\citenamefont {Adame}\ \emph
  {et~al.}(2025{\natexlab{b}})\citenamefont {Adame} \emph
  {et~al.}}]{DESI:2024uvr}%
  \BibitemOpen
  \bibfield  {author} {\bibinfo {author} {\bibfnamefont {A.~G.}\ \bibnamefont
  {Adame}} \emph {et~al.} (\bibinfo {collaboration} {DESI}),\ }\href {\doibase
  10.1088/1475-7516/2025/04/012} {\bibfield  {journal} {\bibinfo  {journal}
  {JCAP}\ }\textbf {\bibinfo {volume} {04}},\ \bibinfo {pages} {012} (\bibinfo
  {year} {2025}{\natexlab{b}})},\ \Eprint {http://arxiv.org/abs/2404.03000}
  {arXiv:2404.03000 [astro-ph.CO]} \BibitemShut {NoStop}%
\bibitem [{\citenamefont {Adame}\ \emph
  {et~al.}(2025{\natexlab{c}})\citenamefont {Adame} \emph
  {et~al.}}]{DESI:2024lzq}%
  \BibitemOpen
  \bibfield  {author} {\bibinfo {author} {\bibfnamefont {A.~G.}\ \bibnamefont
  {Adame}} \emph {et~al.} (\bibinfo {collaboration} {DESI}),\ }\href {\doibase
  10.1088/1475-7516/2025/01/124} {\bibfield  {journal} {\bibinfo  {journal}
  {JCAP}\ }\textbf {\bibinfo {volume} {01}},\ \bibinfo {pages} {124} (\bibinfo
  {year} {2025}{\natexlab{c}})},\ \Eprint {http://arxiv.org/abs/2404.03001}
  {arXiv:2404.03001 [astro-ph.CO]} \BibitemShut {NoStop}%
\bibitem [{\citenamefont {Alam}\ \emph
  {et~al.}(2017{\natexlab{b}})\citenamefont {Alam} \emph
  {et~al.}}]{Alam:2016hwk}%
  \BibitemOpen
  \bibfield  {author} {\bibinfo {author} {\bibfnamefont {S.}~\bibnamefont
  {Alam}} \emph {et~al.} (\bibinfo {collaboration} {BOSS}),\ }\href {\doibase
  10.1093/mnras/stx721} {\bibfield  {journal} {\bibinfo  {journal} {Mon. Not.
  Roy. Astron. Soc.}\ }\textbf {\bibinfo {volume} {470}},\ \bibinfo {pages}
  {2617} (\bibinfo {year} {2017}{\natexlab{b}})},\ \Eprint
  {http://arxiv.org/abs/1607.03155} {arXiv:1607.03155 [astro-ph.CO]}
  \BibitemShut {NoStop}%
\bibitem [{\citenamefont {Alam}\ \emph {et~al.}(2021)\citenamefont {Alam} \emph
  {et~al.}}]{eBOSS:2020yzd}%
  \BibitemOpen
  \bibfield  {author} {\bibinfo {author} {\bibfnamefont {S.}~\bibnamefont
  {Alam}} \emph {et~al.} (\bibinfo {collaboration} {eBOSS}),\ }\href {\doibase
  10.1103/PhysRevD.103.083533} {\bibfield  {journal} {\bibinfo  {journal}
  {Phys. Rev. D}\ }\textbf {\bibinfo {volume} {103}},\ \bibinfo {pages}
  {083533} (\bibinfo {year} {2021})},\ \Eprint
  {http://arxiv.org/abs/2007.08991} {arXiv:2007.08991 [astro-ph.CO]}
  \BibitemShut {NoStop}%
\bibitem [{\citenamefont {Weinberg}\ \emph {et~al.}(2013)\citenamefont
  {Weinberg}, \citenamefont {Mortonson}, \citenamefont {Eisenstein},
  \citenamefont {Hirata}, \citenamefont {Riess},\ and\ \citenamefont
  {Rozo}}]{Weinberg:2013agg}%
  \BibitemOpen
  \bibfield  {author} {\bibinfo {author} {\bibfnamefont {D.~H.}\ \bibnamefont
  {Weinberg}}, \bibinfo {author} {\bibfnamefont {M.~J.}\ \bibnamefont
  {Mortonson}}, \bibinfo {author} {\bibfnamefont {D.~J.}\ \bibnamefont
  {Eisenstein}}, \bibinfo {author} {\bibfnamefont {C.}~\bibnamefont {Hirata}},
  \bibinfo {author} {\bibfnamefont {A.~G.}\ \bibnamefont {Riess}}, \ and\
  \bibinfo {author} {\bibfnamefont {E.}~\bibnamefont {Rozo}},\ }\href {\doibase
  10.1016/j.physrep.2013.05.001} {\bibfield  {journal} {\bibinfo  {journal}
  {Phys. Rept.}\ }\textbf {\bibinfo {volume} {530}},\ \bibinfo {pages} {87}
  (\bibinfo {year} {2013})},\ \Eprint {http://arxiv.org/abs/1201.2434}
  {arXiv:1201.2434 [astro-ph.CO]} \BibitemShut {NoStop}%
\bibitem [{\citenamefont {Mortonson}\ \emph {et~al.}(2013)\citenamefont
  {Mortonson}, \citenamefont {Weinberg},\ and\ \citenamefont
  {White}}]{Mortonson:2013zfa}%
  \BibitemOpen
  \bibfield  {author} {\bibinfo {author} {\bibfnamefont {M.~J.}\ \bibnamefont
  {Mortonson}}, \bibinfo {author} {\bibfnamefont {D.~H.}\ \bibnamefont
  {Weinberg}}, \ and\ \bibinfo {author} {\bibfnamefont {M.}~\bibnamefont
  {White}},\ }\href@noop {} {\  (\bibinfo {year} {2013})},\ \Eprint
  {http://arxiv.org/abs/1401.0046} {arXiv:1401.0046 [astro-ph.CO]} \BibitemShut
  {NoStop}%
\bibitem [{\citenamefont {Frieman}\ \emph {et~al.}(2008)\citenamefont
  {Frieman}, \citenamefont {Turner},\ and\ \citenamefont
  {Huterer}}]{Frieman:2008sn}%
  \BibitemOpen
  \bibfield  {author} {\bibinfo {author} {\bibfnamefont {J.}~\bibnamefont
  {Frieman}}, \bibinfo {author} {\bibfnamefont {M.}~\bibnamefont {Turner}}, \
  and\ \bibinfo {author} {\bibfnamefont {D.}~\bibnamefont {Huterer}},\ }\href
  {\doibase 10.1146/annurev.astro.46.060407.145243} {\bibfield  {journal}
  {\bibinfo  {journal} {Ann. Rev. Astron. Astrophys.}\ }\textbf {\bibinfo
  {volume} {46}},\ \bibinfo {pages} {385} (\bibinfo {year} {2008})},\ \Eprint
  {http://arxiv.org/abs/0803.0982} {arXiv:0803.0982 [astro-ph]} \BibitemShut
  {NoStop}%
\bibitem [{\citenamefont {Copeland}\ \emph {et~al.}(2006)\citenamefont
  {Copeland}, \citenamefont {Sami},\ and\ \citenamefont
  {Tsujikawa}}]{Copeland:2006wr}%
  \BibitemOpen
  \bibfield  {author} {\bibinfo {author} {\bibfnamefont {E.~J.}\ \bibnamefont
  {Copeland}}, \bibinfo {author} {\bibfnamefont {M.}~\bibnamefont {Sami}}, \
  and\ \bibinfo {author} {\bibfnamefont {S.}~\bibnamefont {Tsujikawa}},\ }\href
  {\doibase 10.1142/S021827180600942X} {\bibfield  {journal} {\bibinfo
  {journal} {Int. J. Mod. Phys. D}\ }\textbf {\bibinfo {volume} {15}},\
  \bibinfo {pages} {1753} (\bibinfo {year} {2006})},\ \Eprint
  {http://arxiv.org/abs/hep-th/0603057} {arXiv:hep-th/0603057} \BibitemShut
  {NoStop}%
\bibitem [{\citenamefont {Riess}\ \emph {et~al.}(1998)\citenamefont {Riess}
  \emph {et~al.}}]{SupernovaSearchTeam:1998fmf}%
  \BibitemOpen
  \bibfield  {author} {\bibinfo {author} {\bibfnamefont {A.~G.}\ \bibnamefont
  {Riess}} \emph {et~al.} (\bibinfo {collaboration} {Supernova Search Team}),\
  }\href {\doibase 10.1086/300499} {\bibfield  {journal} {\bibinfo  {journal}
  {Astron. J.}\ }\textbf {\bibinfo {volume} {116}},\ \bibinfo {pages} {1009}
  (\bibinfo {year} {1998})},\ \Eprint {http://arxiv.org/abs/astro-ph/9805201}
  {arXiv:astro-ph/9805201} \BibitemShut {NoStop}%
\bibitem [{\citenamefont {Perlmutter}\ \emph {et~al.}(1999)\citenamefont
  {Perlmutter} \emph {et~al.}}]{SupernovaCosmologyProject:1998vns}%
  \BibitemOpen
  \bibfield  {author} {\bibinfo {author} {\bibfnamefont {S.}~\bibnamefont
  {Perlmutter}} \emph {et~al.} (\bibinfo {collaboration} {Supernova Cosmology
  Project}),\ }\href {\doibase 10.1086/307221} {\bibfield  {journal} {\bibinfo
  {journal} {Astrophys. J.}\ }\textbf {\bibinfo {volume} {517}},\ \bibinfo
  {pages} {565} (\bibinfo {year} {1999})},\ \Eprint
  {http://arxiv.org/abs/astro-ph/9812133} {arXiv:astro-ph/9812133} \BibitemShut
  {NoStop}%
\bibitem [{\citenamefont {Sahni}\ and\ \citenamefont
  {Starobinsky}(2000)}]{Sahni:1999gb}%
  \BibitemOpen
  \bibfield  {author} {\bibinfo {author} {\bibfnamefont {V.}~\bibnamefont
  {Sahni}}\ and\ \bibinfo {author} {\bibfnamefont {A.~A.}\ \bibnamefont
  {Starobinsky}},\ }\href {\doibase 10.1142/S0218271800000542} {\bibfield
  {journal} {\bibinfo  {journal} {Int. J. Mod. Phys. D}\ }\textbf {\bibinfo
  {volume} {9}},\ \bibinfo {pages} {373} (\bibinfo {year} {2000})},\ \Eprint
  {http://arxiv.org/abs/astro-ph/9904398} {arXiv:astro-ph/9904398} \BibitemShut
  {NoStop}%
\bibitem [{\citenamefont {Carroll}(2001)}]{Carroll:2000fy}%
  \BibitemOpen
  \bibfield  {author} {\bibinfo {author} {\bibfnamefont {S.~M.}\ \bibnamefont
  {Carroll}},\ }\href {\doibase 10.12942/lrr-2001-1} {\bibfield  {journal}
  {\bibinfo  {journal} {Living Rev. Rel.}\ }\textbf {\bibinfo {volume} {4}},\
  \bibinfo {pages} {1} (\bibinfo {year} {2001})},\ \Eprint
  {http://arxiv.org/abs/astro-ph/0004075} {arXiv:astro-ph/0004075} \BibitemShut
  {NoStop}%
\bibitem [{\citenamefont {Peebles}\ and\ \citenamefont
  {Ratra}(2003)}]{Peebles:2002gy}%
  \BibitemOpen
  \bibfield  {author} {\bibinfo {author} {\bibfnamefont {P.~J.~E.}\
  \bibnamefont {Peebles}}\ and\ \bibinfo {author} {\bibfnamefont
  {B.}~\bibnamefont {Ratra}},\ }\href {\doibase 10.1103/RevModPhys.75.559}
  {\bibfield  {journal} {\bibinfo  {journal} {Rev. Mod. Phys.}\ }\textbf
  {\bibinfo {volume} {75}},\ \bibinfo {pages} {559} (\bibinfo {year} {2003})},\
  \Eprint {http://arxiv.org/abs/astro-ph/0207347} {arXiv:astro-ph/0207347}
  \BibitemShut {NoStop}%
\bibitem [{\citenamefont {Padmanabhan}(2003)}]{Padmanabhan:2002ji}%
  \BibitemOpen
  \bibfield  {author} {\bibinfo {author} {\bibfnamefont {T.}~\bibnamefont
  {Padmanabhan}},\ }\href {\doibase 10.1016/S0370-1573(03)00120-0} {\bibfield
  {journal} {\bibinfo  {journal} {Phys. Rept.}\ }\textbf {\bibinfo {volume}
  {380}},\ \bibinfo {pages} {235} (\bibinfo {year} {2003})},\ \Eprint
  {http://arxiv.org/abs/hep-th/0212290} {arXiv:hep-th/0212290} \BibitemShut
  {NoStop}%
\bibitem [{\citenamefont {Weinberg}(1989)}]{Weinberg:1988cp}%
  \BibitemOpen
  \bibfield  {author} {\bibinfo {author} {\bibfnamefont {S.}~\bibnamefont
  {Weinberg}},\ }\href {\doibase 10.1103/RevModPhys.61.1} {\bibfield  {journal}
  {\bibinfo  {journal} {Rev. Mod. Phys.}\ }\textbf {\bibinfo {volume} {61}},\
  \bibinfo {pages} {1} (\bibinfo {year} {1989})}\BibitemShut {NoStop}%
\bibitem [{\citenamefont {Martin}(2012)}]{Martin:2012bt}%
  \BibitemOpen
  \bibfield  {author} {\bibinfo {author} {\bibfnamefont {J.}~\bibnamefont
  {Martin}},\ }\href {\doibase 10.1016/j.crhy.2012.04.008} {\bibfield
  {journal} {\bibinfo  {journal} {Comptes Rendus Physique}\ }\textbf {\bibinfo
  {volume} {13}},\ \bibinfo {pages} {566} (\bibinfo {year} {2012})},\ \Eprint
  {http://arxiv.org/abs/1205.3365} {arXiv:1205.3365 [astro-ph.CO]} \BibitemShut
  {NoStop}%
\bibitem [{\citenamefont {Solà}(2013)}]{Sola:2013gha}%
  \BibitemOpen
  \bibfield  {author} {\bibinfo {author} {\bibfnamefont {J.}~\bibnamefont
  {Solà}},\ }\href {\doibase 10.1088/1742-6596/453/1/012015} {\bibfield
  {journal} {\bibinfo  {journal} {J. Phys. Conf. Ser.}\ }\textbf {\bibinfo
  {volume} {453}},\ \bibinfo {pages} {012015} (\bibinfo {year} {2013})},\
  \Eprint {http://arxiv.org/abs/1306.1527} {arXiv:1306.1527 [gr-qc]}
  \BibitemShut {NoStop}%
\bibitem [{\citenamefont {Sol\`a~Peracaula}(2022)}]{SolaPeracaula:2022hpd}%
  \BibitemOpen
  \bibfield  {author} {\bibinfo {author} {\bibfnamefont {J.}~\bibnamefont
  {Sol\`a~Peracaula}},\ }\href {\doibase 10.1098/rsta.2021.0182} {\bibfield
  {journal} {\bibinfo  {journal} {Phil. Trans. Roy. Soc. Lond. A}\ }\textbf
  {\bibinfo {volume} {380}},\ \bibinfo {pages} {20210182} (\bibinfo {year}
  {2022})},\ \Eprint {http://arxiv.org/abs/2203.13757} {arXiv:2203.13757
  [gr-qc]} \BibitemShut {NoStop}%
\bibitem [{\citenamefont {Amendola}\ and\ \citenamefont
  {Tsujikawa}(2015)}]{Amendola:2015ksp}%
  \BibitemOpen
  \bibfield  {author} {\bibinfo {author} {\bibfnamefont {L.}~\bibnamefont
  {Amendola}}\ and\ \bibinfo {author} {\bibfnamefont {S.}~\bibnamefont
  {Tsujikawa}},\ }\href@noop {} {\emph {\bibinfo {title} {{Dark Energy}:
  {Theory and Observations}}}}\ (\bibinfo  {publisher} {Cambridge University
  Press},\ \bibinfo {year} {2015})\BibitemShut {NoStop}%
\bibitem [{\citenamefont {Chevallier}\ and\ \citenamefont
  {Polarski}(2001)}]{Chevallier:2000qy}%
  \BibitemOpen
  \bibfield  {author} {\bibinfo {author} {\bibfnamefont {M.}~\bibnamefont
  {Chevallier}}\ and\ \bibinfo {author} {\bibfnamefont {D.}~\bibnamefont
  {Polarski}},\ }\href {\doibase 10.1142/S0218271801000822} {\bibfield
  {journal} {\bibinfo  {journal} {Int. J. Mod. Phys. D}\ }\textbf {\bibinfo
  {volume} {10}},\ \bibinfo {pages} {213} (\bibinfo {year} {2001})},\ \Eprint
  {http://arxiv.org/abs/gr-qc/0009008} {arXiv:gr-qc/0009008} \BibitemShut
  {NoStop}%
\bibitem [{\citenamefont {Linder}(2003)}]{Linder:2002et}%
  \BibitemOpen
  \bibfield  {author} {\bibinfo {author} {\bibfnamefont {E.~V.}\ \bibnamefont
  {Linder}},\ }\href {\doibase 10.1103/PhysRevLett.90.091301} {\bibfield
  {journal} {\bibinfo  {journal} {Phys. Rev. Lett.}\ }\textbf {\bibinfo
  {volume} {90}},\ \bibinfo {pages} {091301} (\bibinfo {year} {2003})},\
  \Eprint {http://arxiv.org/abs/astro-ph/0208512} {arXiv:astro-ph/0208512}
  \BibitemShut {NoStop}%
\bibitem [{\citenamefont {Giar\`e}\ \emph {et~al.}(2024)\citenamefont
  {Giar\`e}, \citenamefont {Najafi}, \citenamefont {Pan}, \citenamefont
  {Di~Valentino},\ and\ \citenamefont {Firouzjaee}}]{Giare:2024gpk}%
  \BibitemOpen
  \bibfield  {author} {\bibinfo {author} {\bibfnamefont {W.}~\bibnamefont
  {Giar\`e}}, \bibinfo {author} {\bibfnamefont {M.}~\bibnamefont {Najafi}},
  \bibinfo {author} {\bibfnamefont {S.}~\bibnamefont {Pan}}, \bibinfo {author}
  {\bibfnamefont {E.}~\bibnamefont {Di~Valentino}}, \ and\ \bibinfo {author}
  {\bibfnamefont {J.~T.}\ \bibnamefont {Firouzjaee}},\ }\href {\doibase
  10.1088/1475-7516/2024/10/035} {\bibfield  {journal} {\bibinfo  {journal}
  {JCAP}\ }\textbf {\bibinfo {volume} {10}},\ \bibinfo {pages} {035} (\bibinfo
  {year} {2024})},\ \Eprint {http://arxiv.org/abs/2407.16689} {arXiv:2407.16689
  [astro-ph.CO]} \BibitemShut {NoStop}%
\bibitem [{\citenamefont {Najafi}\ \emph {et~al.}(2024)\citenamefont {Najafi},
  \citenamefont {Pan}, \citenamefont {Di~Valentino},\ and\ \citenamefont
  {Firouzjaee}}]{Najafi:2024qzm}%
  \BibitemOpen
  \bibfield  {author} {\bibinfo {author} {\bibfnamefont {M.}~\bibnamefont
  {Najafi}}, \bibinfo {author} {\bibfnamefont {S.}~\bibnamefont {Pan}},
  \bibinfo {author} {\bibfnamefont {E.}~\bibnamefont {Di~Valentino}}, \ and\
  \bibinfo {author} {\bibfnamefont {J.~T.}\ \bibnamefont {Firouzjaee}},\ }\href
  {\doibase 10.1016/j.dark.2024.101539} {\bibfield  {journal} {\bibinfo
  {journal} {Phys. Dark Univ.}\ }\textbf {\bibinfo {volume} {45}},\ \bibinfo
  {pages} {101539} (\bibinfo {year} {2024})},\ \Eprint
  {http://arxiv.org/abs/2407.14939} {arXiv:2407.14939 [astro-ph.CO]}
  \BibitemShut {NoStop}%
\bibitem [{\citenamefont {Scherrer}(2015)}]{Scherrer:2015tra}%
  \BibitemOpen
  \bibfield  {author} {\bibinfo {author} {\bibfnamefont {R.~J.}\ \bibnamefont
  {Scherrer}},\ }\href {\doibase 10.1103/PhysRevD.92.043001} {\bibfield
  {journal} {\bibinfo  {journal} {Phys. Rev. D}\ }\textbf {\bibinfo {volume}
  {92}},\ \bibinfo {pages} {043001} (\bibinfo {year} {2015})},\ \Eprint
  {http://arxiv.org/abs/1505.05781} {arXiv:1505.05781 [astro-ph.CO]}
  \BibitemShut {NoStop}%
\bibitem [{\citenamefont {Park}\ \emph
  {et~al.}(2024{\natexlab{a}})\citenamefont {Park}, \citenamefont
  {de~Cruz~P\'erez},\ and\ \citenamefont {Ratra}}]{Chan-GyungPark:2024mlx}%
  \BibitemOpen
  \bibfield  {author} {\bibinfo {author} {\bibfnamefont {C.-G.}\ \bibnamefont
  {Park}}, \bibinfo {author} {\bibfnamefont {J.}~\bibnamefont
  {de~Cruz~P\'erez}}, \ and\ \bibinfo {author} {\bibfnamefont {B.}~\bibnamefont
  {Ratra}},\ }\href {\doibase 10.1103/PhysRevD.110.123533} {\bibfield
  {journal} {\bibinfo  {journal} {Phys. Rev. D}\ }\textbf {\bibinfo {volume}
  {110}},\ \bibinfo {pages} {123533} (\bibinfo {year} {2024}{\natexlab{a}})},\
  \Eprint {http://arxiv.org/abs/2405.00502} {arXiv:2405.00502 [astro-ph.CO]}
  \BibitemShut {NoStop}%
\bibitem [{\citenamefont {Liu}\ \emph {et~al.}(2024{\natexlab{b}})\citenamefont
  {Liu}, \citenamefont {Wang},\ and\ \citenamefont {Zhao}}]{Liu:2024gfy}%
  \BibitemOpen
  \bibfield  {author} {\bibinfo {author} {\bibfnamefont {G.}~\bibnamefont
  {Liu}}, \bibinfo {author} {\bibfnamefont {Y.}~\bibnamefont {Wang}}, \ and\
  \bibinfo {author} {\bibfnamefont {W.}~\bibnamefont {Zhao}},\ }\href@noop {}
  {\  (\bibinfo {year} {2024}{\natexlab{b}})},\ \Eprint
  {http://arxiv.org/abs/2407.04385} {arXiv:2407.04385 [astro-ph.CO]}
  \BibitemShut {NoStop}%
\bibitem [{\citenamefont {Cort\^es}\ and\ \citenamefont
  {Liddle}(2024)}]{Cortes:2024lgw}%
  \BibitemOpen
  \bibfield  {author} {\bibinfo {author} {\bibfnamefont {M.}~\bibnamefont
  {Cort\^es}}\ and\ \bibinfo {author} {\bibfnamefont {A.~R.}\ \bibnamefont
  {Liddle}},\ }\href {\doibase 10.1088/1475-7516/2024/12/007} {\bibfield
  {journal} {\bibinfo  {journal} {JCAP}\ }\textbf {\bibinfo {volume} {12}},\
  \bibinfo {pages} {007} (\bibinfo {year} {2024})},\ \Eprint
  {http://arxiv.org/abs/2404.08056} {arXiv:2404.08056 [astro-ph.CO]}
  \BibitemShut {NoStop}%
\bibitem [{\citenamefont {Malekjani}\ \emph {et~al.}(2025)\citenamefont
  {Malekjani}, \citenamefont {Davari},\ and\ \citenamefont
  {Pourojaghi}}]{Pourojaghi:2024tmw}%
  \BibitemOpen
  \bibfield  {author} {\bibinfo {author} {\bibfnamefont {M.}~\bibnamefont
  {Malekjani}}, \bibinfo {author} {\bibfnamefont {Z.}~\bibnamefont {Davari}}, \
  and\ \bibinfo {author} {\bibfnamefont {S.}~\bibnamefont {Pourojaghi}}
  (\bibinfo {collaboration} {DESI}),\ }\href {\doibase
  10.1103/PhysRevD.111.083547} {\bibfield  {journal} {\bibinfo  {journal}
  {Phys. Rev. D}\ }\textbf {\bibinfo {volume} {111}},\ \bibinfo {pages}
  {083547} (\bibinfo {year} {2025})},\ \Eprint
  {http://arxiv.org/abs/2407.09767} {arXiv:2407.09767 [astro-ph.CO]}
  \BibitemShut {NoStop}%
\bibitem [{\citenamefont {Shlivko}\ and\ \citenamefont
  {Steinhardt}(2024)}]{Shlivko:2024llw}%
  \BibitemOpen
  \bibfield  {author} {\bibinfo {author} {\bibfnamefont {D.}~\bibnamefont
  {Shlivko}}\ and\ \bibinfo {author} {\bibfnamefont {P.~J.}\ \bibnamefont
  {Steinhardt}},\ }\href {\doibase 10.1016/j.physletb.2024.138826} {\bibfield
  {journal} {\bibinfo  {journal} {Phys. Lett. B}\ }\textbf {\bibinfo {volume}
  {855}},\ \bibinfo {pages} {138826} (\bibinfo {year} {2024})},\ \Eprint
  {http://arxiv.org/abs/2405.03933} {arXiv:2405.03933 [astro-ph.CO]}
  \BibitemShut {NoStop}%
\bibitem [{\citenamefont {Wang}\ and\ \citenamefont
  {Piao}(2024)}]{Wang:2024dka}%
  \BibitemOpen
  \bibfield  {author} {\bibinfo {author} {\bibfnamefont {H.}~\bibnamefont
  {Wang}}\ and\ \bibinfo {author} {\bibfnamefont {Y.-S.}\ \bibnamefont
  {Piao}},\ }\href@noop {} {\  (\bibinfo {year} {2024})},\ \Eprint
  {http://arxiv.org/abs/2404.18579} {arXiv:2404.18579 [astro-ph.CO]}
  \BibitemShut {NoStop}%
\bibitem [{\citenamefont {Gialamas}\ \emph {et~al.}(2025)\citenamefont
  {Gialamas}, \citenamefont {H\"utsi}, \citenamefont {Kannike}, \citenamefont
  {Racioppi}, \citenamefont {Raidal}, \citenamefont {Vasar},\ and\
  \citenamefont {Veerm\"ae}}]{Gialamas:2024lyw}%
  \BibitemOpen
  \bibfield  {author} {\bibinfo {author} {\bibfnamefont {I.~D.}\ \bibnamefont
  {Gialamas}}, \bibinfo {author} {\bibfnamefont {G.}~\bibnamefont {H\"utsi}},
  \bibinfo {author} {\bibfnamefont {K.}~\bibnamefont {Kannike}}, \bibinfo
  {author} {\bibfnamefont {A.}~\bibnamefont {Racioppi}}, \bibinfo {author}
  {\bibfnamefont {M.}~\bibnamefont {Raidal}}, \bibinfo {author} {\bibfnamefont
  {M.}~\bibnamefont {Vasar}}, \ and\ \bibinfo {author} {\bibfnamefont
  {H.}~\bibnamefont {Veerm\"ae}},\ }\href {\doibase
  10.1103/PhysRevD.111.043540} {\bibfield  {journal} {\bibinfo  {journal}
  {Phys. Rev. D}\ }\textbf {\bibinfo {volume} {111}},\ \bibinfo {pages}
  {043540} (\bibinfo {year} {2025})},\ \Eprint
  {http://arxiv.org/abs/2406.07533} {arXiv:2406.07533 [astro-ph.CO]}
  \BibitemShut {NoStop}%
\bibitem [{\citenamefont {Notari}\ \emph {et~al.}(2024)\citenamefont {Notari},
  \citenamefont {Redi},\ and\ \citenamefont {Tesi}}]{Notari:2024rti}%
  \BibitemOpen
  \bibfield  {author} {\bibinfo {author} {\bibfnamefont {A.}~\bibnamefont
  {Notari}}, \bibinfo {author} {\bibfnamefont {M.}~\bibnamefont {Redi}}, \ and\
  \bibinfo {author} {\bibfnamefont {A.}~\bibnamefont {Tesi}},\ }\href {\doibase
  10.1088/1475-7516/2024/11/025} {\bibfield  {journal} {\bibinfo  {journal}
  {JCAP}\ }\textbf {\bibinfo {volume} {11}},\ \bibinfo {pages} {025} (\bibinfo
  {year} {2024})},\ \Eprint {http://arxiv.org/abs/2406.08459} {arXiv:2406.08459
  [astro-ph.CO]} \BibitemShut {NoStop}%
\bibitem [{\citenamefont {Park}\ \emph
  {et~al.}(2024{\natexlab{b}})\citenamefont {Park}, \citenamefont
  {de~Cruz~P\'erez},\ and\ \citenamefont {Ratra}}]{Chan-GyungPark:2024brx}%
  \BibitemOpen
  \bibfield  {author} {\bibinfo {author} {\bibfnamefont {C.-G.}\ \bibnamefont
  {Park}}, \bibinfo {author} {\bibfnamefont {J.}~\bibnamefont
  {de~Cruz~P\'erez}}, \ and\ \bibinfo {author} {\bibfnamefont {B.}~\bibnamefont
  {Ratra}},\ }\href@noop {} {\  (\bibinfo {year} {2024}{\natexlab{b}})},\
  \Eprint {http://arxiv.org/abs/2410.13627} {arXiv:2410.13627 [astro-ph.CO]}
  \BibitemShut {NoStop}%
\bibitem [{\citenamefont {Wolf}\ \emph {et~al.}(2024)\citenamefont {Wolf},
  \citenamefont {Garc\'\i{}a-Garc\'\i{}a}, \citenamefont {Bartlett},\ and\
  \citenamefont {Ferreira}}]{Wolf:2024eph}%
  \BibitemOpen
  \bibfield  {author} {\bibinfo {author} {\bibfnamefont {W.~J.}\ \bibnamefont
  {Wolf}}, \bibinfo {author} {\bibfnamefont {C.}~\bibnamefont
  {Garc\'\i{}a-Garc\'\i{}a}}, \bibinfo {author} {\bibfnamefont {D.~J.}\
  \bibnamefont {Bartlett}}, \ and\ \bibinfo {author} {\bibfnamefont {P.~G.}\
  \bibnamefont {Ferreira}},\ }\href {\doibase 10.1103/PhysRevD.110.083528}
  {\bibfield  {journal} {\bibinfo  {journal} {Phys. Rev. D}\ }\textbf {\bibinfo
  {volume} {110}},\ \bibinfo {pages} {083528} (\bibinfo {year} {2024})},\
  \Eprint {http://arxiv.org/abs/2408.17318} {arXiv:2408.17318 [astro-ph.CO]}
  \BibitemShut {NoStop}%
\bibitem [{\citenamefont {Wolf}\ \emph {et~al.}(2025)\citenamefont {Wolf},
  \citenamefont {Ferreira},\ and\ \citenamefont
  {Garc\'\i{}a-Garc\'\i{}a}}]{Wolf:2024stt}%
  \BibitemOpen
  \bibfield  {author} {\bibinfo {author} {\bibfnamefont {W.~J.}\ \bibnamefont
  {Wolf}}, \bibinfo {author} {\bibfnamefont {P.~G.}\ \bibnamefont {Ferreira}},
  \ and\ \bibinfo {author} {\bibfnamefont {C.}~\bibnamefont
  {Garc\'\i{}a-Garc\'\i{}a}},\ }\href {\doibase 10.1103/PhysRevD.111.L041303}
  {\bibfield  {journal} {\bibinfo  {journal} {Phys. Rev. D}\ }\textbf {\bibinfo
  {volume} {111}},\ \bibinfo {pages} {L041303} (\bibinfo {year} {2025})},\
  \Eprint {http://arxiv.org/abs/2409.17019} {arXiv:2409.17019 [astro-ph.CO]}
  \BibitemShut {NoStop}%
\bibitem [{\citenamefont {Sahni}\ \emph {et~al.}(2014)\citenamefont {Sahni},
  \citenamefont {Shafieloo},\ and\ \citenamefont
  {Starobinsky}}]{Sahni:2014ooa}%
  \BibitemOpen
  \bibfield  {author} {\bibinfo {author} {\bibfnamefont {V.}~\bibnamefont
  {Sahni}}, \bibinfo {author} {\bibfnamefont {A.}~\bibnamefont {Shafieloo}}, \
  and\ \bibinfo {author} {\bibfnamefont {A.~A.}\ \bibnamefont {Starobinsky}},\
  }\href {\doibase 10.1088/2041-8205/793/2/L40} {\bibfield  {journal} {\bibinfo
   {journal} {Astrophys. J. Lett.}\ }\textbf {\bibinfo {volume} {793}},\
  \bibinfo {pages} {L40} (\bibinfo {year} {2014})},\ \Eprint
  {http://arxiv.org/abs/1406.2209} {arXiv:1406.2209 [astro-ph.CO]} \BibitemShut
  {NoStop}%
\bibitem [{\citenamefont {Sol\`a}\ \emph {et~al.}(2015)\citenamefont {Sol\`a},
  \citenamefont {G\'omez-Valent},\ and\ \citenamefont
  {de~Cruz~P\'erez}}]{Sola:2015wwa}%
  \BibitemOpen
  \bibfield  {author} {\bibinfo {author} {\bibfnamefont {J.}~\bibnamefont
  {Sol\`a}}, \bibinfo {author} {\bibfnamefont {A.}~\bibnamefont
  {G\'omez-Valent}}, \ and\ \bibinfo {author} {\bibfnamefont {J.}~\bibnamefont
  {de~Cruz~P\'erez}},\ }\href {\doibase 10.1088/2041-8205/811/1/L14} {\bibfield
   {journal} {\bibinfo  {journal} {Astrophys. J. Lett.}\ }\textbf {\bibinfo
  {volume} {811}},\ \bibinfo {pages} {L14} (\bibinfo {year} {2015})},\ \Eprint
  {http://arxiv.org/abs/1506.05793} {arXiv:1506.05793 [gr-qc]} \BibitemShut
  {NoStop}%
\bibitem [{\citenamefont {Sol\`a}\ \emph {et~al.}(2017)\citenamefont {Sol\`a},
  \citenamefont {G\'omez-Valent},\ and\ \citenamefont
  {de~Cruz~P\'erez}}]{Sola:2016jky}%
  \BibitemOpen
  \bibfield  {author} {\bibinfo {author} {\bibfnamefont {J.}~\bibnamefont
  {Sol\`a}}, \bibinfo {author} {\bibfnamefont {A.}~\bibnamefont
  {G\'omez-Valent}}, \ and\ \bibinfo {author} {\bibfnamefont {J.}~\bibnamefont
  {de~Cruz~P\'erez}},\ }\href {\doibase 10.3847/1538-4357/836/1/43} {\bibfield
  {journal} {\bibinfo  {journal} {Astrophys. J.}\ }\textbf {\bibinfo {volume}
  {836}},\ \bibinfo {pages} {43} (\bibinfo {year} {2017})},\ \Eprint
  {http://arxiv.org/abs/1602.02103} {arXiv:1602.02103 [astro-ph.CO]}
  \BibitemShut {NoStop}%
\bibitem [{\citenamefont {Sol\`a~Peracaula}\ \emph
  {et~al.}(2018{\natexlab{a}})\citenamefont {Sol\`a~Peracaula}, \citenamefont
  {de~Cruz~P\'erez},\ and\ \citenamefont
  {G\'omez-Valent}}]{SolaPeracaula:2016qlq}%
  \BibitemOpen
  \bibfield  {author} {\bibinfo {author} {\bibfnamefont {J.}~\bibnamefont
  {Sol\`a~Peracaula}}, \bibinfo {author} {\bibfnamefont {J.}~\bibnamefont
  {de~Cruz~P\'erez}}, \ and\ \bibinfo {author} {\bibfnamefont {A.}~\bibnamefont
  {G\'omez-Valent}},\ }\href {\doibase 10.1209/0295-5075/121/39001} {\bibfield
  {journal} {\bibinfo  {journal} {EPL}\ }\textbf {\bibinfo {volume} {121}},\
  \bibinfo {pages} {39001} (\bibinfo {year} {2018}{\natexlab{a}})},\ \Eprint
  {http://arxiv.org/abs/1606.00450} {arXiv:1606.00450 [gr-qc]} \BibitemShut
  {NoStop}%
\bibitem [{\citenamefont {Sol\`a~Peracaula}\ \emph
  {et~al.}(2018{\natexlab{b}})\citenamefont {Sol\`a~Peracaula}, \citenamefont
  {de~Cruz~P\'erez},\ and\ \citenamefont
  {G\'omez-Valent}}]{SolaPeracaula:2017esw}%
  \BibitemOpen
  \bibfield  {author} {\bibinfo {author} {\bibfnamefont {J.}~\bibnamefont
  {Sol\`a~Peracaula}}, \bibinfo {author} {\bibfnamefont {J.}~\bibnamefont
  {de~Cruz~P\'erez}}, \ and\ \bibinfo {author} {\bibfnamefont {A.}~\bibnamefont
  {G\'omez-Valent}},\ }\href {\doibase 10.1093/mnras/sty1253} {\bibfield
  {journal} {\bibinfo  {journal} {Mon. Not. Roy. Astron. Soc.}\ }\textbf
  {\bibinfo {volume} {478}},\ \bibinfo {pages} {4357} (\bibinfo {year}
  {2018}{\natexlab{b}})},\ \Eprint {http://arxiv.org/abs/1703.08218}
  {arXiv:1703.08218 [astro-ph.CO]} \BibitemShut {NoStop}%
\bibitem [{\citenamefont {Zhao}\ \emph {et~al.}(2017)\citenamefont {Zhao} \emph
  {et~al.}}]{Zhao:2017cud}%
  \BibitemOpen
  \bibfield  {author} {\bibinfo {author} {\bibfnamefont {G.-B.}\ \bibnamefont
  {Zhao}} \emph {et~al.},\ }\href {\doibase 10.1038/s41550-017-0216-z}
  {\bibfield  {journal} {\bibinfo  {journal} {Nature Astron.}\ }\textbf
  {\bibinfo {volume} {1}},\ \bibinfo {pages} {627} (\bibinfo {year} {2017})},\
  \Eprint {http://arxiv.org/abs/1701.08165} {arXiv:1701.08165 [astro-ph.CO]}
  \BibitemShut {NoStop}%
\bibitem [{\citenamefont {Solà~Peracaula}\ \emph {et~al.}(2019)\citenamefont
  {Solà~Peracaula}, \citenamefont {G\'omez-Valent},\ and\ \citenamefont
  {de~Cruz~P\'erez}}]{SolaPeracaula:2018wwm}%
  \BibitemOpen
  \bibfield  {author} {\bibinfo {author} {\bibfnamefont {J.}~\bibnamefont
  {Solà~Peracaula}}, \bibinfo {author} {\bibfnamefont {A.}~\bibnamefont
  {G\'omez-Valent}}, \ and\ \bibinfo {author} {\bibfnamefont {J.}~\bibnamefont
  {de~Cruz~P\'erez}},\ }\href {\doibase 10.1016/j.dark.2019.100311} {\bibfield
  {journal} {\bibinfo  {journal} {Phys. Dark Univ.}\ }\textbf {\bibinfo
  {volume} {25}},\ \bibinfo {pages} {100311} (\bibinfo {year} {2019})},\
  \Eprint {http://arxiv.org/abs/1811.03505} {arXiv:1811.03505 [astro-ph.CO]}
  \BibitemShut {NoStop}%
\bibitem [{\citenamefont {Mangano}\ \emph {et~al.}(2002)\citenamefont
  {Mangano}, \citenamefont {Miele}, \citenamefont {Pastor},\ and\ \citenamefont
  {Peloso}}]{Mangano:2001iu}%
  \BibitemOpen
  \bibfield  {author} {\bibinfo {author} {\bibfnamefont {G.}~\bibnamefont
  {Mangano}}, \bibinfo {author} {\bibfnamefont {G.}~\bibnamefont {Miele}},
  \bibinfo {author} {\bibfnamefont {S.}~\bibnamefont {Pastor}}, \ and\ \bibinfo
  {author} {\bibfnamefont {M.}~\bibnamefont {Peloso}},\ }\href {\doibase
  10.1016/S0370-2693(02)01622-2} {\bibfield  {journal} {\bibinfo  {journal}
  {Phys. Lett. B}\ }\textbf {\bibinfo {volume} {534}},\ \bibinfo {pages} {8}
  (\bibinfo {year} {2002})},\ \Eprint {http://arxiv.org/abs/astro-ph/0111408}
  {arXiv:astro-ph/0111408} \BibitemShut {NoStop}%
\bibitem [{\citenamefont {Bennett}\ \emph {et~al.}(2020)\citenamefont
  {Bennett}, \citenamefont {Buldgen}, \citenamefont {Drewes},\ and\
  \citenamefont {Wong}}]{Bennett:2019ewm}%
  \BibitemOpen
  \bibfield  {author} {\bibinfo {author} {\bibfnamefont {J.~J.}\ \bibnamefont
  {Bennett}}, \bibinfo {author} {\bibfnamefont {G.}~\bibnamefont {Buldgen}},
  \bibinfo {author} {\bibfnamefont {M.}~\bibnamefont {Drewes}}, \ and\ \bibinfo
  {author} {\bibfnamefont {Y.~Y.~Y.}\ \bibnamefont {Wong}},\ }\href {\doibase
  10.1088/1475-7516/2020/03/003} {\bibfield  {journal} {\bibinfo  {journal}
  {JCAP}\ }\textbf {\bibinfo {volume} {03}},\ \bibinfo {pages} {003} (\bibinfo
  {year} {2020})},\ \bibinfo {note} {[Addendum: JCAP 03, A01 (2021)]},\ \Eprint
  {http://arxiv.org/abs/1911.04504} {arXiv:1911.04504 [hep-ph]} \BibitemShut
  {NoStop}%
\bibitem [{\citenamefont {Mangano}\ \emph {et~al.}(2005)\citenamefont
  {Mangano}, \citenamefont {Miele}, \citenamefont {Pastor}, \citenamefont
  {Pinto}, \citenamefont {Pisanti},\ and\ \citenamefont
  {Serpico}}]{Mangano:2005cc}%
  \BibitemOpen
  \bibfield  {author} {\bibinfo {author} {\bibfnamefont {G.}~\bibnamefont
  {Mangano}}, \bibinfo {author} {\bibfnamefont {G.}~\bibnamefont {Miele}},
  \bibinfo {author} {\bibfnamefont {S.}~\bibnamefont {Pastor}}, \bibinfo
  {author} {\bibfnamefont {T.}~\bibnamefont {Pinto}}, \bibinfo {author}
  {\bibfnamefont {O.}~\bibnamefont {Pisanti}}, \ and\ \bibinfo {author}
  {\bibfnamefont {P.~D.}\ \bibnamefont {Serpico}},\ }\href {\doibase
  10.1016/j.nuclphysb.2005.09.041} {\bibfield  {journal} {\bibinfo  {journal}
  {Nucl. Phys. B}\ }\textbf {\bibinfo {volume} {729}},\ \bibinfo {pages} {221}
  (\bibinfo {year} {2005})},\ \Eprint {http://arxiv.org/abs/hep-ph/0506164}
  {arXiv:hep-ph/0506164} \BibitemShut {NoStop}%
\bibitem [{\citenamefont {de~Salas}\ and\ \citenamefont
  {Pastor}(2016)}]{deSalas:2016ztq}%
  \BibitemOpen
  \bibfield  {author} {\bibinfo {author} {\bibfnamefont {P.~F.}\ \bibnamefont
  {de~Salas}}\ and\ \bibinfo {author} {\bibfnamefont {S.}~\bibnamefont
  {Pastor}},\ }\href {\doibase 10.1088/1475-7516/2016/07/051} {\bibfield
  {journal} {\bibinfo  {journal} {JCAP}\ }\textbf {\bibinfo {volume} {07}},\
  \bibinfo {pages} {051} (\bibinfo {year} {2016})},\ \Eprint
  {http://arxiv.org/abs/1606.06986} {arXiv:1606.06986 [hep-ph]} \BibitemShut
  {NoStop}%
\bibitem [{\citenamefont {Akita}\ and\ \citenamefont
  {Yamaguchi}(2020)}]{Akita:2020szl}%
  \BibitemOpen
  \bibfield  {author} {\bibinfo {author} {\bibfnamefont {K.}~\bibnamefont
  {Akita}}\ and\ \bibinfo {author} {\bibfnamefont {M.}~\bibnamefont
  {Yamaguchi}},\ }\href {\doibase 10.1088/1475-7516/2020/08/012} {\bibfield
  {journal} {\bibinfo  {journal} {JCAP}\ }\textbf {\bibinfo {volume} {08}},\
  \bibinfo {pages} {012} (\bibinfo {year} {2020})},\ \Eprint
  {http://arxiv.org/abs/2005.07047} {arXiv:2005.07047 [hep-ph]} \BibitemShut
  {NoStop}%
\bibitem [{\citenamefont {Froustey}\ \emph {et~al.}(2020)\citenamefont
  {Froustey}, \citenamefont {Pitrou},\ and\ \citenamefont
  {Volpe}}]{Froustey:2020mcq}%
  \BibitemOpen
  \bibfield  {author} {\bibinfo {author} {\bibfnamefont {J.}~\bibnamefont
  {Froustey}}, \bibinfo {author} {\bibfnamefont {C.}~\bibnamefont {Pitrou}}, \
  and\ \bibinfo {author} {\bibfnamefont {M.~C.}\ \bibnamefont {Volpe}},\ }\href
  {\doibase 10.1088/1475-7516/2020/12/015} {\bibfield  {journal} {\bibinfo
  {journal} {JCAP}\ }\textbf {\bibinfo {volume} {12}},\ \bibinfo {pages} {015}
  (\bibinfo {year} {2020})},\ \Eprint {http://arxiv.org/abs/2008.01074}
  {arXiv:2008.01074 [hep-ph]} \BibitemShut {NoStop}%
\bibitem [{\citenamefont {Bennett}\ \emph {et~al.}(2021)\citenamefont
  {Bennett}, \citenamefont {Buldgen}, \citenamefont {De~Salas}, \citenamefont
  {Drewes}, \citenamefont {Gariazzo}, \citenamefont {Pastor},\ and\
  \citenamefont {Wong}}]{Bennett:2020zkv}%
  \BibitemOpen
  \bibfield  {author} {\bibinfo {author} {\bibfnamefont {J.~J.}\ \bibnamefont
  {Bennett}}, \bibinfo {author} {\bibfnamefont {G.}~\bibnamefont {Buldgen}},
  \bibinfo {author} {\bibfnamefont {P.~F.}\ \bibnamefont {De~Salas}}, \bibinfo
  {author} {\bibfnamefont {M.}~\bibnamefont {Drewes}}, \bibinfo {author}
  {\bibfnamefont {S.}~\bibnamefont {Gariazzo}}, \bibinfo {author}
  {\bibfnamefont {S.}~\bibnamefont {Pastor}}, \ and\ \bibinfo {author}
  {\bibfnamefont {Y.~Y.~Y.}\ \bibnamefont {Wong}},\ }\href {\doibase
  10.1088/1475-7516/2021/04/073} {\bibfield  {journal} {\bibinfo  {journal}
  {JCAP}\ }\textbf {\bibinfo {volume} {04}},\ \bibinfo {pages} {073} (\bibinfo
  {year} {2021})},\ \Eprint {http://arxiv.org/abs/2012.02726} {arXiv:2012.02726
  [hep-ph]} \BibitemShut {NoStop}%
\bibitem [{\citenamefont {Cielo}\ \emph {et~al.}(2023)\citenamefont {Cielo},
  \citenamefont {Escudero}, \citenamefont {Mangano},\ and\ \citenamefont
  {Pisanti}}]{Cielo:2023bqp}%
  \BibitemOpen
  \bibfield  {author} {\bibinfo {author} {\bibfnamefont {M.}~\bibnamefont
  {Cielo}}, \bibinfo {author} {\bibfnamefont {M.}~\bibnamefont {Escudero}},
  \bibinfo {author} {\bibfnamefont {G.}~\bibnamefont {Mangano}}, \ and\
  \bibinfo {author} {\bibfnamefont {O.}~\bibnamefont {Pisanti}},\ }\href
  {\doibase 10.1103/PhysRevD.108.L121301} {\bibfield  {journal} {\bibinfo
  {journal} {Phys. Rev. D}\ }\textbf {\bibinfo {volume} {108}},\ \bibinfo
  {pages} {L121301} (\bibinfo {year} {2023})},\ \Eprint
  {http://arxiv.org/abs/2306.05460} {arXiv:2306.05460 [hep-ph]} \BibitemShut
  {NoStop}%
\bibitem [{\citenamefont {Maggiore}(2000)}]{Maggiore:1999vm}%
  \BibitemOpen
  \bibfield  {author} {\bibinfo {author} {\bibfnamefont {M.}~\bibnamefont
  {Maggiore}},\ }\href {\doibase 10.1016/S0370-1573(99)00102-7} {\bibfield
  {journal} {\bibinfo  {journal} {Phys. Rept.}\ }\textbf {\bibinfo {volume}
  {331}},\ \bibinfo {pages} {283} (\bibinfo {year} {2000})},\ \Eprint
  {http://arxiv.org/abs/gr-qc/9909001} {arXiv:gr-qc/9909001} \BibitemShut
  {NoStop}%
\bibitem [{\citenamefont {de~Salas}\ \emph {et~al.}(2015)\citenamefont
  {de~Salas}, \citenamefont {Lattanzi}, \citenamefont {Mangano}, \citenamefont
  {Miele}, \citenamefont {Pastor},\ and\ \citenamefont
  {Pisanti}}]{deSalas:2015glj}%
  \BibitemOpen
  \bibfield  {author} {\bibinfo {author} {\bibfnamefont {P.~F.}\ \bibnamefont
  {de~Salas}}, \bibinfo {author} {\bibfnamefont {M.}~\bibnamefont {Lattanzi}},
  \bibinfo {author} {\bibfnamefont {G.}~\bibnamefont {Mangano}}, \bibinfo
  {author} {\bibfnamefont {G.}~\bibnamefont {Miele}}, \bibinfo {author}
  {\bibfnamefont {S.}~\bibnamefont {Pastor}}, \ and\ \bibinfo {author}
  {\bibfnamefont {O.}~\bibnamefont {Pisanti}},\ }\href {\doibase
  10.1103/PhysRevD.92.123534} {\bibfield  {journal} {\bibinfo  {journal} {Phys.
  Rev. D}\ }\textbf {\bibinfo {volume} {92}},\ \bibinfo {pages} {123534}
  (\bibinfo {year} {2015})},\ \Eprint {http://arxiv.org/abs/1511.00672}
  {arXiv:1511.00672 [astro-ph.CO]} \BibitemShut {NoStop}%
\bibitem [{\citenamefont {Giar\`e}\ \emph {et~al.}(2023)\citenamefont
  {Giar\`e}, \citenamefont {Forconi}, \citenamefont {Di~Valentino},\ and\
  \citenamefont {Melchiorri}}]{Giare:2022wxq}%
  \BibitemOpen
  \bibfield  {author} {\bibinfo {author} {\bibfnamefont {W.}~\bibnamefont
  {Giar\`e}}, \bibinfo {author} {\bibfnamefont {M.}~\bibnamefont {Forconi}},
  \bibinfo {author} {\bibfnamefont {E.}~\bibnamefont {Di~Valentino}}, \ and\
  \bibinfo {author} {\bibfnamefont {A.}~\bibnamefont {Melchiorri}},\ }\href
  {\doibase 10.1093/mnras/stad258} {\bibfield  {journal} {\bibinfo  {journal}
  {Mon. Not. Roy. Astron. Soc.}\ }\textbf {\bibinfo {volume} {520}},\ \bibinfo
  {pages} {2} (\bibinfo {year} {2023})},\ \Eprint
  {http://arxiv.org/abs/2210.14159} {arXiv:2210.14159 [astro-ph.CO]}
  \BibitemShut {NoStop}%
\bibitem [{\citenamefont {Barenboim}\ \emph {et~al.}(2024)\citenamefont
  {Barenboim}, \citenamefont {Sanchis}, \citenamefont {Kinney},\ and\
  \citenamefont {Rios}}]{Barenboim:2024wek}%
  \BibitemOpen
  \bibfield  {author} {\bibinfo {author} {\bibfnamefont {G.}~\bibnamefont
  {Barenboim}}, \bibinfo {author} {\bibfnamefont {H.}~\bibnamefont {Sanchis}},
  \bibinfo {author} {\bibfnamefont {W.~H.}\ \bibnamefont {Kinney}}, \ and\
  \bibinfo {author} {\bibfnamefont {D.}~\bibnamefont {Rios}},\ }\href {\doibase
  10.1103/PhysRevD.110.123535} {\bibfield  {journal} {\bibinfo  {journal}
  {Phys. Rev. D}\ }\textbf {\bibinfo {volume} {110}},\ \bibinfo {pages}
  {123535} (\bibinfo {year} {2024})},\ \Eprint
  {http://arxiv.org/abs/2407.18102} {arXiv:2407.18102 [astro-ph.CO]}
  \BibitemShut {NoStop}%
\bibitem [{\citenamefont {Lesgourgues}\ and\ \citenamefont
  {Pastor}(2006)}]{Lesgourgues:2006nd}%
  \BibitemOpen
  \bibfield  {author} {\bibinfo {author} {\bibfnamefont {J.}~\bibnamefont
  {Lesgourgues}}\ and\ \bibinfo {author} {\bibfnamefont {S.}~\bibnamefont
  {Pastor}},\ }\href {\doibase 10.1016/j.physrep.2006.04.001} {\bibfield
  {journal} {\bibinfo  {journal} {Phys. Rept.}\ }\textbf {\bibinfo {volume}
  {429}},\ \bibinfo {pages} {307} (\bibinfo {year} {2006})},\ \Eprint
  {http://arxiv.org/abs/astro-ph/0603494} {arXiv:astro-ph/0603494} \BibitemShut
  {NoStop}%
\bibitem [{\citenamefont {Lattanzi}\ and\ \citenamefont
  {Gerbino}(2018)}]{Lattanzi:2017ubx}%
  \BibitemOpen
  \bibfield  {author} {\bibinfo {author} {\bibfnamefont {M.}~\bibnamefont
  {Lattanzi}}\ and\ \bibinfo {author} {\bibfnamefont {M.}~\bibnamefont
  {Gerbino}},\ }\href {\doibase 10.3389/fphy.2017.00070} {\bibfield  {journal}
  {\bibinfo  {journal} {Front. in Phys.}\ }\textbf {\bibinfo {volume} {5}},\
  \bibinfo {pages} {70} (\bibinfo {year} {2018})},\ \Eprint
  {http://arxiv.org/abs/1712.07109} {arXiv:1712.07109 [astro-ph.CO]}
  \BibitemShut {NoStop}%
\bibitem [{\citenamefont {Loureiro}\ \emph {et~al.}(2019)\citenamefont
  {Loureiro} \emph {et~al.}}]{Loureiro:2018pdz}%
  \BibitemOpen
  \bibfield  {author} {\bibinfo {author} {\bibfnamefont {A.}~\bibnamefont
  {Loureiro}} \emph {et~al.},\ }\href {\doibase 10.1103/PhysRevLett.123.081301}
  {\bibfield  {journal} {\bibinfo  {journal} {Phys. Rev. Lett.}\ }\textbf
  {\bibinfo {volume} {123}},\ \bibinfo {pages} {081301} (\bibinfo {year}
  {2019})},\ \Eprint {http://arxiv.org/abs/1811.02578} {arXiv:1811.02578
  [astro-ph.CO]} \BibitemShut {NoStop}%
\bibitem [{\citenamefont {Roy~Choudhury}\ and\ \citenamefont
  {Choubey}(2018)}]{RoyChoudhury:2018gay}%
  \BibitemOpen
  \bibfield  {author} {\bibinfo {author} {\bibfnamefont {S.}~\bibnamefont
  {Roy~Choudhury}}\ and\ \bibinfo {author} {\bibfnamefont {S.}~\bibnamefont
  {Choubey}},\ }\href {\doibase 10.1088/1475-7516/2018/09/017} {\bibfield
  {journal} {\bibinfo  {journal} {JCAP}\ }\textbf {\bibinfo {volume} {09}},\
  \bibinfo {pages} {017} (\bibinfo {year} {2018})},\ \Eprint
  {http://arxiv.org/abs/1806.10832} {arXiv:1806.10832 [astro-ph.CO]}
  \BibitemShut {NoStop}%
\bibitem [{\citenamefont {De~Salas}\ \emph {et~al.}(2018)\citenamefont
  {De~Salas}, \citenamefont {Gariazzo}, \citenamefont {Mena}, \citenamefont
  {Ternes},\ and\ \citenamefont {T\'ortola}}]{DeSalas:2018rby}%
  \BibitemOpen
  \bibfield  {author} {\bibinfo {author} {\bibfnamefont {P.~F.}\ \bibnamefont
  {De~Salas}}, \bibinfo {author} {\bibfnamefont {S.}~\bibnamefont {Gariazzo}},
  \bibinfo {author} {\bibfnamefont {O.}~\bibnamefont {Mena}}, \bibinfo {author}
  {\bibfnamefont {C.~A.}\ \bibnamefont {Ternes}}, \ and\ \bibinfo {author}
  {\bibfnamefont {M.}~\bibnamefont {T\'ortola}},\ }\href {\doibase
  10.3389/fspas.2018.00036} {\bibfield  {journal} {\bibinfo  {journal} {Front.
  Astron. Space Sci.}\ }\textbf {\bibinfo {volume} {5}},\ \bibinfo {pages} {36}
  (\bibinfo {year} {2018})},\ \Eprint {http://arxiv.org/abs/1806.11051}
  {arXiv:1806.11051 [hep-ph]} \BibitemShut {NoStop}%
\bibitem [{\citenamefont {Roy~Choudhury}\ and\ \citenamefont
  {Hannestad}(2020)}]{RoyChoudhury:2019hls}%
  \BibitemOpen
  \bibfield  {author} {\bibinfo {author} {\bibfnamefont {S.}~\bibnamefont
  {Roy~Choudhury}}\ and\ \bibinfo {author} {\bibfnamefont {S.}~\bibnamefont
  {Hannestad}},\ }\href {\doibase 10.1088/1475-7516/2020/07/037} {\bibfield
  {journal} {\bibinfo  {journal} {JCAP}\ }\textbf {\bibinfo {volume} {07}},\
  \bibinfo {pages} {037} (\bibinfo {year} {2020})},\ \Eprint
  {http://arxiv.org/abs/1907.12598} {arXiv:1907.12598 [astro-ph.CO]}
  \BibitemShut {NoStop}%
\bibitem [{\citenamefont {Di~Valentino}\ \emph
  {et~al.}(2021{\natexlab{b}})\citenamefont {Di~Valentino}, \citenamefont
  {Gariazzo},\ and\ \citenamefont {Mena}}]{DiValentino:2021hoh}%
  \BibitemOpen
  \bibfield  {author} {\bibinfo {author} {\bibfnamefont {E.}~\bibnamefont
  {Di~Valentino}}, \bibinfo {author} {\bibfnamefont {S.}~\bibnamefont
  {Gariazzo}}, \ and\ \bibinfo {author} {\bibfnamefont {O.}~\bibnamefont
  {Mena}},\ }\href {\doibase 10.1103/PhysRevD.104.083504} {\bibfield  {journal}
  {\bibinfo  {journal} {Phys. Rev. D}\ }\textbf {\bibinfo {volume} {104}},\
  \bibinfo {pages} {083504} (\bibinfo {year} {2021}{\natexlab{b}})},\ \Eprint
  {http://arxiv.org/abs/2106.15267} {arXiv:2106.15267 [astro-ph.CO]}
  \BibitemShut {NoStop}%
\bibitem [{\citenamefont {Capozzi}\ \emph {et~al.}(2021)\citenamefont
  {Capozzi}, \citenamefont {Di~Valentino}, \citenamefont {Lisi}, \citenamefont
  {Marrone}, \citenamefont {Melchiorri},\ and\ \citenamefont
  {Palazzo}}]{Capozzi:2021fjo}%
  \BibitemOpen
  \bibfield  {author} {\bibinfo {author} {\bibfnamefont {F.}~\bibnamefont
  {Capozzi}}, \bibinfo {author} {\bibfnamefont {E.}~\bibnamefont
  {Di~Valentino}}, \bibinfo {author} {\bibfnamefont {E.}~\bibnamefont {Lisi}},
  \bibinfo {author} {\bibfnamefont {A.}~\bibnamefont {Marrone}}, \bibinfo
  {author} {\bibfnamefont {A.}~\bibnamefont {Melchiorri}}, \ and\ \bibinfo
  {author} {\bibfnamefont {A.}~\bibnamefont {Palazzo}},\ }\href {\doibase
  10.1103/PhysRevD.104.083031} {\bibfield  {journal} {\bibinfo  {journal}
  {Phys. Rev. D}\ }\textbf {\bibinfo {volume} {104}},\ \bibinfo {pages}
  {083031} (\bibinfo {year} {2021})},\ \Eprint
  {http://arxiv.org/abs/2107.00532} {arXiv:2107.00532 [hep-ph]} \BibitemShut
  {NoStop}%
\bibitem [{\citenamefont {Di~Valentino}\ and\ \citenamefont
  {Melchiorri}(2022)}]{DiValentino:2021imh}%
  \BibitemOpen
  \bibfield  {author} {\bibinfo {author} {\bibfnamefont {E.}~\bibnamefont
  {Di~Valentino}}\ and\ \bibinfo {author} {\bibfnamefont {A.}~\bibnamefont
  {Melchiorri}},\ }\href {\doibase 10.3847/2041-8213/ac6ef5} {\bibfield
  {journal} {\bibinfo  {journal} {Astrophys. J. Lett.}\ }\textbf {\bibinfo
  {volume} {931}},\ \bibinfo {pages} {L18} (\bibinfo {year} {2022})},\ \Eprint
  {http://arxiv.org/abs/2112.02993} {arXiv:2112.02993 [astro-ph.CO]}
  \BibitemShut {NoStop}%
\bibitem [{\citenamefont {Tanseri}\ \emph {et~al.}(2022)\citenamefont
  {Tanseri}, \citenamefont {Hagstotz}, \citenamefont {Vagnozzi}, \citenamefont
  {Giusarma},\ and\ \citenamefont {Freese}}]{Tanseri:2022zfe}%
  \BibitemOpen
  \bibfield  {author} {\bibinfo {author} {\bibfnamefont {I.}~\bibnamefont
  {Tanseri}}, \bibinfo {author} {\bibfnamefont {S.}~\bibnamefont {Hagstotz}},
  \bibinfo {author} {\bibfnamefont {S.}~\bibnamefont {Vagnozzi}}, \bibinfo
  {author} {\bibfnamefont {E.}~\bibnamefont {Giusarma}}, \ and\ \bibinfo
  {author} {\bibfnamefont {K.}~\bibnamefont {Freese}},\ }\href {\doibase
  10.1016/j.jheap.2022.07.002} {\bibfield  {journal} {\bibinfo  {journal}
  {JHEAp}\ }\textbf {\bibinfo {volume} {36}},\ \bibinfo {pages} {1} (\bibinfo
  {year} {2022})},\ \Eprint {http://arxiv.org/abs/2207.01913} {arXiv:2207.01913
  [astro-ph.CO]} \BibitemShut {NoStop}%
\bibitem [{\citenamefont {di~Valentino}\ \emph {et~al.}(2022)\citenamefont
  {di~Valentino}, \citenamefont {Gariazzo},\ and\ \citenamefont
  {Mena}}]{diValentino:2022njd}%
  \BibitemOpen
  \bibfield  {author} {\bibinfo {author} {\bibfnamefont {E.}~\bibnamefont
  {di~Valentino}}, \bibinfo {author} {\bibfnamefont {S.}~\bibnamefont
  {Gariazzo}}, \ and\ \bibinfo {author} {\bibfnamefont {O.}~\bibnamefont
  {Mena}},\ }\href {\doibase 10.1103/PhysRevD.106.043540} {\bibfield  {journal}
  {\bibinfo  {journal} {Phys. Rev. D}\ }\textbf {\bibinfo {volume} {106}},\
  \bibinfo {pages} {043540} (\bibinfo {year} {2022})},\ \Eprint
  {http://arxiv.org/abs/2207.05167} {arXiv:2207.05167 [astro-ph.CO]}
  \BibitemShut {NoStop}%
\bibitem [{\citenamefont {Di~Valentino}\ \emph {et~al.}(2023)\citenamefont
  {Di~Valentino}, \citenamefont {Gariazzo}, \citenamefont {Giar\`e},\ and\
  \citenamefont {Mena}}]{DiValentino:2023fei}%
  \BibitemOpen
  \bibfield  {author} {\bibinfo {author} {\bibfnamefont {E.}~\bibnamefont
  {Di~Valentino}}, \bibinfo {author} {\bibfnamefont {S.}~\bibnamefont
  {Gariazzo}}, \bibinfo {author} {\bibfnamefont {W.}~\bibnamefont {Giar\`e}}, \
  and\ \bibinfo {author} {\bibfnamefont {O.}~\bibnamefont {Mena}},\ }\href
  {\doibase 10.1103/PhysRevD.108.083509} {\bibfield  {journal} {\bibinfo
  {journal} {Phys. Rev. D}\ }\textbf {\bibinfo {volume} {108}},\ \bibinfo
  {pages} {083509} (\bibinfo {year} {2023})},\ \Eprint
  {http://arxiv.org/abs/2305.12989} {arXiv:2305.12989 [astro-ph.CO]}
  \BibitemShut {NoStop}%
\bibitem [{\citenamefont {Gariazzo}\ \emph {et~al.}(2023)\citenamefont
  {Gariazzo}, \citenamefont {Mena},\ and\ \citenamefont
  {Schwetz}}]{Gariazzo:2023joe}%
  \BibitemOpen
  \bibfield  {author} {\bibinfo {author} {\bibfnamefont {S.}~\bibnamefont
  {Gariazzo}}, \bibinfo {author} {\bibfnamefont {O.}~\bibnamefont {Mena}}, \
  and\ \bibinfo {author} {\bibfnamefont {T.}~\bibnamefont {Schwetz}},\ }\href
  {\doibase 10.1016/j.dark.2023.101226} {\bibfield  {journal} {\bibinfo
  {journal} {Phys. Dark Univ.}\ }\textbf {\bibinfo {volume} {40}},\ \bibinfo
  {pages} {101226} (\bibinfo {year} {2023})},\ \Eprint
  {http://arxiv.org/abs/2302.14159} {arXiv:2302.14159 [hep-ph]} \BibitemShut
  {NoStop}%
\bibitem [{\citenamefont {Craig}\ \emph {et~al.}(2024)\citenamefont {Craig},
  \citenamefont {Green}, \citenamefont {Meyers},\ and\ \citenamefont
  {Rajendran}}]{Craig:2024tky}%
  \BibitemOpen
  \bibfield  {author} {\bibinfo {author} {\bibfnamefont {N.}~\bibnamefont
  {Craig}}, \bibinfo {author} {\bibfnamefont {D.}~\bibnamefont {Green}},
  \bibinfo {author} {\bibfnamefont {J.}~\bibnamefont {Meyers}}, \ and\ \bibinfo
  {author} {\bibfnamefont {S.}~\bibnamefont {Rajendran}},\ }\href {\doibase
  10.1007/JHEP09(2024)097} {\bibfield  {journal} {\bibinfo  {journal} {JHEP}\
  }\textbf {\bibinfo {volume} {09}},\ \bibinfo {pages} {097} (\bibinfo {year}
  {2024})},\ \Eprint {http://arxiv.org/abs/2405.00836} {arXiv:2405.00836
  [astro-ph.CO]} \BibitemShut {NoStop}%
\bibitem [{\citenamefont {Bert\'olez-Mart\'\i{}nez}\ \emph
  {et~al.}(2024)\citenamefont {Bert\'olez-Mart\'\i{}nez}, \citenamefont
  {Esteban}, \citenamefont {Hajjar}, \citenamefont {Mena},\ and\ \citenamefont
  {Salvad\'o}}]{Bertolez-Martinez:2024wez}%
  \BibitemOpen
  \bibfield  {author} {\bibinfo {author} {\bibfnamefont {T.}~\bibnamefont
  {Bert\'olez-Mart\'\i{}nez}}, \bibinfo {author} {\bibfnamefont
  {I.}~\bibnamefont {Esteban}}, \bibinfo {author} {\bibfnamefont
  {R.}~\bibnamefont {Hajjar}}, \bibinfo {author} {\bibfnamefont
  {O.}~\bibnamefont {Mena}}, \ and\ \bibinfo {author} {\bibfnamefont
  {J.}~\bibnamefont {Salvad\'o}},\ }\href@noop {} {\  (\bibinfo {year}
  {2024})},\ \Eprint {http://arxiv.org/abs/2411.14524} {arXiv:2411.14524
  [astro-ph.CO]} \BibitemShut {NoStop}%
\bibitem [{\citenamefont {Shao}\ \emph {et~al.}(2025)\citenamefont {Shao},
  \citenamefont {Givans}, \citenamefont {Dunkley}, \citenamefont
  {Madhavacheril}, \citenamefont {Qu}, \citenamefont {Farren},\ and\
  \citenamefont {Sherwin}}]{Shao:2024mag}%
  \BibitemOpen
  \bibfield  {author} {\bibinfo {author} {\bibfnamefont {H.}~\bibnamefont
  {Shao}}, \bibinfo {author} {\bibfnamefont {J.~J.}\ \bibnamefont {Givans}},
  \bibinfo {author} {\bibfnamefont {J.}~\bibnamefont {Dunkley}}, \bibinfo
  {author} {\bibfnamefont {M.}~\bibnamefont {Madhavacheril}}, \bibinfo {author}
  {\bibfnamefont {F.~J.}\ \bibnamefont {Qu}}, \bibinfo {author} {\bibfnamefont
  {G.}~\bibnamefont {Farren}}, \ and\ \bibinfo {author} {\bibfnamefont
  {B.}~\bibnamefont {Sherwin}},\ }\href {\doibase 10.1103/PhysRevD.111.083535}
  {\bibfield  {journal} {\bibinfo  {journal} {Phys. Rev. D}\ }\textbf {\bibinfo
  {volume} {111}},\ \bibinfo {pages} {083535} (\bibinfo {year} {2025})},\
  \Eprint {http://arxiv.org/abs/2409.02295} {arXiv:2409.02295 [astro-ph.CO]}
  \BibitemShut {NoStop}%
\bibitem [{\citenamefont {Esteban}\ \emph {et~al.}(2024)\citenamefont
  {Esteban}, \citenamefont {Gonz\'alez-Garc\'ia}, \citenamefont {Maltoni},
  \citenamefont {Mart\'inez-Soler}, \citenamefont {Pinheiro},\ and\
  \citenamefont {Schwetz}}]{Esteban:2024eli}%
  \BibitemOpen
  \bibfield  {author} {\bibinfo {author} {\bibfnamefont {I.}~\bibnamefont
  {Esteban}}, \bibinfo {author} {\bibfnamefont {M.~C.}\ \bibnamefont
  {Gonz\'alez-Garc\'ia}}, \bibinfo {author} {\bibfnamefont {M.}~\bibnamefont
  {Maltoni}}, \bibinfo {author} {\bibfnamefont {I.}~\bibnamefont
  {Mart\'inez-Soler}}, \bibinfo {author} {\bibfnamefont {J.~a.~P.}\
  \bibnamefont {Pinheiro}}, \ and\ \bibinfo {author} {\bibfnamefont
  {T.}~\bibnamefont {Schwetz}},\ }\href {\doibase 10.1007/JHEP12(2024)216}
  {\bibfield  {journal} {\bibinfo  {journal} {JHEP}\ }\textbf {\bibinfo
  {volume} {12}},\ \bibinfo {pages} {216} (\bibinfo {year} {2024})},\ \Eprint
  {http://arxiv.org/abs/2410.05380} {arXiv:2410.05380 [hep-ph]} \BibitemShut
  {NoStop}%
\bibitem [{\citenamefont {Doran}\ and\ \citenamefont
  {Robbers}(2006)}]{Doran:2006kp}%
  \BibitemOpen
  \bibfield  {author} {\bibinfo {author} {\bibfnamefont {M.}~\bibnamefont
  {Doran}}\ and\ \bibinfo {author} {\bibfnamefont {G.}~\bibnamefont
  {Robbers}},\ }\href {\doibase 10.1088/1475-7516/2006/06/026} {\bibfield
  {journal} {\bibinfo  {journal} {JCAP}\ }\textbf {\bibinfo {volume} {06}},\
  \bibinfo {pages} {026} (\bibinfo {year} {2006})},\ \Eprint
  {http://arxiv.org/abs/astro-ph/0601544} {arXiv:astro-ph/0601544} \BibitemShut
  {NoStop}%
\bibitem [{\citenamefont {Hollenstein}\ \emph {et~al.}(2009)\citenamefont
  {Hollenstein}, \citenamefont {Sapone}, \citenamefont {Crittenden},\ and\
  \citenamefont {Schaefer}}]{Hollenstein:2009ph}%
  \BibitemOpen
  \bibfield  {author} {\bibinfo {author} {\bibfnamefont {L.}~\bibnamefont
  {Hollenstein}}, \bibinfo {author} {\bibfnamefont {D.}~\bibnamefont {Sapone}},
  \bibinfo {author} {\bibfnamefont {R.}~\bibnamefont {Crittenden}}, \ and\
  \bibinfo {author} {\bibfnamefont {B.~M.}\ \bibnamefont {Schaefer}},\ }\href
  {\doibase 10.1088/1475-7516/2009/04/012} {\bibfield  {journal} {\bibinfo
  {journal} {JCAP}\ }\textbf {\bibinfo {volume} {04}},\ \bibinfo {pages} {012}
  (\bibinfo {year} {2009})},\ \Eprint {http://arxiv.org/abs/0902.1494}
  {arXiv:0902.1494 [astro-ph.CO]} \BibitemShut {NoStop}%
\bibitem [{\citenamefont {Calabrese}\ \emph
  {et~al.}(2011{\natexlab{a}})\citenamefont {Calabrese}, \citenamefont
  {de~Putter}, \citenamefont {Huterer}, \citenamefont {Linder},\ and\
  \citenamefont {Melchiorri}}]{Calabrese:2010uf}%
  \BibitemOpen
  \bibfield  {author} {\bibinfo {author} {\bibfnamefont {E.}~\bibnamefont
  {Calabrese}}, \bibinfo {author} {\bibfnamefont {R.}~\bibnamefont
  {de~Putter}}, \bibinfo {author} {\bibfnamefont {D.}~\bibnamefont {Huterer}},
  \bibinfo {author} {\bibfnamefont {E.~V.}\ \bibnamefont {Linder}}, \ and\
  \bibinfo {author} {\bibfnamefont {A.}~\bibnamefont {Melchiorri}},\ }\href
  {\doibase 10.1103/PhysRevD.83.023011} {\bibfield  {journal} {\bibinfo
  {journal} {Phys. Rev. D}\ }\textbf {\bibinfo {volume} {83}},\ \bibinfo
  {pages} {023011} (\bibinfo {year} {2011}{\natexlab{a}})},\ \Eprint
  {http://arxiv.org/abs/1010.5612} {arXiv:1010.5612 [astro-ph.CO]} \BibitemShut
  {NoStop}%
\bibitem [{\citenamefont {Calabrese}\ \emph
  {et~al.}(2011{\natexlab{b}})\citenamefont {Calabrese}, \citenamefont
  {Huterer}, \citenamefont {Linder}, \citenamefont {Melchiorri},\ and\
  \citenamefont {Pagano}}]{Calabrese:2011hg}%
  \BibitemOpen
  \bibfield  {author} {\bibinfo {author} {\bibfnamefont {E.}~\bibnamefont
  {Calabrese}}, \bibinfo {author} {\bibfnamefont {D.}~\bibnamefont {Huterer}},
  \bibinfo {author} {\bibfnamefont {E.~V.}\ \bibnamefont {Linder}}, \bibinfo
  {author} {\bibfnamefont {A.}~\bibnamefont {Melchiorri}}, \ and\ \bibinfo
  {author} {\bibfnamefont {L.}~\bibnamefont {Pagano}},\ }\href {\doibase
  10.1103/PhysRevD.83.123504} {\bibfield  {journal} {\bibinfo  {journal} {Phys.
  Rev. D}\ }\textbf {\bibinfo {volume} {83}},\ \bibinfo {pages} {123504}
  (\bibinfo {year} {2011}{\natexlab{b}})},\ \Eprint
  {http://arxiv.org/abs/1103.4132} {arXiv:1103.4132 [astro-ph.CO]} \BibitemShut
  {NoStop}%
\bibitem [{\citenamefont {Pettorino}\ \emph {et~al.}(2013)\citenamefont
  {Pettorino}, \citenamefont {Amendola},\ and\ \citenamefont
  {Wetterich}}]{Pettorino:2013ia}%
  \BibitemOpen
  \bibfield  {author} {\bibinfo {author} {\bibfnamefont {V.}~\bibnamefont
  {Pettorino}}, \bibinfo {author} {\bibfnamefont {L.}~\bibnamefont {Amendola}},
  \ and\ \bibinfo {author} {\bibfnamefont {C.}~\bibnamefont {Wetterich}},\
  }\href {\doibase 10.1103/PhysRevD.87.083009} {\bibfield  {journal} {\bibinfo
  {journal} {Phys. Rev. D}\ }\textbf {\bibinfo {volume} {87}},\ \bibinfo
  {pages} {083009} (\bibinfo {year} {2013})},\ \Eprint
  {http://arxiv.org/abs/1301.5279} {arXiv:1301.5279 [astro-ph.CO]} \BibitemShut
  {NoStop}%
\bibitem [{\citenamefont {Archidiacono}\ \emph {et~al.}(2014)\citenamefont
  {Archidiacono}, \citenamefont {Lopez-Honorez},\ and\ \citenamefont
  {Mena}}]{Archidiacono:2014msa}%
  \BibitemOpen
  \bibfield  {author} {\bibinfo {author} {\bibfnamefont {M.}~\bibnamefont
  {Archidiacono}}, \bibinfo {author} {\bibfnamefont {L.}~\bibnamefont
  {Lopez-Honorez}}, \ and\ \bibinfo {author} {\bibfnamefont {O.}~\bibnamefont
  {Mena}},\ }\href {\doibase 10.1103/PhysRevD.90.123016} {\bibfield  {journal}
  {\bibinfo  {journal} {Phys. Rev. D}\ }\textbf {\bibinfo {volume} {90}},\
  \bibinfo {pages} {123016} (\bibinfo {year} {2014})},\ \Eprint
  {http://arxiv.org/abs/1409.1802} {arXiv:1409.1802 [astro-ph.CO]} \BibitemShut
  {NoStop}%
\bibitem [{\citenamefont {Poulin}\ \emph {et~al.}(2019)\citenamefont {Poulin},
  \citenamefont {Smith}, \citenamefont {Karwal},\ and\ \citenamefont
  {Kamionkowski}}]{Poulin:2018cxd}%
  \BibitemOpen
  \bibfield  {author} {\bibinfo {author} {\bibfnamefont {V.}~\bibnamefont
  {Poulin}}, \bibinfo {author} {\bibfnamefont {T.~L.}\ \bibnamefont {Smith}},
  \bibinfo {author} {\bibfnamefont {T.}~\bibnamefont {Karwal}}, \ and\ \bibinfo
  {author} {\bibfnamefont {M.}~\bibnamefont {Kamionkowski}},\ }\href {\doibase
  10.1103/PhysRevLett.122.221301} {\bibfield  {journal} {\bibinfo  {journal}
  {Phys. Rev. Lett.}\ }\textbf {\bibinfo {volume} {122}},\ \bibinfo {pages}
  {221301} (\bibinfo {year} {2019})},\ \Eprint
  {http://arxiv.org/abs/1811.04083} {arXiv:1811.04083 [astro-ph.CO]}
  \BibitemShut {NoStop}%
\bibitem [{\citenamefont {Poulin}\ \emph
  {et~al.}(2018{\natexlab{a}})\citenamefont {Poulin}, \citenamefont {Smith},
  \citenamefont {Grin}, \citenamefont {Karwal},\ and\ \citenamefont
  {Kamionkowski}}]{Poulin:2018dzj}%
  \BibitemOpen
  \bibfield  {author} {\bibinfo {author} {\bibfnamefont {V.}~\bibnamefont
  {Poulin}}, \bibinfo {author} {\bibfnamefont {T.~L.}\ \bibnamefont {Smith}},
  \bibinfo {author} {\bibfnamefont {D.}~\bibnamefont {Grin}}, \bibinfo {author}
  {\bibfnamefont {T.}~\bibnamefont {Karwal}}, \ and\ \bibinfo {author}
  {\bibfnamefont {M.}~\bibnamefont {Kamionkowski}},\ }\href {\doibase
  10.1103/PhysRevD.98.083525} {\bibfield  {journal} {\bibinfo  {journal} {Phys.
  Rev. D}\ }\textbf {\bibinfo {volume} {98}},\ \bibinfo {pages} {083525}
  (\bibinfo {year} {2018}{\natexlab{a}})},\ \Eprint
  {http://arxiv.org/abs/1806.10608} {arXiv:1806.10608 [astro-ph.CO]}
  \BibitemShut {NoStop}%
\bibitem [{\citenamefont {Poulin}\ \emph
  {et~al.}(2018{\natexlab{b}})\citenamefont {Poulin}, \citenamefont {Boddy},
  \citenamefont {Bird},\ and\ \citenamefont {Kamionkowski}}]{Poulin:2018zxs}%
  \BibitemOpen
  \bibfield  {author} {\bibinfo {author} {\bibfnamefont {V.}~\bibnamefont
  {Poulin}}, \bibinfo {author} {\bibfnamefont {K.~K.}\ \bibnamefont {Boddy}},
  \bibinfo {author} {\bibfnamefont {S.}~\bibnamefont {Bird}}, \ and\ \bibinfo
  {author} {\bibfnamefont {M.}~\bibnamefont {Kamionkowski}},\ }\href {\doibase
  10.1103/PhysRevD.97.123504} {\bibfield  {journal} {\bibinfo  {journal} {Phys.
  Rev. D}\ }\textbf {\bibinfo {volume} {97}},\ \bibinfo {pages} {123504}
  (\bibinfo {year} {2018}{\natexlab{b}})},\ \Eprint
  {http://arxiv.org/abs/1803.02474} {arXiv:1803.02474 [astro-ph.CO]}
  \BibitemShut {NoStop}%
\bibitem [{\citenamefont {Niedermann}\ and\ \citenamefont
  {Sloth}(2021)}]{Niedermann:2019olb}%
  \BibitemOpen
  \bibfield  {author} {\bibinfo {author} {\bibfnamefont {F.}~\bibnamefont
  {Niedermann}}\ and\ \bibinfo {author} {\bibfnamefont {M.~S.}\ \bibnamefont
  {Sloth}},\ }\href {\doibase 10.1103/PhysRevD.103.L041303} {\bibfield
  {journal} {\bibinfo  {journal} {Phys. Rev. D}\ }\textbf {\bibinfo {volume}
  {103}},\ \bibinfo {pages} {L041303} (\bibinfo {year} {2021})},\ \Eprint
  {http://arxiv.org/abs/1910.10739} {arXiv:1910.10739 [astro-ph.CO]}
  \BibitemShut {NoStop}%
\bibitem [{\citenamefont {Smith}\ \emph {et~al.}(2021)\citenamefont {Smith},
  \citenamefont {Poulin}, \citenamefont {Bernal}, \citenamefont {Boddy},
  \citenamefont {Kamionkowski},\ and\ \citenamefont {Murgia}}]{Smith:2020rxx}%
  \BibitemOpen
  \bibfield  {author} {\bibinfo {author} {\bibfnamefont {T.~L.}\ \bibnamefont
  {Smith}}, \bibinfo {author} {\bibfnamefont {V.}~\bibnamefont {Poulin}},
  \bibinfo {author} {\bibfnamefont {J.~L.}\ \bibnamefont {Bernal}}, \bibinfo
  {author} {\bibfnamefont {K.~K.}\ \bibnamefont {Boddy}}, \bibinfo {author}
  {\bibfnamefont {M.}~\bibnamefont {Kamionkowski}}, \ and\ \bibinfo {author}
  {\bibfnamefont {R.}~\bibnamefont {Murgia}},\ }\href {\doibase
  10.1103/PhysRevD.103.123542} {\bibfield  {journal} {\bibinfo  {journal}
  {Phys. Rev. D}\ }\textbf {\bibinfo {volume} {103}},\ \bibinfo {pages}
  {123542} (\bibinfo {year} {2021})},\ \Eprint
  {http://arxiv.org/abs/2009.10740} {arXiv:2009.10740 [astro-ph.CO]}
  \BibitemShut {NoStop}%
\bibitem [{\citenamefont {Niedermann}\ and\ \citenamefont
  {Sloth}(2020)}]{Niedermann:2020dwg}%
  \BibitemOpen
  \bibfield  {author} {\bibinfo {author} {\bibfnamefont {F.}~\bibnamefont
  {Niedermann}}\ and\ \bibinfo {author} {\bibfnamefont {M.~S.}\ \bibnamefont
  {Sloth}},\ }\href {\doibase 10.1103/PhysRevD.102.063527} {\bibfield
  {journal} {\bibinfo  {journal} {Phys. Rev. D}\ }\textbf {\bibinfo {volume}
  {102}},\ \bibinfo {pages} {063527} (\bibinfo {year} {2020})},\ \Eprint
  {http://arxiv.org/abs/2006.06686} {arXiv:2006.06686 [astro-ph.CO]}
  \BibitemShut {NoStop}%
\bibitem [{\citenamefont {Murgia}\ \emph {et~al.}(2021)\citenamefont {Murgia},
  \citenamefont {Abell\'an},\ and\ \citenamefont {Poulin}}]{Murgia:2020ryi}%
  \BibitemOpen
  \bibfield  {author} {\bibinfo {author} {\bibfnamefont {R.}~\bibnamefont
  {Murgia}}, \bibinfo {author} {\bibfnamefont {G.~F.}\ \bibnamefont
  {Abell\'an}}, \ and\ \bibinfo {author} {\bibfnamefont {V.}~\bibnamefont
  {Poulin}},\ }\href {\doibase 10.1103/PhysRevD.103.063502} {\bibfield
  {journal} {\bibinfo  {journal} {Phys. Rev. D}\ }\textbf {\bibinfo {volume}
  {103}},\ \bibinfo {pages} {063502} (\bibinfo {year} {2021})},\ \Eprint
  {http://arxiv.org/abs/2009.10733} {arXiv:2009.10733 [astro-ph.CO]}
  \BibitemShut {NoStop}%
\bibitem [{\citenamefont {Klypin}\ \emph {et~al.}(2021)\citenamefont {Klypin},
  \citenamefont {Poulin}, \citenamefont {Prada}, \citenamefont {Primack},
  \citenamefont {Kamionkowski}, \citenamefont {Avila-Reese}, \citenamefont
  {Rodriguez-Puebla}, \citenamefont {Behroozi}, \citenamefont {Hellinger},\
  and\ \citenamefont {Smith}}]{Klypin:2020tud}%
  \BibitemOpen
  \bibfield  {author} {\bibinfo {author} {\bibfnamefont {A.}~\bibnamefont
  {Klypin}}, \bibinfo {author} {\bibfnamefont {V.}~\bibnamefont {Poulin}},
  \bibinfo {author} {\bibfnamefont {F.}~\bibnamefont {Prada}}, \bibinfo
  {author} {\bibfnamefont {J.}~\bibnamefont {Primack}}, \bibinfo {author}
  {\bibfnamefont {M.}~\bibnamefont {Kamionkowski}}, \bibinfo {author}
  {\bibfnamefont {V.}~\bibnamefont {Avila-Reese}}, \bibinfo {author}
  {\bibfnamefont {A.}~\bibnamefont {Rodriguez-Puebla}}, \bibinfo {author}
  {\bibfnamefont {P.}~\bibnamefont {Behroozi}}, \bibinfo {author}
  {\bibfnamefont {D.}~\bibnamefont {Hellinger}}, \ and\ \bibinfo {author}
  {\bibfnamefont {T.~L.}\ \bibnamefont {Smith}},\ }\href {\doibase
  10.1093/mnras/stab769} {\bibfield  {journal} {\bibinfo  {journal} {Mon. Not.
  Roy. Astron. Soc.}\ }\textbf {\bibinfo {volume} {504}},\ \bibinfo {pages}
  {769} (\bibinfo {year} {2021})},\ \Eprint {http://arxiv.org/abs/2006.14910}
  {arXiv:2006.14910 [astro-ph.CO]} \BibitemShut {NoStop}%
\bibitem [{\citenamefont {Hill}\ \emph {et~al.}(2020)\citenamefont {Hill},
  \citenamefont {McDonough}, \citenamefont {Toomey},\ and\ \citenamefont
  {Alexander}}]{Hill:2020osr}%
  \BibitemOpen
  \bibfield  {author} {\bibinfo {author} {\bibfnamefont {J.~C.}\ \bibnamefont
  {Hill}}, \bibinfo {author} {\bibfnamefont {E.}~\bibnamefont {McDonough}},
  \bibinfo {author} {\bibfnamefont {M.~W.}\ \bibnamefont {Toomey}}, \ and\
  \bibinfo {author} {\bibfnamefont {S.}~\bibnamefont {Alexander}},\ }\href
  {\doibase 10.1103/PhysRevD.102.043507} {\bibfield  {journal} {\bibinfo
  {journal} {Phys. Rev. D}\ }\textbf {\bibinfo {volume} {102}},\ \bibinfo
  {pages} {043507} (\bibinfo {year} {2020})},\ \Eprint
  {http://arxiv.org/abs/2003.07355} {arXiv:2003.07355 [astro-ph.CO]}
  \BibitemShut {NoStop}%
\bibitem [{\citenamefont {Herold}\ \emph {et~al.}(2022)\citenamefont {Herold},
  \citenamefont {Ferreira},\ and\ \citenamefont {Komatsu}}]{Herold:2021ksg}%
  \BibitemOpen
  \bibfield  {author} {\bibinfo {author} {\bibfnamefont {L.}~\bibnamefont
  {Herold}}, \bibinfo {author} {\bibfnamefont {E.~G.~M.}\ \bibnamefont
  {Ferreira}}, \ and\ \bibinfo {author} {\bibfnamefont {E.}~\bibnamefont
  {Komatsu}},\ }\href {\doibase 10.3847/2041-8213/ac63a3} {\bibfield  {journal}
  {\bibinfo  {journal} {Astrophys. J. Lett.}\ }\textbf {\bibinfo {volume}
  {929}},\ \bibinfo {pages} {L16} (\bibinfo {year} {2022})},\ \Eprint
  {http://arxiv.org/abs/2112.12140} {arXiv:2112.12140 [astro-ph.CO]}
  \BibitemShut {NoStop}%
\bibitem [{\citenamefont {Reeves}\ \emph {et~al.}(2023)\citenamefont {Reeves},
  \citenamefont {Herold}, \citenamefont {Vagnozzi}, \citenamefont {Sherwin},\
  and\ \citenamefont {Ferreira}}]{Reeves:2022aoi}%
  \BibitemOpen
  \bibfield  {author} {\bibinfo {author} {\bibfnamefont {A.}~\bibnamefont
  {Reeves}}, \bibinfo {author} {\bibfnamefont {L.}~\bibnamefont {Herold}},
  \bibinfo {author} {\bibfnamefont {S.}~\bibnamefont {Vagnozzi}}, \bibinfo
  {author} {\bibfnamefont {B.~D.}\ \bibnamefont {Sherwin}}, \ and\ \bibinfo
  {author} {\bibfnamefont {E.~G.~M.}\ \bibnamefont {Ferreira}},\ }\href
  {\doibase 10.1093/mnras/stad317} {\bibfield  {journal} {\bibinfo  {journal}
  {Mon. Not. Roy. Astron. Soc.}\ }\textbf {\bibinfo {volume} {520}},\ \bibinfo
  {pages} {3688} (\bibinfo {year} {2023})},\ \Eprint
  {http://arxiv.org/abs/2207.01501} {arXiv:2207.01501 [astro-ph.CO]}
  \BibitemShut {NoStop}%
\bibitem [{\citenamefont {Jiang}\ and\ \citenamefont
  {Piao}(2022)}]{Jiang:2022uyg}%
  \BibitemOpen
  \bibfield  {author} {\bibinfo {author} {\bibfnamefont {J.-Q.}\ \bibnamefont
  {Jiang}}\ and\ \bibinfo {author} {\bibfnamefont {Y.-S.}\ \bibnamefont
  {Piao}},\ }\href {\doibase 10.1103/PhysRevD.105.103514} {\bibfield  {journal}
  {\bibinfo  {journal} {Phys. Rev. D}\ }\textbf {\bibinfo {volume} {105}},\
  \bibinfo {pages} {103514} (\bibinfo {year} {2022})},\ \Eprint
  {http://arxiv.org/abs/2202.13379} {arXiv:2202.13379 [astro-ph.CO]}
  \BibitemShut {NoStop}%
\bibitem [{\citenamefont {Simon}\ \emph {et~al.}(2023)\citenamefont {Simon},
  \citenamefont {Zhang}, \citenamefont {Poulin},\ and\ \citenamefont
  {Smith}}]{Simon:2022adh}%
  \BibitemOpen
  \bibfield  {author} {\bibinfo {author} {\bibfnamefont {T.}~\bibnamefont
  {Simon}}, \bibinfo {author} {\bibfnamefont {P.}~\bibnamefont {Zhang}},
  \bibinfo {author} {\bibfnamefont {V.}~\bibnamefont {Poulin}}, \ and\ \bibinfo
  {author} {\bibfnamefont {T.~L.}\ \bibnamefont {Smith}},\ }\href {\doibase
  10.1103/PhysRevD.107.063505} {\bibfield  {journal} {\bibinfo  {journal}
  {Phys. Rev. D}\ }\textbf {\bibinfo {volume} {107}},\ \bibinfo {pages}
  {063505} (\bibinfo {year} {2023})},\ \Eprint
  {http://arxiv.org/abs/2208.05930} {arXiv:2208.05930 [astro-ph.CO]}
  \BibitemShut {NoStop}%
\bibitem [{\citenamefont {Smith}\ \emph {et~al.}(2022)\citenamefont {Smith},
  \citenamefont {Lucca}, \citenamefont {Poulin}, \citenamefont {Abellan},
  \citenamefont {Balkenhol}, \citenamefont {Benabed}, \citenamefont {Galli},\
  and\ \citenamefont {Murgia}}]{Smith:2022hwi}%
  \BibitemOpen
  \bibfield  {author} {\bibinfo {author} {\bibfnamefont {T.~L.}\ \bibnamefont
  {Smith}}, \bibinfo {author} {\bibfnamefont {M.}~\bibnamefont {Lucca}},
  \bibinfo {author} {\bibfnamefont {V.}~\bibnamefont {Poulin}}, \bibinfo
  {author} {\bibfnamefont {G.~F.}\ \bibnamefont {Abellan}}, \bibinfo {author}
  {\bibfnamefont {L.}~\bibnamefont {Balkenhol}}, \bibinfo {author}
  {\bibfnamefont {K.}~\bibnamefont {Benabed}}, \bibinfo {author} {\bibfnamefont
  {S.}~\bibnamefont {Galli}}, \ and\ \bibinfo {author} {\bibfnamefont
  {R.}~\bibnamefont {Murgia}},\ }\href {\doibase 10.1103/PhysRevD.106.043526}
  {\bibfield  {journal} {\bibinfo  {journal} {Phys. Rev. D}\ }\textbf {\bibinfo
  {volume} {106}},\ \bibinfo {pages} {043526} (\bibinfo {year} {2022})},\
  \Eprint {http://arxiv.org/abs/2202.09379} {arXiv:2202.09379 [astro-ph.CO]}
  \BibitemShut {NoStop}%
\bibitem [{\citenamefont {Kamionkowski}\ and\ \citenamefont
  {Riess}(2023)}]{Kamionkowski:2022pkx}%
  \BibitemOpen
  \bibfield  {author} {\bibinfo {author} {\bibfnamefont {M.}~\bibnamefont
  {Kamionkowski}}\ and\ \bibinfo {author} {\bibfnamefont {A.~G.}\ \bibnamefont
  {Riess}},\ }\href {\doibase 10.1146/annurev-nucl-111422-024107} {\bibfield
  {journal} {\bibinfo  {journal} {Ann. Rev. Nucl. Part. Sci.}\ }\textbf
  {\bibinfo {volume} {73}},\ \bibinfo {pages} {153} (\bibinfo {year} {2023})},\
  \Eprint {http://arxiv.org/abs/2211.04492} {arXiv:2211.04492 [astro-ph.CO]}
  \BibitemShut {NoStop}%
\bibitem [{\citenamefont {Niedermann}\ and\ \citenamefont
  {Sloth}(2023)}]{Niedermann:2023ssr}%
  \BibitemOpen
  \bibfield  {author} {\bibinfo {author} {\bibfnamefont {F.}~\bibnamefont
  {Niedermann}}\ and\ \bibinfo {author} {\bibfnamefont {M.~S.}\ \bibnamefont
  {Sloth}},\ }\href {\doibase 10.1007/978-981-99-0177-7\_23} {\  (\bibinfo
  {year} {2023}),\ 10.1007/978-981-99-0177-7\_23},\ \Eprint
  {http://arxiv.org/abs/2307.03481} {arXiv:2307.03481 [hep-ph]} \BibitemShut
  {NoStop}%
\bibitem [{\citenamefont {Poulin}\ \emph {et~al.}(2023)\citenamefont {Poulin},
  \citenamefont {Smith},\ and\ \citenamefont {Karwal}}]{Poulin:2023lkg}%
  \BibitemOpen
  \bibfield  {author} {\bibinfo {author} {\bibfnamefont {V.}~\bibnamefont
  {Poulin}}, \bibinfo {author} {\bibfnamefont {T.~L.}\ \bibnamefont {Smith}}, \
  and\ \bibinfo {author} {\bibfnamefont {T.}~\bibnamefont {Karwal}},\ }\href
  {\doibase 10.1016/j.dark.2023.101348} {\bibfield  {journal} {\bibinfo
  {journal} {Phys. Dark Univ.}\ }\textbf {\bibinfo {volume} {42}},\ \bibinfo
  {pages} {101348} (\bibinfo {year} {2023})},\ \Eprint
  {http://arxiv.org/abs/2302.09032} {arXiv:2302.09032 [astro-ph.CO]}
  \BibitemShut {NoStop}%
\bibitem [{\citenamefont {Smith}\ and\ \citenamefont
  {Poulin}(2024)}]{Smith:2023oop}%
  \BibitemOpen
  \bibfield  {author} {\bibinfo {author} {\bibfnamefont {T.~L.}\ \bibnamefont
  {Smith}}\ and\ \bibinfo {author} {\bibfnamefont {V.}~\bibnamefont {Poulin}},\
  }\href {\doibase 10.1103/PhysRevD.109.103506} {\bibfield  {journal} {\bibinfo
   {journal} {Phys. Rev. D}\ }\textbf {\bibinfo {volume} {109}},\ \bibinfo
  {pages} {103506} (\bibinfo {year} {2024})},\ \Eprint
  {http://arxiv.org/abs/2309.03265} {arXiv:2309.03265 [astro-ph.CO]}
  \BibitemShut {NoStop}%
\bibitem [{\citenamefont {Cruz}\ \emph {et~al.}(2023)\citenamefont {Cruz},
  \citenamefont {Niedermann},\ and\ \citenamefont {Sloth}}]{Cruz:2023lmn}%
  \BibitemOpen
  \bibfield  {author} {\bibinfo {author} {\bibfnamefont {J.~S.}\ \bibnamefont
  {Cruz}}, \bibinfo {author} {\bibfnamefont {F.}~\bibnamefont {Niedermann}}, \
  and\ \bibinfo {author} {\bibfnamefont {M.~S.}\ \bibnamefont {Sloth}},\ }\href
  {\doibase 10.1088/1475-7516/2023/11/033} {\bibfield  {journal} {\bibinfo
  {journal} {JCAP}\ }\textbf {\bibinfo {volume} {11}},\ \bibinfo {pages} {033}
  (\bibinfo {year} {2023})},\ \Eprint {http://arxiv.org/abs/2305.08895}
  {arXiv:2305.08895 [astro-ph.CO]} \BibitemShut {NoStop}%
\bibitem [{\citenamefont {Eskilt}\ \emph {et~al.}(2023)\citenamefont {Eskilt},
  \citenamefont {Herold}, \citenamefont {Komatsu}, \citenamefont {Murai},
  \citenamefont {Namikawa},\ and\ \citenamefont {Naokawa}}]{Eskilt:2023nxm}%
  \BibitemOpen
  \bibfield  {author} {\bibinfo {author} {\bibfnamefont {J.~R.}\ \bibnamefont
  {Eskilt}}, \bibinfo {author} {\bibfnamefont {L.}~\bibnamefont {Herold}},
  \bibinfo {author} {\bibfnamefont {E.}~\bibnamefont {Komatsu}}, \bibinfo
  {author} {\bibfnamefont {K.}~\bibnamefont {Murai}}, \bibinfo {author}
  {\bibfnamefont {T.}~\bibnamefont {Namikawa}}, \ and\ \bibinfo {author}
  {\bibfnamefont {F.}~\bibnamefont {Naokawa}},\ }\href {\doibase
  10.1103/PhysRevLett.131.121001} {\bibfield  {journal} {\bibinfo  {journal}
  {Phys. Rev. Lett.}\ }\textbf {\bibinfo {volume} {131}},\ \bibinfo {pages}
  {121001} (\bibinfo {year} {2023})},\ \Eprint
  {http://arxiv.org/abs/2303.15369} {arXiv:2303.15369 [astro-ph.CO]}
  \BibitemShut {NoStop}%
\bibitem [{\citenamefont {Sharma}\ \emph {et~al.}(2024)\citenamefont {Sharma},
  \citenamefont {Das},\ and\ \citenamefont {Poulin}}]{Sharma:2023kzr}%
  \BibitemOpen
  \bibfield  {author} {\bibinfo {author} {\bibfnamefont {R.~K.}\ \bibnamefont
  {Sharma}}, \bibinfo {author} {\bibfnamefont {S.}~\bibnamefont {Das}}, \ and\
  \bibinfo {author} {\bibfnamefont {V.}~\bibnamefont {Poulin}},\ }\href
  {\doibase 10.1103/PhysRevD.109.043530} {\bibfield  {journal} {\bibinfo
  {journal} {Phys. Rev. D}\ }\textbf {\bibinfo {volume} {109}},\ \bibinfo
  {pages} {043530} (\bibinfo {year} {2024})},\ \Eprint
  {http://arxiv.org/abs/2309.00401} {arXiv:2309.00401 [astro-ph.CO]}
  \BibitemShut {NoStop}%
\bibitem [{\citenamefont {Efstathiou}\ \emph {et~al.}(2024)\citenamefont
  {Efstathiou}, \citenamefont {Rosenberg},\ and\ \citenamefont
  {Poulin}}]{Efstathiou:2023fbn}%
  \BibitemOpen
  \bibfield  {author} {\bibinfo {author} {\bibfnamefont {G.}~\bibnamefont
  {Efstathiou}}, \bibinfo {author} {\bibfnamefont {E.}~\bibnamefont
  {Rosenberg}}, \ and\ \bibinfo {author} {\bibfnamefont {V.}~\bibnamefont
  {Poulin}},\ }\href {\doibase 10.1103/PhysRevLett.132.221002} {\bibfield
  {journal} {\bibinfo  {journal} {Phys. Rev. Lett.}\ }\textbf {\bibinfo
  {volume} {132}},\ \bibinfo {pages} {221002} (\bibinfo {year} {2024})},\
  \Eprint {http://arxiv.org/abs/2311.00524} {arXiv:2311.00524 [astro-ph.CO]}
  \BibitemShut {NoStop}%
\bibitem [{\citenamefont {Gsponer}\ \emph {et~al.}(2024)\citenamefont
  {Gsponer}, \citenamefont {Zhao}, \citenamefont {Donald-McCann}, \citenamefont
  {Bacon}, \citenamefont {Koyama}, \citenamefont {Crittenden}, \citenamefont
  {Simon},\ and\ \citenamefont {Mueller}}]{Gsponer:2023wpm}%
  \BibitemOpen
  \bibfield  {author} {\bibinfo {author} {\bibfnamefont {R.}~\bibnamefont
  {Gsponer}}, \bibinfo {author} {\bibfnamefont {R.}~\bibnamefont {Zhao}},
  \bibinfo {author} {\bibfnamefont {J.}~\bibnamefont {Donald-McCann}}, \bibinfo
  {author} {\bibfnamefont {D.}~\bibnamefont {Bacon}}, \bibinfo {author}
  {\bibfnamefont {K.}~\bibnamefont {Koyama}}, \bibinfo {author} {\bibfnamefont
  {R.}~\bibnamefont {Crittenden}}, \bibinfo {author} {\bibfnamefont
  {T.}~\bibnamefont {Simon}}, \ and\ \bibinfo {author} {\bibfnamefont {E.-M.}\
  \bibnamefont {Mueller}},\ }\href {\doibase 10.1093/mnras/stae992} {\bibfield
  {journal} {\bibinfo  {journal} {Mon. Not. Roy. Astron. Soc.}\ }\textbf
  {\bibinfo {volume} {530}},\ \bibinfo {pages} {3075} (\bibinfo {year}
  {2024})},\ \Eprint {http://arxiv.org/abs/2312.01977} {arXiv:2312.01977
  [astro-ph.CO]} \BibitemShut {NoStop}%
\bibitem [{\citenamefont {Goldstein}\ \emph {et~al.}(2023)\citenamefont
  {Goldstein}, \citenamefont {Hill}, \citenamefont {Ir\v{s}i\v{c}},\ and\
  \citenamefont {Sherwin}}]{Goldstein:2023gnw}%
  \BibitemOpen
  \bibfield  {author} {\bibinfo {author} {\bibfnamefont {S.}~\bibnamefont
  {Goldstein}}, \bibinfo {author} {\bibfnamefont {J.~C.}\ \bibnamefont {Hill}},
  \bibinfo {author} {\bibfnamefont {V.}~\bibnamefont {Ir\v{s}i\v{c}}}, \ and\
  \bibinfo {author} {\bibfnamefont {B.~D.}\ \bibnamefont {Sherwin}},\ }\href
  {\doibase 10.1103/PhysRevLett.131.201001} {\bibfield  {journal} {\bibinfo
  {journal} {Phys. Rev. Lett.}\ }\textbf {\bibinfo {volume} {131}},\ \bibinfo
  {pages} {201001} (\bibinfo {year} {2023})},\ \Eprint
  {http://arxiv.org/abs/2303.00746} {arXiv:2303.00746 [astro-ph.CO]}
  \BibitemShut {NoStop}%
\bibitem [{\citenamefont {Toda}\ \emph
  {et~al.}(2024{\natexlab{b}})\citenamefont {Toda}, \citenamefont {Giar\`e},
  \citenamefont {\"Oz\"ulker}, \citenamefont {Di~Valentino},\ and\
  \citenamefont {Vagnozzi}}]{Toda:2024ncp}%
  \BibitemOpen
  \bibfield  {author} {\bibinfo {author} {\bibfnamefont {Y.}~\bibnamefont
  {Toda}}, \bibinfo {author} {\bibfnamefont {W.}~\bibnamefont {Giar\`e}},
  \bibinfo {author} {\bibfnamefont {E.}~\bibnamefont {\"Oz\"ulker}}, \bibinfo
  {author} {\bibfnamefont {E.}~\bibnamefont {Di~Valentino}}, \ and\ \bibinfo
  {author} {\bibfnamefont {S.}~\bibnamefont {Vagnozzi}},\ }\href {\doibase
  10.1016/j.dark.2024.101676} {\bibfield  {journal} {\bibinfo  {journal} {Phys.
  Dark Univ.}\ }\textbf {\bibinfo {volume} {46}},\ \bibinfo {pages} {101676}
  (\bibinfo {year} {2024}{\natexlab{b}})},\ \Eprint
  {http://arxiv.org/abs/2407.01173} {arXiv:2407.01173 [astro-ph.CO]}
  \BibitemShut {NoStop}%
\bibitem [{\citenamefont {Sol\`a~Peracaula}\ \emph {et~al.}(2019)\citenamefont
  {Sol\`a~Peracaula}, \citenamefont {G\'omez-Valent}, \citenamefont
  {de~Cruz~P\'erez},\ and\ \citenamefont
  {Moreno-Pulido}}]{SolaPeracaula:2019zsl}%
  \BibitemOpen
  \bibfield  {author} {\bibinfo {author} {\bibfnamefont {J.}~\bibnamefont
  {Sol\`a~Peracaula}}, \bibinfo {author} {\bibfnamefont {A.}~\bibnamefont
  {G\'omez-Valent}}, \bibinfo {author} {\bibfnamefont {J.}~\bibnamefont
  {de~Cruz~P\'erez}}, \ and\ \bibinfo {author} {\bibfnamefont {C.}~\bibnamefont
  {Moreno-Pulido}},\ }\href {\doibase 10.3847/2041-8213/ab53e9} {\bibfield
  {journal} {\bibinfo  {journal} {Astrophys. J. Lett.}\ }\textbf {\bibinfo
  {volume} {886}},\ \bibinfo {pages} {L6} (\bibinfo {year} {2019})},\ \Eprint
  {http://arxiv.org/abs/1909.02554} {arXiv:1909.02554 [astro-ph.CO]}
  \BibitemShut {NoStop}%
\bibitem [{\citenamefont {Braglia}\ \emph {et~al.}(2021)\citenamefont
  {Braglia}, \citenamefont {Ballardini}, \citenamefont {Finelli},\ and\
  \citenamefont {Koyama}}]{Braglia:2020auw}%
  \BibitemOpen
  \bibfield  {author} {\bibinfo {author} {\bibfnamefont {M.}~\bibnamefont
  {Braglia}}, \bibinfo {author} {\bibfnamefont {M.}~\bibnamefont {Ballardini}},
  \bibinfo {author} {\bibfnamefont {F.}~\bibnamefont {Finelli}}, \ and\
  \bibinfo {author} {\bibfnamefont {K.}~\bibnamefont {Koyama}},\ }\href
  {\doibase 10.1103/PhysRevD.103.043528} {\bibfield  {journal} {\bibinfo
  {journal} {Phys. Rev. D}\ }\textbf {\bibinfo {volume} {103}},\ \bibinfo
  {pages} {043528} (\bibinfo {year} {2021})},\ \Eprint
  {http://arxiv.org/abs/2011.12934} {arXiv:2011.12934 [astro-ph.CO]}
  \BibitemShut {NoStop}%
\bibitem [{\citenamefont {Benevento}\ \emph {et~al.}(2022)\citenamefont
  {Benevento}, \citenamefont {Kable}, \citenamefont {Addison},\ and\
  \citenamefont {Bennett}}]{Benevento:2022cql}%
  \BibitemOpen
  \bibfield  {author} {\bibinfo {author} {\bibfnamefont {G.}~\bibnamefont
  {Benevento}}, \bibinfo {author} {\bibfnamefont {J.~A.}\ \bibnamefont
  {Kable}}, \bibinfo {author} {\bibfnamefont {G.~E.}\ \bibnamefont {Addison}},
  \ and\ \bibinfo {author} {\bibfnamefont {C.~L.}\ \bibnamefont {Bennett}},\
  }\href {\doibase 10.3847/1538-4357/ac80fd} {\bibfield  {journal} {\bibinfo
  {journal} {Astrophys. J.}\ }\textbf {\bibinfo {volume} {935}},\ \bibinfo
  {pages} {156} (\bibinfo {year} {2022})},\ \Eprint
  {http://arxiv.org/abs/2202.09356} {arXiv:2202.09356 [astro-ph.CO]}
  \BibitemShut {NoStop}%
\bibitem [{\citenamefont {Kable}\ \emph {et~al.}(2023)\citenamefont {Kable},
  \citenamefont {Benevento}, \citenamefont {Addison},\ and\ \citenamefont
  {Bennett}}]{Kable:2023bsg}%
  \BibitemOpen
  \bibfield  {author} {\bibinfo {author} {\bibfnamefont {J.~A.}\ \bibnamefont
  {Kable}}, \bibinfo {author} {\bibfnamefont {G.}~\bibnamefont {Benevento}},
  \bibinfo {author} {\bibfnamefont {G.~E.}\ \bibnamefont {Addison}}, \ and\
  \bibinfo {author} {\bibfnamefont {C.~L.}\ \bibnamefont {Bennett}},\ }\href
  {\doibase 10.3847/1538-4357/acfed0} {\bibfield  {journal} {\bibinfo
  {journal} {Astrophys. J.}\ }\textbf {\bibinfo {volume} {959}},\ \bibinfo
  {pages} {143} (\bibinfo {year} {2023})},\ \Eprint
  {http://arxiv.org/abs/2307.12174} {arXiv:2307.12174 [astro-ph.CO]}
  \BibitemShut {NoStop}%
\bibitem [{\citenamefont {Franco~Abell\'an}\ \emph {et~al.}(2023)\citenamefont
  {Franco~Abell\'an}, \citenamefont {Braglia}, \citenamefont {Ballardini},
  \citenamefont {Finelli},\ and\ \citenamefont
  {Poulin}}]{FrancoAbellan:2023gec}%
  \BibitemOpen
  \bibfield  {author} {\bibinfo {author} {\bibfnamefont {G.}~\bibnamefont
  {Franco~Abell\'an}}, \bibinfo {author} {\bibfnamefont {M.}~\bibnamefont
  {Braglia}}, \bibinfo {author} {\bibfnamefont {M.}~\bibnamefont {Ballardini}},
  \bibinfo {author} {\bibfnamefont {F.}~\bibnamefont {Finelli}}, \ and\
  \bibinfo {author} {\bibfnamefont {V.}~\bibnamefont {Poulin}},\ }\href
  {\doibase 10.1088/1475-7516/2023/12/017} {\bibfield  {journal} {\bibinfo
  {journal} {JCAP}\ }\textbf {\bibinfo {volume} {12}},\ \bibinfo {pages} {017}
  (\bibinfo {year} {2023})},\ \Eprint {http://arxiv.org/abs/2308.12345}
  {arXiv:2308.12345 [astro-ph.CO]} \BibitemShut {NoStop}%
\bibitem [{\citenamefont {Erdem}(2024)}]{Erdem:2024vsr}%
  \BibitemOpen
  \bibfield  {author} {\bibinfo {author} {\bibfnamefont {R.}~\bibnamefont
  {Erdem}},\ }\href {\doibase 10.3390/universe10090338} {\bibfield  {journal}
  {\bibinfo  {journal} {Universe}\ }\textbf {\bibinfo {volume} {10}},\ \bibinfo
  {pages} {338} (\bibinfo {year} {2024})},\ \Eprint
  {http://arxiv.org/abs/2402.16791} {arXiv:2402.16791 [gr-qc]} \BibitemShut
  {NoStop}%
\bibitem [{\citenamefont {Vagnozzi}(2021)}]{Vagnozzi:2021gjh}%
  \BibitemOpen
  \bibfield  {author} {\bibinfo {author} {\bibfnamefont {S.}~\bibnamefont
  {Vagnozzi}},\ }\href {\doibase 10.1103/PhysRevD.104.063524} {\bibfield
  {journal} {\bibinfo  {journal} {Phys. Rev. D}\ }\textbf {\bibinfo {volume}
  {104}},\ \bibinfo {pages} {063524} (\bibinfo {year} {2021})},\ \Eprint
  {http://arxiv.org/abs/2105.10425} {arXiv:2105.10425 [astro-ph.CO]}
  \BibitemShut {NoStop}%
\bibitem [{\citenamefont {Shen}\ \emph {et~al.}(2024)\citenamefont {Shen},
  \citenamefont {Vogelsberger}, \citenamefont {Boylan-Kolchin}, \citenamefont
  {Tacchella},\ and\ \citenamefont {Naidu}}]{Shen:2024hpx}%
  \BibitemOpen
  \bibfield  {author} {\bibinfo {author} {\bibfnamefont {X.}~\bibnamefont
  {Shen}}, \bibinfo {author} {\bibfnamefont {M.}~\bibnamefont {Vogelsberger}},
  \bibinfo {author} {\bibfnamefont {M.}~\bibnamefont {Boylan-Kolchin}},
  \bibinfo {author} {\bibfnamefont {S.}~\bibnamefont {Tacchella}}, \ and\
  \bibinfo {author} {\bibfnamefont {R.~P.}\ \bibnamefont {Naidu}},\ }\href
  {\doibase 10.1093/mnras/stae1932} {\bibfield  {journal} {\bibinfo  {journal}
  {Mon. Not. Roy. Astron. Soc.}\ }\textbf {\bibinfo {volume} {533}},\ \bibinfo
  {pages} {3923} (\bibinfo {year} {2024})},\ \Eprint
  {http://arxiv.org/abs/2406.15548} {arXiv:2406.15548 [astro-ph.GA]}
  \BibitemShut {NoStop}%
\bibitem [{\citenamefont {Akaike}(1974)}]{Akaike}%
  \BibitemOpen
  \bibfield  {author} {\bibinfo {author} {\bibfnamefont {H.}~\bibnamefont
  {Akaike}},\ }\href {\doibase 10.1109/TAC.1974.1100705} {\bibfield  {journal}
  {\bibinfo  {journal} {IEEE Trans. Autom. Control}\ }\textbf {\bibinfo
  {volume} {19}},\ \bibinfo {pages} {716} (\bibinfo {year} {1974})}\BibitemShut
  {NoStop}%
\bibitem [{\citenamefont {Verdinelli}\ and\ \citenamefont
  {W.}(1995)}]{Verdineli:1995}%
  \BibitemOpen
  \bibfield  {author} {\bibinfo {author} {\bibfnamefont {I.}~\bibnamefont
  {Verdinelli}}\ and\ \bibinfo {author} {\bibfnamefont {L.}~\bibnamefont
  {W.}},\ }\href {\doibase 10.1080/00107510802066753} {\bibfield  {journal}
  {\bibinfo  {journal} {J. Am. Statist. Assoc.}\ }\textbf {\bibinfo {volume}
  {90}},\ \bibinfo {pages} {614} (\bibinfo {year} {1995})},\ \Eprint
  {http://arxiv.org/abs/0803.4089} {arXiv:0803.4089 [astro-ph]} \BibitemShut
  {NoStop}%
\bibitem [{\citenamefont {Kass}\ and\ \citenamefont
  {Raftery}(1995)}]{Kass:1995loi}%
  \BibitemOpen
  \bibfield  {author} {\bibinfo {author} {\bibfnamefont {R.~E.}\ \bibnamefont
  {Kass}}\ and\ \bibinfo {author} {\bibfnamefont {A.~E.}\ \bibnamefont
  {Raftery}},\ }\href {\doibase 10.1080/01621459.1995.10476572} {\bibfield
  {journal} {\bibinfo  {journal} {J. Am. Statist. Assoc.}\ }\textbf {\bibinfo
  {volume} {90}},\ \bibinfo {pages} {773} (\bibinfo {year} {1995})}\BibitemShut
  {NoStop}%
\bibitem [{\citenamefont {Marshall}\ \emph {et~al.}(2006)\citenamefont
  {Marshall}, \citenamefont {Rajguru},\ and\ \citenamefont
  {Slosar}}]{Marshall:2004zd}%
  \BibitemOpen
  \bibfield  {author} {\bibinfo {author} {\bibfnamefont {P.}~\bibnamefont
  {Marshall}}, \bibinfo {author} {\bibfnamefont {N.}~\bibnamefont {Rajguru}}, \
  and\ \bibinfo {author} {\bibfnamefont {A.}~\bibnamefont {Slosar}},\ }\href
  {\doibase 10.1103/PhysRevD.73.067302} {\bibfield  {journal} {\bibinfo
  {journal} {Phys. Rev. D}\ }\textbf {\bibinfo {volume} {73}},\ \bibinfo
  {pages} {067302} (\bibinfo {year} {2006})},\ \Eprint
  {http://arxiv.org/abs/astro-ph/0412535} {arXiv:astro-ph/0412535} \BibitemShut
  {NoStop}%
\bibitem [{\citenamefont {Spiegelhalter}\ \emph {et~al.}(2002)\citenamefont
  {Spiegelhalter}, \citenamefont {Best}, \citenamefont {Carlin},\ and\
  \citenamefont {van~der Linde}}]{DIC}%
  \BibitemOpen
  \bibfield  {author} {\bibinfo {author} {\bibfnamefont {D.~J.}\ \bibnamefont
  {Spiegelhalter}}, \bibinfo {author} {\bibfnamefont {N.~G.}\ \bibnamefont
  {Best}}, \bibinfo {author} {\bibfnamefont {B.~P.}\ \bibnamefont {Carlin}}, \
  and\ \bibinfo {author} {\bibfnamefont {A.}~\bibnamefont {van~der Linde}},\
  }\href {\doibase 10.1111/1467-9868.00353} {\bibfield  {journal} {\bibinfo
  {journal} {J. Roy. Stat. Soc.}\ }\textbf {\bibinfo {volume} {64}},\ \bibinfo
  {pages} {583} (\bibinfo {year} {2002})}\BibitemShut {NoStop}%
\bibitem [{\citenamefont {Trotta}(2008)}]{Trotta:2008qt}%
  \BibitemOpen
  \bibfield  {author} {\bibinfo {author} {\bibfnamefont {R.}~\bibnamefont
  {Trotta}},\ }\href {\doibase 10.1080/00107510802066753} {\bibfield  {journal}
  {\bibinfo  {journal} {Contemp. Phys.}\ }\textbf {\bibinfo {volume} {49}},\
  \bibinfo {pages} {71} (\bibinfo {year} {2008})},\ \Eprint
  {http://arxiv.org/abs/0803.4089} {arXiv:0803.4089 [astro-ph]} \BibitemShut
  {NoStop}%
\bibitem [{\citenamefont {Keeley}\ and\ \citenamefont
  {Shafieloo}(2022)}]{Keeley:2021dmx}%
  \BibitemOpen
  \bibfield  {author} {\bibinfo {author} {\bibfnamefont {R.~E.}\ \bibnamefont
  {Keeley}}\ and\ \bibinfo {author} {\bibfnamefont {A.}~\bibnamefont
  {Shafieloo}},\ }\href {\doibase 10.1093/mnras/stac1851} {\bibfield  {journal}
  {\bibinfo  {journal} {Mon. Not. Roy. Astron. Soc.}\ }\textbf {\bibinfo
  {volume} {515}},\ \bibinfo {pages} {293} (\bibinfo {year} {2022})},\ \Eprint
  {http://arxiv.org/abs/2111.04231} {arXiv:2111.04231 [astro-ph.CO]}
  \BibitemShut {NoStop}%
\bibitem [{\citenamefont {Vagnozzi}\ \emph {et~al.}(2018)\citenamefont
  {Vagnozzi}, \citenamefont {Dhawan}, \citenamefont {Gerbino}, \citenamefont
  {Freese}, \citenamefont {Goobar},\ and\ \citenamefont
  {Mena}}]{Vagnozzi:2018jhn}%
  \BibitemOpen
  \bibfield  {author} {\bibinfo {author} {\bibfnamefont {S.}~\bibnamefont
  {Vagnozzi}}, \bibinfo {author} {\bibfnamefont {S.}~\bibnamefont {Dhawan}},
  \bibinfo {author} {\bibfnamefont {M.}~\bibnamefont {Gerbino}}, \bibinfo
  {author} {\bibfnamefont {K.}~\bibnamefont {Freese}}, \bibinfo {author}
  {\bibfnamefont {A.}~\bibnamefont {Goobar}}, \ and\ \bibinfo {author}
  {\bibfnamefont {O.}~\bibnamefont {Mena}},\ }\href {\doibase
  10.1103/PhysRevD.98.083501} {\bibfield  {journal} {\bibinfo  {journal} {Phys.
  Rev. D}\ }\textbf {\bibinfo {volume} {98}},\ \bibinfo {pages} {083501}
  (\bibinfo {year} {2018})},\ \Eprint {http://arxiv.org/abs/1801.08553}
  {arXiv:1801.08553 [astro-ph.CO]} \BibitemShut {NoStop}%
\bibitem [{\citenamefont {G\'omez-Valent}\ and\ \citenamefont
  {Amendola}(2018)}]{Gomez-Valent:2018hwc}%
  \BibitemOpen
  \bibfield  {author} {\bibinfo {author} {\bibfnamefont {A.}~\bibnamefont
  {G\'omez-Valent}}\ and\ \bibinfo {author} {\bibfnamefont {L.}~\bibnamefont
  {Amendola}},\ }\href {\doibase 10.1088/1475-7516/2018/04/051} {\bibfield
  {journal} {\bibinfo  {journal} {JCAP}\ }\textbf {\bibinfo {volume} {04}},\
  \bibinfo {pages} {051} (\bibinfo {year} {2018})},\ \Eprint
  {http://arxiv.org/abs/1802.01505} {arXiv:1802.01505 [astro-ph.CO]}
  \BibitemShut {NoStop}%
\bibitem [{\citenamefont {Kreisch}\ \emph {et~al.}(2020)\citenamefont
  {Kreisch}, \citenamefont {Cyr-Racine},\ and\ \citenamefont
  {Dor\'e}}]{Kreisch:2019yzn}%
  \BibitemOpen
  \bibfield  {author} {\bibinfo {author} {\bibfnamefont {C.~D.}\ \bibnamefont
  {Kreisch}}, \bibinfo {author} {\bibfnamefont {F.-Y.}\ \bibnamefont
  {Cyr-Racine}}, \ and\ \bibinfo {author} {\bibfnamefont {O.}~\bibnamefont
  {Dor\'e}},\ }\href {\doibase 10.1103/PhysRevD.101.123505} {\bibfield
  {journal} {\bibinfo  {journal} {Phys. Rev. D}\ }\textbf {\bibinfo {volume}
  {101}},\ \bibinfo {pages} {123505} (\bibinfo {year} {2020})},\ \Eprint
  {http://arxiv.org/abs/1902.00534} {arXiv:1902.00534 [astro-ph.CO]}
  \BibitemShut {NoStop}%
\bibitem [{\citenamefont {Agrawal}\ \emph {et~al.}(2023)\citenamefont
  {Agrawal}, \citenamefont {Cyr-Racine}, \citenamefont {Pinner},\ and\
  \citenamefont {Randall}}]{Agrawal:2019lmo}%
  \BibitemOpen
  \bibfield  {author} {\bibinfo {author} {\bibfnamefont {P.}~\bibnamefont
  {Agrawal}}, \bibinfo {author} {\bibfnamefont {F.-Y.}\ \bibnamefont
  {Cyr-Racine}}, \bibinfo {author} {\bibfnamefont {D.}~\bibnamefont {Pinner}},
  \ and\ \bibinfo {author} {\bibfnamefont {L.}~\bibnamefont {Randall}},\ }\href
  {\doibase 10.1016/j.dark.2023.101347} {\bibfield  {journal} {\bibinfo
  {journal} {Phys. Dark Univ.}\ }\textbf {\bibinfo {volume} {42}},\ \bibinfo
  {pages} {101347} (\bibinfo {year} {2023})},\ \Eprint
  {http://arxiv.org/abs/1904.01016} {arXiv:1904.01016 [astro-ph.CO]}
  \BibitemShut {NoStop}%
\bibitem [{\citenamefont {Visinelli}\ \emph {et~al.}(2019)\citenamefont
  {Visinelli}, \citenamefont {Vagnozzi},\ and\ \citenamefont
  {Danielsson}}]{Visinelli:2019qqu}%
  \BibitemOpen
  \bibfield  {author} {\bibinfo {author} {\bibfnamefont {L.}~\bibnamefont
  {Visinelli}}, \bibinfo {author} {\bibfnamefont {S.}~\bibnamefont {Vagnozzi}},
  \ and\ \bibinfo {author} {\bibfnamefont {U.}~\bibnamefont {Danielsson}},\
  }\href {\doibase 10.3390/sym11081035} {\bibfield  {journal} {\bibinfo
  {journal} {Symmetry}\ }\textbf {\bibinfo {volume} {11}},\ \bibinfo {pages}
  {1035} (\bibinfo {year} {2019})},\ \Eprint {http://arxiv.org/abs/1907.07953}
  {arXiv:1907.07953 [astro-ph.CO]} \BibitemShut {NoStop}%
\bibitem [{\citenamefont {Jeffreys}(1939)}]{Jeffreys:1939xee}%
  \BibitemOpen
  \bibfield  {author} {\bibinfo {author} {\bibfnamefont {H.}~\bibnamefont
  {Jeffreys}},\ }\href@noop {} {\emph {\bibinfo {title} {{The Theory of
  Probability}}}},\ Oxford Classic Texts in the Physical Sciences\ (\bibinfo
  {year} {1939})\BibitemShut {NoStop}%
\bibitem [{\citenamefont {Knox}\ and\ \citenamefont
  {Millea}(2020)}]{Knox:2019rjx}%
  \BibitemOpen
  \bibfield  {author} {\bibinfo {author} {\bibfnamefont {L.}~\bibnamefont
  {Knox}}\ and\ \bibinfo {author} {\bibfnamefont {M.}~\bibnamefont {Millea}},\
  }\href {\doibase 10.1103/PhysRevD.101.043533} {\bibfield  {journal} {\bibinfo
   {journal} {Phys. Rev. D}\ }\textbf {\bibinfo {volume} {101}},\ \bibinfo
  {pages} {043533} (\bibinfo {year} {2020})},\ \Eprint
  {http://arxiv.org/abs/1908.03663} {arXiv:1908.03663 [astro-ph.CO]}
  \BibitemShut {NoStop}%
\bibitem [{\citenamefont {G\'omez-Valent}\ \emph
  {et~al.}(2024{\natexlab{b}})\citenamefont {G\'omez-Valent}, \citenamefont
  {Favale}, \citenamefont {Migliaccio},\ and\ \citenamefont
  {Sen}}]{Gomez-Valent:2023uof}%
  \BibitemOpen
  \bibfield  {author} {\bibinfo {author} {\bibfnamefont {A.}~\bibnamefont
  {G\'omez-Valent}}, \bibinfo {author} {\bibfnamefont {A.}~\bibnamefont
  {Favale}}, \bibinfo {author} {\bibfnamefont {M.}~\bibnamefont {Migliaccio}},
  \ and\ \bibinfo {author} {\bibfnamefont {A.~A.}\ \bibnamefont {Sen}},\ }\href
  {\doibase 10.1103/PhysRevD.109.023525} {\bibfield  {journal} {\bibinfo
  {journal} {Phys. Rev. D}\ }\textbf {\bibinfo {volume} {109}},\ \bibinfo
  {pages} {023525} (\bibinfo {year} {2024}{\natexlab{b}})},\ \Eprint
  {http://arxiv.org/abs/2309.07795} {arXiv:2309.07795 [astro-ph.CO]}
  \BibitemShut {NoStop}%
\bibitem [{\citenamefont {Bernui}\ \emph {et~al.}(2023)\citenamefont {Bernui},
  \citenamefont {Di~Valentino}, \citenamefont {Giar\`e}, \citenamefont
  {Kumar},\ and\ \citenamefont {Nunes}}]{Bernui:2023byc}%
  \BibitemOpen
  \bibfield  {author} {\bibinfo {author} {\bibfnamefont {A.}~\bibnamefont
  {Bernui}}, \bibinfo {author} {\bibfnamefont {E.}~\bibnamefont
  {Di~Valentino}}, \bibinfo {author} {\bibfnamefont {W.}~\bibnamefont
  {Giar\`e}}, \bibinfo {author} {\bibfnamefont {S.}~\bibnamefont {Kumar}}, \
  and\ \bibinfo {author} {\bibfnamefont {R.~C.}\ \bibnamefont {Nunes}},\ }\href
  {\doibase 10.1103/PhysRevD.107.103531} {\bibfield  {journal} {\bibinfo
  {journal} {Phys. Rev. D}\ }\textbf {\bibinfo {volume} {107}},\ \bibinfo
  {pages} {103531} (\bibinfo {year} {2023})},\ \Eprint
  {http://arxiv.org/abs/2301.06097} {arXiv:2301.06097 [astro-ph.CO]}
  \BibitemShut {NoStop}%
\bibitem [{\citenamefont {Akarsu}\ \emph {et~al.}(2023)\citenamefont {Akarsu},
  \citenamefont {Di~Valentino}, \citenamefont {Kumar}, \citenamefont {Nunes},
  \citenamefont {Vazquez},\ and\ \citenamefont {Yadav}}]{Akarsu:2023mfb}%
  \BibitemOpen
  \bibfield  {author} {\bibinfo {author} {\bibfnamefont {O.}~\bibnamefont
  {Akarsu}}, \bibinfo {author} {\bibfnamefont {E.}~\bibnamefont
  {Di~Valentino}}, \bibinfo {author} {\bibfnamefont {S.}~\bibnamefont {Kumar}},
  \bibinfo {author} {\bibfnamefont {R.~C.}\ \bibnamefont {Nunes}}, \bibinfo
  {author} {\bibfnamefont {J.~A.}\ \bibnamefont {Vazquez}}, \ and\ \bibinfo
  {author} {\bibfnamefont {A.}~\bibnamefont {Yadav}},\ }\href@noop {} {\
  (\bibinfo {year} {2023})},\ \Eprint {http://arxiv.org/abs/2307.10899}
  {arXiv:2307.10899 [astro-ph.CO]} \BibitemShut {NoStop}%
\bibitem [{\citenamefont {Anchordoqui}\ \emph {et~al.}(2024)\citenamefont
  {Anchordoqui}, \citenamefont {Antoniadis}, \citenamefont {Lust},
  \citenamefont {Noble},\ and\ \citenamefont {Soriano}}]{Anchordoqui:2024gfa}%
  \BibitemOpen
  \bibfield  {author} {\bibinfo {author} {\bibfnamefont {L.~A.}\ \bibnamefont
  {Anchordoqui}}, \bibinfo {author} {\bibfnamefont {I.}~\bibnamefont
  {Antoniadis}}, \bibinfo {author} {\bibfnamefont {D.}~\bibnamefont {Lust}},
  \bibinfo {author} {\bibfnamefont {N.~T.}\ \bibnamefont {Noble}}, \ and\
  \bibinfo {author} {\bibfnamefont {J.~F.}\ \bibnamefont {Soriano}},\ }\href
  {\doibase 10.1016/j.dark.2024.101715} {\bibfield  {journal} {\bibinfo
  {journal} {Phys. Dark Univ.}\ }\textbf {\bibinfo {volume} {46}},\ \bibinfo
  {pages} {101715} (\bibinfo {year} {2024})},\ \Eprint
  {http://arxiv.org/abs/2404.17334} {arXiv:2404.17334 [astro-ph.CO]}
  \BibitemShut {NoStop}%
\bibitem [{\citenamefont {Yadav}\ \emph {et~al.}(2025)\citenamefont {Yadav},
  \citenamefont {Kumar}, \citenamefont {Kibris},\ and\ \citenamefont
  {Akarsu}}]{Yadav:2024duq}%
  \BibitemOpen
  \bibfield  {author} {\bibinfo {author} {\bibfnamefont {A.}~\bibnamefont
  {Yadav}}, \bibinfo {author} {\bibfnamefont {S.}~\bibnamefont {Kumar}},
  \bibinfo {author} {\bibfnamefont {C.}~\bibnamefont {Kibris}}, \ and\ \bibinfo
  {author} {\bibfnamefont {O.}~\bibnamefont {Akarsu}},\ }\href {\doibase
  10.1088/1475-7516/2025/01/042} {\bibfield  {journal} {\bibinfo  {journal}
  {JCAP}\ }\textbf {\bibinfo {volume} {01}},\ \bibinfo {pages} {042} (\bibinfo
  {year} {2025})},\ \Eprint {http://arxiv.org/abs/2406.18496} {arXiv:2406.18496
  [astro-ph.CO]} \BibitemShut {NoStop}%
\bibitem [{\citenamefont {Dwivedi}\ and\ \citenamefont
  {H\"og\r{a}s}(2024)}]{Dwivedi:2024okk}%
  \BibitemOpen
  \bibfield  {author} {\bibinfo {author} {\bibfnamefont {S.}~\bibnamefont
  {Dwivedi}}\ and\ \bibinfo {author} {\bibfnamefont {M.}~\bibnamefont
  {H\"og\r{a}s}},\ }\href {\doibase 10.3390/universe10110406} {\bibfield
  {journal} {\bibinfo  {journal} {Universe}\ }\textbf {\bibinfo {volume}
  {10}},\ \bibinfo {pages} {406} (\bibinfo {year} {2024})},\ \Eprint
  {http://arxiv.org/abs/2407.04322} {arXiv:2407.04322 [astro-ph.CO]}
  \BibitemShut {NoStop}%
\bibitem [{\citenamefont {Vagnozzi}(2020)}]{Vagnozzi:2019ezj}%
  \BibitemOpen
  \bibfield  {author} {\bibinfo {author} {\bibfnamefont {S.}~\bibnamefont
  {Vagnozzi}},\ }\href {\doibase 10.1103/PhysRevD.102.023518} {\bibfield
  {journal} {\bibinfo  {journal} {Phys. Rev. D}\ }\textbf {\bibinfo {volume}
  {102}},\ \bibinfo {pages} {023518} (\bibinfo {year} {2020})},\ \Eprint
  {http://arxiv.org/abs/1907.07569} {arXiv:1907.07569 [astro-ph.CO]}
  \BibitemShut {NoStop}%
\bibitem [{\citenamefont {Ballesteros}\ \emph {et~al.}(2020)\citenamefont
  {Ballesteros}, \citenamefont {Notari},\ and\ \citenamefont
  {Rompineve}}]{Ballesteros:2020sik}%
  \BibitemOpen
  \bibfield  {author} {\bibinfo {author} {\bibfnamefont {G.}~\bibnamefont
  {Ballesteros}}, \bibinfo {author} {\bibfnamefont {A.}~\bibnamefont {Notari}},
  \ and\ \bibinfo {author} {\bibfnamefont {F.}~\bibnamefont {Rompineve}},\
  }\href {\doibase 10.1088/1475-7516/2020/11/024} {\bibfield  {journal}
  {\bibinfo  {journal} {JCAP}\ }\textbf {\bibinfo {volume} {11}},\ \bibinfo
  {pages} {024} (\bibinfo {year} {2020})},\ \Eprint
  {http://arxiv.org/abs/2004.05049} {arXiv:2004.05049 [astro-ph.CO]}
  \BibitemShut {NoStop}%
\bibitem [{\citenamefont {Jiang}\ \emph {et~al.}(2025)\citenamefont {Jiang},
  \citenamefont {Giar\`e}, \citenamefont {Gariazzo}, \citenamefont {Dainotti},
  \citenamefont {Di~Valentino}, \citenamefont {Mena}, \citenamefont {Pedrotti},
  \citenamefont {da~Costa},\ and\ \citenamefont {Vagnozzi}}]{Jiang:2024viw}%
  \BibitemOpen
  \bibfield  {author} {\bibinfo {author} {\bibfnamefont {J.-Q.}\ \bibnamefont
  {Jiang}}, \bibinfo {author} {\bibfnamefont {W.}~\bibnamefont {Giar\`e}},
  \bibinfo {author} {\bibfnamefont {S.}~\bibnamefont {Gariazzo}}, \bibinfo
  {author} {\bibfnamefont {M.~G.}\ \bibnamefont {Dainotti}}, \bibinfo {author}
  {\bibfnamefont {E.}~\bibnamefont {Di~Valentino}}, \bibinfo {author}
  {\bibfnamefont {O.}~\bibnamefont {Mena}}, \bibinfo {author} {\bibfnamefont
  {D.}~\bibnamefont {Pedrotti}}, \bibinfo {author} {\bibfnamefont {S.~S.}\
  \bibnamefont {da~Costa}}, \ and\ \bibinfo {author} {\bibfnamefont
  {S.}~\bibnamefont {Vagnozzi}},\ }\href {\doibase
  10.1088/1475-7516/2025/01/153} {\bibfield  {journal} {\bibinfo  {journal}
  {JCAP}\ }\textbf {\bibinfo {volume} {01}},\ \bibinfo {pages} {153} (\bibinfo
  {year} {2025})},\ \Eprint {http://arxiv.org/abs/2407.18047} {arXiv:2407.18047
  [astro-ph.CO]} \BibitemShut {NoStop}%
\bibitem [{\citenamefont {Escudero}\ and\ \citenamefont
  {Abazajian}(2025)}]{Escudero:2024uea}%
  \BibitemOpen
  \bibfield  {author} {\bibinfo {author} {\bibfnamefont {H.~G.}\ \bibnamefont
  {Escudero}}\ and\ \bibinfo {author} {\bibfnamefont {K.~N.}\ \bibnamefont
  {Abazajian}},\ }\href {\doibase 10.1103/PhysRevD.111.043520} {\bibfield
  {journal} {\bibinfo  {journal} {Phys. Rev. D}\ }\textbf {\bibinfo {volume}
  {111}},\ \bibinfo {pages} {043520} (\bibinfo {year} {2025})},\ \Eprint
  {http://arxiv.org/abs/2412.05451} {arXiv:2412.05451 [astro-ph.CO]}
  \BibitemShut {NoStop}%
\bibitem [{\citenamefont {Herold}\ and\ \citenamefont
  {Kamionkowski}(2025)}]{Herold:2024nvk}%
  \BibitemOpen
  \bibfield  {author} {\bibinfo {author} {\bibfnamefont {L.}~\bibnamefont
  {Herold}}\ and\ \bibinfo {author} {\bibfnamefont {M.}~\bibnamefont
  {Kamionkowski}},\ }\href {\doibase 10.1103/PhysRevD.111.083518} {\bibfield
  {journal} {\bibinfo  {journal} {Phys. Rev. D}\ }\textbf {\bibinfo {volume}
  {111}},\ \bibinfo {pages} {083518} (\bibinfo {year} {2025})},\ \Eprint
  {http://arxiv.org/abs/2412.03546} {arXiv:2412.03546 [astro-ph.CO]}
  \BibitemShut {NoStop}%
\bibitem [{\citenamefont {Loverde}\ and\ \citenamefont
  {Weiner}(2024)}]{Loverde:2024nfi}%
  \BibitemOpen
  \bibfield  {author} {\bibinfo {author} {\bibfnamefont {M.}~\bibnamefont
  {Loverde}}\ and\ \bibinfo {author} {\bibfnamefont {Z.~J.}\ \bibnamefont
  {Weiner}},\ }\href {\doibase 10.1088/1475-7516/2024/12/048} {\bibfield
  {journal} {\bibinfo  {journal} {JCAP}\ }\textbf {\bibinfo {volume} {12}},\
  \bibinfo {pages} {048} (\bibinfo {year} {2024})},\ \Eprint
  {http://arxiv.org/abs/2410.00090} {arXiv:2410.00090 [astro-ph.CO]}
  \BibitemShut {NoStop}%
\bibitem [{\citenamefont {Allali}\ and\ \citenamefont
  {Notari}(2024)}]{Allali:2024aiv}%
  \BibitemOpen
  \bibfield  {author} {\bibinfo {author} {\bibfnamefont {I.~J.}\ \bibnamefont
  {Allali}}\ and\ \bibinfo {author} {\bibfnamefont {A.}~\bibnamefont
  {Notari}},\ }\href {\doibase 10.1088/1475-7516/2024/12/020} {\bibfield
  {journal} {\bibinfo  {journal} {JCAP}\ }\textbf {\bibinfo {volume} {12}},\
  \bibinfo {pages} {020} (\bibinfo {year} {2024})},\ \Eprint
  {http://arxiv.org/abs/2406.14554} {arXiv:2406.14554 [astro-ph.CO]}
  \BibitemShut {NoStop}%
\bibitem [{\citenamefont {Ivanov}\ \emph {et~al.}(2020)\citenamefont {Ivanov},
  \citenamefont {McDonough}, \citenamefont {Hill}, \citenamefont {Simonovi\'c},
  \citenamefont {Toomey}, \citenamefont {Alexander},\ and\ \citenamefont
  {Zaldarriaga}}]{Ivanov:2020ril}%
  \BibitemOpen
  \bibfield  {author} {\bibinfo {author} {\bibfnamefont {M.~M.}\ \bibnamefont
  {Ivanov}}, \bibinfo {author} {\bibfnamefont {E.}~\bibnamefont {McDonough}},
  \bibinfo {author} {\bibfnamefont {J.~C.}\ \bibnamefont {Hill}}, \bibinfo
  {author} {\bibfnamefont {M.}~\bibnamefont {Simonovi\'c}}, \bibinfo {author}
  {\bibfnamefont {M.~W.}\ \bibnamefont {Toomey}}, \bibinfo {author}
  {\bibfnamefont {S.}~\bibnamefont {Alexander}}, \ and\ \bibinfo {author}
  {\bibfnamefont {M.}~\bibnamefont {Zaldarriaga}},\ }\href {\doibase
  10.1103/PhysRevD.102.103502} {\bibfield  {journal} {\bibinfo  {journal}
  {Phys. Rev. D}\ }\textbf {\bibinfo {volume} {102}},\ \bibinfo {pages}
  {103502} (\bibinfo {year} {2020})},\ \Eprint
  {http://arxiv.org/abs/2006.11235} {arXiv:2006.11235 [astro-ph.CO]}
  \BibitemShut {NoStop}%
\bibitem [{\citenamefont {Giar\`e}(2024)}]{Giare:2024ocw}%
  \BibitemOpen
  \bibfield  {author} {\bibinfo {author} {\bibfnamefont {W.}~\bibnamefont
  {Giar\`e}},\ }\href@noop {} {\  (\bibinfo {year} {2024})},\ \Eprint
  {http://arxiv.org/abs/2409.17074} {arXiv:2409.17074 [astro-ph.CO]}
  \BibitemShut {NoStop}%
\bibitem [{\citenamefont {Khalife}\ \emph {et~al.}(2024)\citenamefont
  {Khalife}, \citenamefont {Zanjani}, \citenamefont {Galli}, \citenamefont
  {G\"unther}, \citenamefont {Lesgourgues},\ and\ \citenamefont
  {Benabed}}]{Khalife:2023qbu}%
  \BibitemOpen
  \bibfield  {author} {\bibinfo {author} {\bibfnamefont {A.~R.}\ \bibnamefont
  {Khalife}}, \bibinfo {author} {\bibfnamefont {M.~B.}\ \bibnamefont
  {Zanjani}}, \bibinfo {author} {\bibfnamefont {S.}~\bibnamefont {Galli}},
  \bibinfo {author} {\bibfnamefont {S.}~\bibnamefont {G\"unther}}, \bibinfo
  {author} {\bibfnamefont {J.}~\bibnamefont {Lesgourgues}}, \ and\ \bibinfo
  {author} {\bibfnamefont {K.}~\bibnamefont {Benabed}},\ }\href {\doibase
  10.1088/1475-7516/2024/04/059} {\bibfield  {journal} {\bibinfo  {journal}
  {JCAP}\ }\textbf {\bibinfo {volume} {04}},\ \bibinfo {pages} {059} (\bibinfo
  {year} {2024})},\ \Eprint {http://arxiv.org/abs/2312.09814} {arXiv:2312.09814
  [astro-ph.CO]} \BibitemShut {NoStop}%
\bibitem [{\citenamefont {Herold}\ and\ \citenamefont
  {Ferreira}(2023)}]{Herold:2022iib}%
  \BibitemOpen
  \bibfield  {author} {\bibinfo {author} {\bibfnamefont {L.}~\bibnamefont
  {Herold}}\ and\ \bibinfo {author} {\bibfnamefont {E.~G.~M.}\ \bibnamefont
  {Ferreira}},\ }\href {\doibase 10.1103/PhysRevD.108.043513} {\bibfield
  {journal} {\bibinfo  {journal} {Phys. Rev. D}\ }\textbf {\bibinfo {volume}
  {108}},\ \bibinfo {pages} {043513} (\bibinfo {year} {2023})},\ \Eprint
  {http://arxiv.org/abs/2210.16296} {arXiv:2210.16296 [astro-ph.CO]}
  \BibitemShut {NoStop}%
\end{thebibliography}%

\begin{figure*}[t!] 
	\centering
	\includegraphics[width=\textwidth]{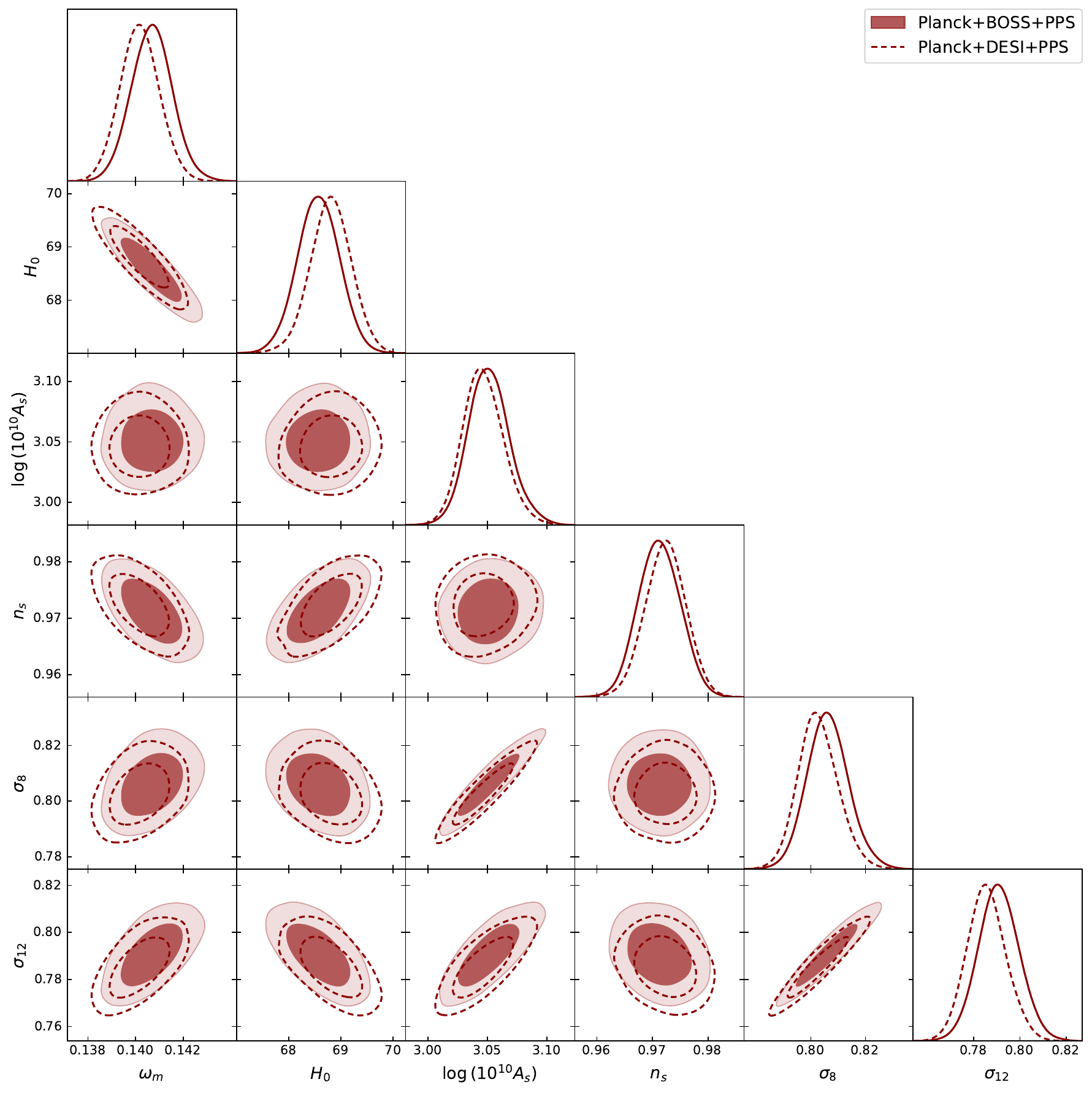}
	\caption{Triangular plot for $\Lambda$CDM obtained from the fitting analysis with Planck+BOSS+PPS (solid contours) and Planck+DESI+PPS (dashed contours), cf. Sec. \ref{sec:data}.  The parameter $\omega_m=\Omega_m^0 h^2$ is the sum of the baryon and CDM reduced density parameters, $\omega_b$ and $\omega_{cdm}$ respectively. The results for $\sigma_8$ and $\sigma_{12}$ are those computed at $z=0$.}
	\label{fig:LCDM}
\end{figure*}

\begin{figure*}[htp]
	\centering
	\includegraphics[width=\textwidth]{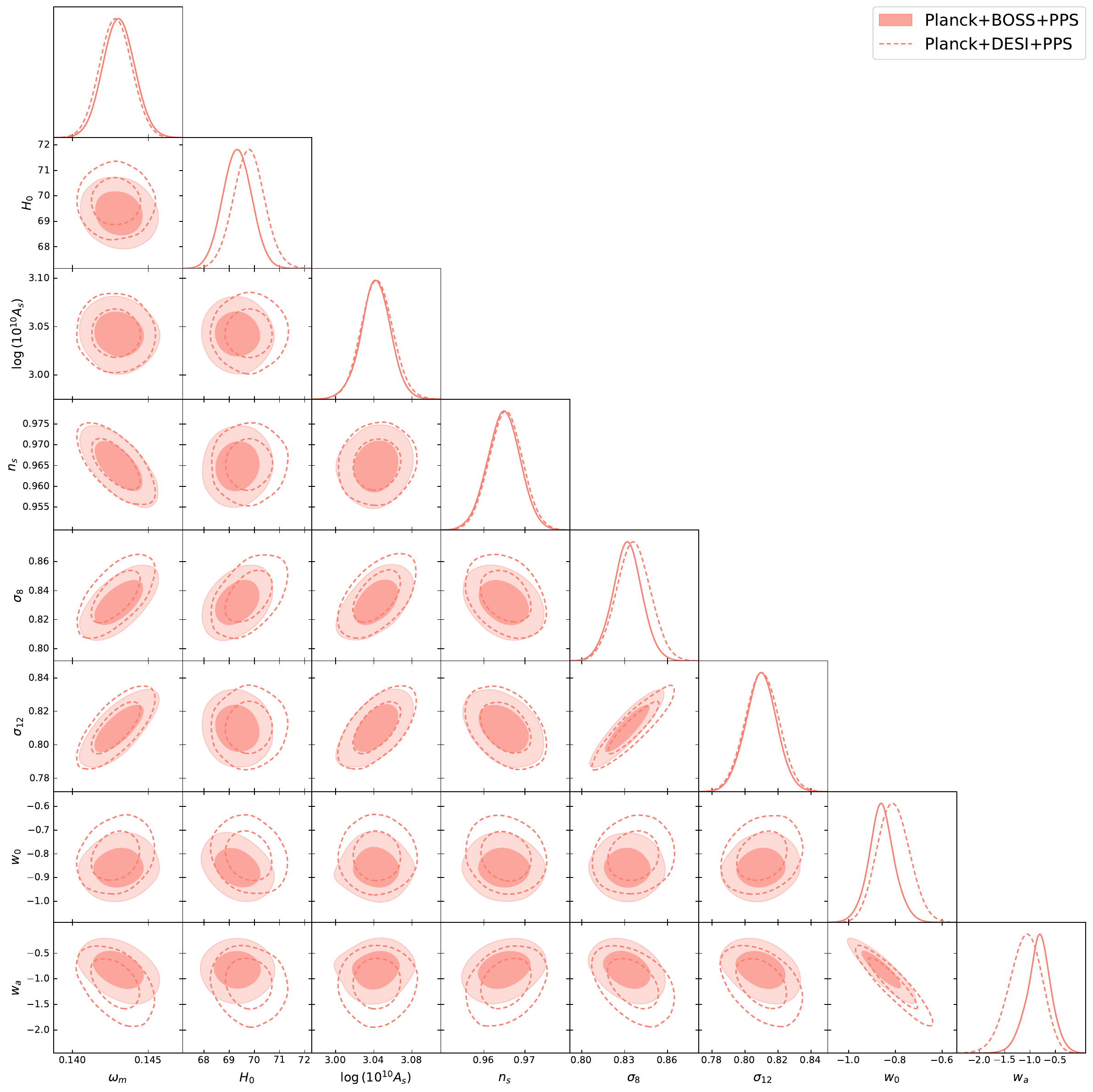}
	\caption{Triangular plot for $w_0w_a$CDM obtained from the fitting analysis with Planck+BOSS+PPS (solid contours) and Planck+DESI+PPS (dashed contours), cf. Sec. \ref{sec:data}.  The parameter $\omega_m=\Omega_m^0 h^2$ is the sum of the baryon and CDM reduced density parameters, $\omega_b$ and $\omega_{cdm}$ respectively. The results for $\sigma_8$ and $\sigma_{12}$ are those computed at $z=0$.}
	\label{fig:wwa}
\end{figure*}

\begin{figure*}[htp]
	\centering
	\includegraphics[width=\textwidth]{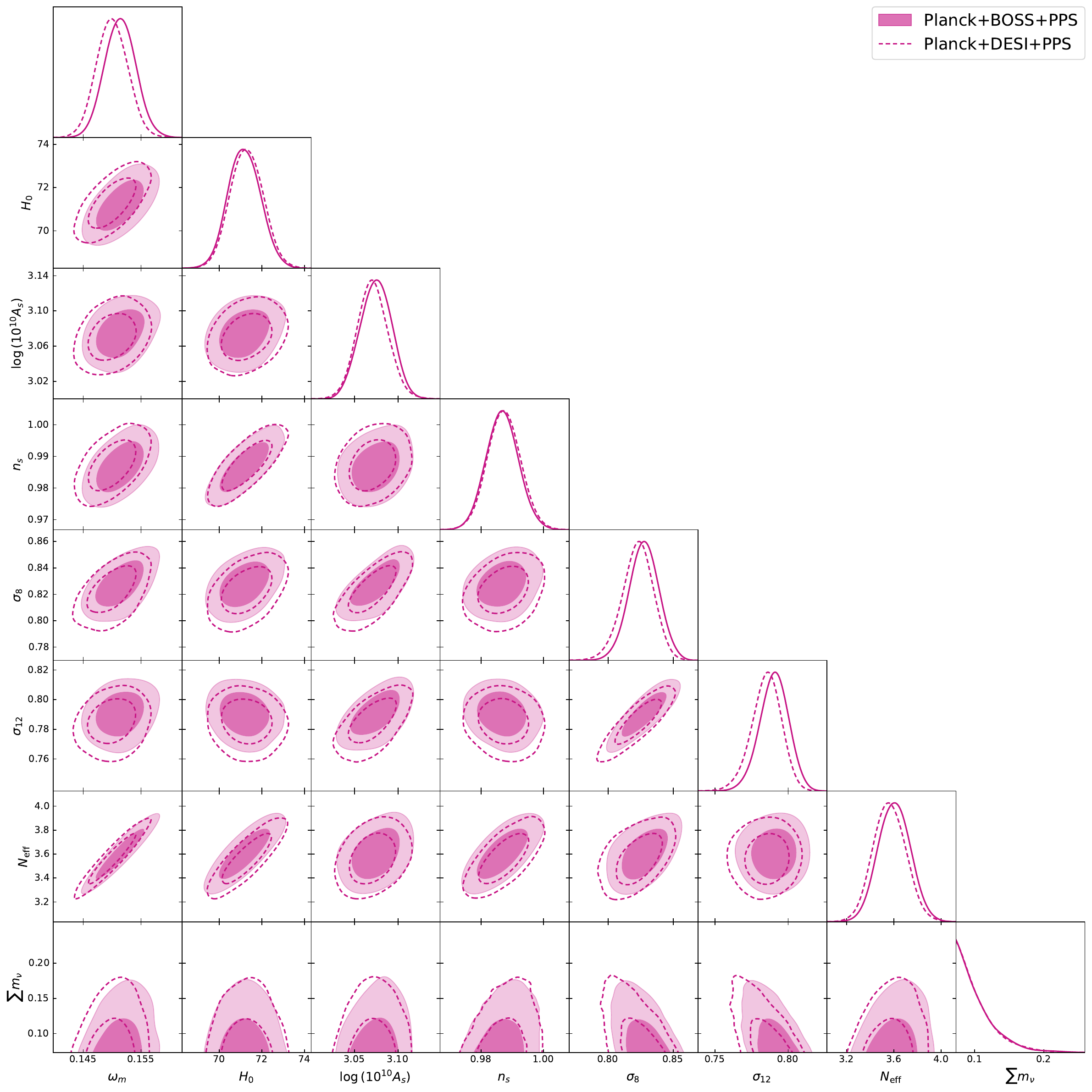}
	\caption{Triangular plot for $\Lambda$CDM$+N_{\rm eff}+\Sigma m_\nu$ obtained from the fitting analysis with Planck+BOSS+PPS (solid contours) and Planck+DESI+PPS (dashed contours), cf. Sec. \ref{sec:data}.  The parameter $\omega_m=\Omega_m^0 h^2$ is the sum of the baryon and CDM reduced density parameters, $\omega_b$ and $\omega_{cdm}$ respectively. The results for $\sigma_8$ and $\sigma_{12}$ are those computed at $z=0$.}
	\label{fig:nnumnu}
\end{figure*}

\end{document}